  \def\versionno{ OMC -- version 12.3 -- by jf -- 28.1.98  }
\def\messs
{\message{ ********************************************************** }}
\let\dl=\bf
\catcode`\@=11

\newif\if@fewtab\@fewtabtrue
{\count255=\time\divide\count255 by 60
\xdef\hourmin{\number\count255}
\multiply\count255 by-60\advance\count255 by\time
\xdef\hourmin{\hourmin:\ifnum\count255<10 0\fi\the\count255}}
\def\ps@draft{\let\@mkboth\@gobbletwo
    \def\@oddhead{}
    \def\@oddfoot{\hbox to 7 cm{\tiny \versionno
       \hfil}\hskip -7cm\hfil\rm\thepage \hfil}
    \def\@evenhead{}\let\@evenfoot\@oddfoot}

\def\draftcite#1{\ifnum\draftcontrol=1#1\else{}\fi}
\def\@lbibitem[#1]#2{\item{}\hskip -3cm \hbox to 2cm
{\hfil$\scriptstyle\draftcite{#2}$}\hskip
1cm[\@biblabel{#1}]\if@filesw
     {\def\protect##1{\string ##1\space}\immediate
      \write\@auxout{\string\bibcite{#2}{#1}}}\fi\ignorespaces}

\def\@bibitem#1{\item\hskip -3cm \hbox to 2cm
{\hfil {\footnotesize\draftcite{#1}}}\hskip 1cm
\if@filesw \immediate\write\@auxout
       {\string\bibcite{#1}{\the\value{\@listctr}}}\fi\ignorespaces}

\newenvironment{internal}
{\par\bigskip\noindent \underline{\sc Technical details:}
\newline\noindent\footnotesize}{\par
\normalsize\noindent{\sc End of technical details.}\bigskip}

                    \global\def\extension{1}    %  1 = extended version

\ifnum\extension=0
\newcommand\binternal[1]{}\def\einternal{}
\fi
\ifnum\extension=1
\def\binternal{\begin{internal}}\def\einternal{\end{internal}} \fi

\newcommand\Ac[2]  {A^{#1,#2}_{}}
\newcommand\Acb[2] {\bar A^{#1,#2}_{}}

\def\aff           {affine Lie algebra}
\newcommand\alpa[3]{{\alpha^{(#1,#2)}_{#3}}}
\newcommand\alpab[3]{{\alphab^{(#1,#2)}_{#3}}}
\def\alphab        {{\bar\alpha}}
\def\alphaO        {\brev{\alpha}}
\newcommand\als[1] {\alpha^{(#1)}}
\newcommand\alv[1] {\alpha^{(#1)^\Vee}}
\def\AO            {\brev{A}}
\def\aO            {\brev{a}}
\newcommand\AOc[2] {\brev A^{#1,#2}_{}}
\newcommand\ASect[2] {\appendix\sect{#1}\label{s.#2}}
\def\atcha         {automorphism-twined character}
\newcommand\av[1]{a^{\Vee}_{#1}}
\def\avi           {a^{\Vee}_{i}}
\def\avj           {a^{\Vee}_{j}}
\newcommand\avO[1]{\brev a^{\Vee}_{#1}}
\newcommand\awO[1]{\brev a_{#1}}
\def\betab         {{\bar\beta}}
\def\BB            {{\hat B}}
\def\BO            {\brev{B}}
\def\be            {\begin{equation}}

\def\bfe           {{\bf1}}
\def\bminus        {\mbox{${\cal B}_-$}}
\def\Bntilde       {{\tilde B_n\twtw}}
\def\brev          {\breve}

\def\calh          {{\cal H}}

\def\cd            {\Delta}
\def\cdo           {\brev\Delta}
\def\cft           {conformal field theory}
\def\cfts          {conformal field theories}
\def\Chi           {{\cal V}}
\def\CHI           {\Chi_{}^{(\omega)}}
\def\chii          {\raisebox{.15em}{$\chi$}}
\def\chil          {\chii_\Lambda^{(\omega)}}
\def\chIl          {\chi_\Lambda^{(\omega)}}
\def\Chil          {\Chi_\Lambda^{(\omega)}}
\def\CHil          {\Chi_\lambda^{(\omega)}}
\def\Chili         {\Chi_{\lambda_i}^{(\omega)}}
\def\chilj         {\chii_{\lambda_j}^{(\omega)}}

\def\Chiltau       {\Chi^{(\omega)}(\tau)}
\def\ChilTau       {\Chi^{(\omega)}(q)}

\def\ChilnTau      {\Chi_{}^{(\omega)}(q^N)}
\def\chiO          {\raisebox{.15em}{$\brev\chi$}}
\def\ChiO          {\brev\Chi}

\newcommand\Cite[1]{[\,{#1}\,]}
\def\cO            {{\brev{c}}}
\def\CO            {{\bar C_2}}
\def\coo           {\mbox{$\Gamma_{\!00}$}}
\def\complex       {{\dl C}}
\def\conj          {\mbox{$\gamma$}}
\def\csa           {Cartan subalgebra}
\def\ctwo          {{\cal C}_2}
\def\ctwob         {\overline{\cal C}_2}

\def\deltaO        {{\brev\delta}}

\newcommand\dfrac[2] {{\displaystyle\frac{#1}{#2}}}
\def\diag          {{\rm diag}}
\def\dO            {\brev{d}}
\def\DO            {\brev{D}}
\def\dpth          {{\rm dp}}
\def\dprod         {\displaystyle\prod}
\def\drac          {\displaystyle\frac }
\def\dsum          {\displaystyle\sum}
\def\dyd           {Dynkin diagram}
\def\ee            {\end{equation}}
\def\eE            {{\rm e}}
\newcommand\EE[2] {E^{#1}_{#2}}
\def\eins          {\id} 
\def\el            {\ell}
\def\ela           {{\ell_{\bar\alpha}}}

\def\Ene           {{\cal E}_1^{(\vec n)}}
\def\Enm           {{\cal E}_{}^{(\vec n,\vec m)}}
\def\Emz           {{\cal E}_2^{(\vec m)}}
\def\Ete           {\tilde{\cal E}_1^{(n_0,n_1,n')}}
\def\eps           {\epsilon}
\def\epsh          {\hat\epsilon}
\def\epsO          {\brev\epsilon}

\def\eq            {\,{=}\,}
\newcommand\erf[1]{(\ref{#1})}
\def\etaa          {{\eta_\alphab^{}}}
\newcommand\fcft[3]{{{#1}^{\mskip-#3 mu\raise #2 pt\hbox{${\scriptstyle\circ}$}}}}
\def\findim        {finite-dimensional}
\newcommand{\fline}[1]{\vfill\noindent ------------------\\[1 mm]}
\newcommand\Frac[2]{\mbox{\large$\frac{#1}{#2}$}}
\def\fS            {{\fcft{S}{3.5}{9.5}}}
\def\fT            {{\fcft{T}{3.5}{12}}}
\def\futnot#1      {\ifnum\draftcontrol=1
                   \footnote{~{\sc internal footnote:} #1}\ \fi}
\def\futnote#1     {\footnote{~#1}\ }
\def\g             {{\sf g}}

\def\gb            {\mbox{$\bar\g$}}

\newcommand\Gb[2]  {\bar G_{#1,#2}}
\def\gd            {\mbox{$\hat\g$}}
\def\gD            {\g_D^{}}
\def\gdeins        {\g_D^{(0)}}
\def\geins         {\g_\circ^{(0)}}
\def\gf            {\mbox{$\fcft{\g}{1.15}{11.1}$}}
\def\gfb           {\mbox{$\bar{\fcft\g{1.7}{11.2}}$}}
\newcommand\GG[2]  {B^{#1,#2}}
\newcommand\GGb[2] {\bar B^{#1,#2}}
\newcommand\GGO[2] {\brev B^{#1,#2}}
\def\gh            {\mbox{$\hat\g$}}
\def\gho           {\mbox{$\hat\g_\circ$}}
\def\gH            {\g_\J^{}}
\def\gK            {\g_K^{}}
\def\gkeins        {\g_K^{(0)}}
\def\gnull         {\mbox{$\overline{\brev{\g}}$}}
\def\go            {\g_\circ}
\def\goeins        {\g_\circ^{(0)}}
\def\gO            {\mbox{$\brev{\g}$}}
\def\GO            {\brev G}
\def\gOb           {\mbox{$\overline{\brev{\g}}$}}

\newcommand\Gobb[2]{\overline{\brev G}_{#1,#2}}
\def\goh           {\hat\g_\circ}
\def\goheins       {\hat\g_\circ^{(0)}}

\def\gOo           {\brev\g_\circ}
\def\gOoh          {\hat{\brev\g}_\circ}
\newcommand\gstar[1] {\g_\circ^{\star (#1)}}
\def\gv            {\mbox{$g_{}^\Vee$}}
\def\gvf           {{\fcft{g}{1.25}{9.4}{}^{\Vee}}}
\def\gvO           {\mbox{$\brev g_{}^\Vee$}}
\def\gvo           {{\brev g^{\Vee}}}
\def\h             {{\sf h}}
\def\half          {\mbox{$\frac12\,$}}
\newcommand\HH[2]{H^{#1}_{#2}}
\def\hil           {\mbox{$\cal H$}}
\def\hill          {\mbox{${\cal H}_\Lambda$}}
\def\hilL          {{{\cal H}_\Lambda}}
\def\hillo         {\mbox{${\cal H}_{\omtLa}$}}

\def\hillt         {\mbox{${\cal H}^\omega_\Lambda$}}

\def\HO            {\brev{H}}

\def\hsa           {horizontal subalgebra}
\newcommand\hsp[1] {\mbox{\hspace{#1 mm}}}
\def\hw            {highest weight}
\def\hwm           {highest weight module}

\def\hwv           {highest weight vector}
\def\hy            {$\mbox{-\hspace{-.66 mm}-}$}
\def\id            {\sl id}
\def\ii            {{\rm i}}
\def\ihwm          {irreducible highest weight module}

\def\I             {I}
\def\Ib            {\bar I}
\def\ini           {\in\I}
\def\inib          {\in\Ib}
\def\inio          {\in\IO}
\def\IO            {{\brev{I}}}
\def\irmod         {irreducible module}

\def\J             {J}
\newcommand\JJ[1]  {\mbox{$\J^{#1}$}}
\def\Jc            {\mbox{$\sigma_{\rm c}$}}
\def\Js            {\mbox{$\sigma_{\rm s}$}}
\def\js            {\mbox{$\sigma_{\rm s}$}}
\def\jv            {\mbox{$\sigma_{\rm v}$}}
\def\Jv            {\mbox{$\sigma_{\rm v}$}}
\def\K             {\kappa}
\def\kma           {Kac\hy Moo\-dy algebra}
\def\KO            {{\brev K}}
\def\KOa           {{\brev\kappa}}
\def\kpf           {Kac\hy Peterson formula}
\def\kv            {\mbox{$k_{}^{\Vee}$}}
\def\kvf           {{\fcft{k}{3.3}{10}{}^{\Vee}}}
\def\kvl           {\mbox{$k_\lambda^{\Vee}$}}
\def\kvL           {\mbox{$k_\Lambda^{\Vee}$}}
\def\kvlo          {\mbox{$\brev k_\lambda^{\Vee}$}}
\def\kvO           {\mbox{$\brev k_{}^{\Vee}$}}
\def\kvo           {{\brev k^{\Vee}}}
\def\kvol          {{\brev k^{\Vee}_{\lambda}}}
\def\kvom          {{\brev k^{\Vee}_{\mu}}}

\newcommand\lab[1] {\mbox{$\Lambdab_{(#1)}$}}
\def\labda         {\xi}
\long\def\labl#1   {\label{#1}\ee \ifnum\draftcontrol=1
                   \mbox{ }\\[-12 mm]\query{#1}\\[5 mm] \fi}
\newcommand\lambdab{\bar\lambda}
\newcommand\Lambdab{\bar\Lambda}
\def\Lambdaf       {{\fcft\Lambda{3.4}{11.2}}}
\def\lambdaf       {{\fcft\lambda{3.2}{11.2}}}

\def\lambdafb      {\bar{\fcft\lambda{3.4}{11.2}}}
\def\Lambdafp      {{\fcft\Lambda{3.4}{11.2}{}'}}
\def\LambdaO       {{\brev\Lambda}}
\def\lambdaO       {{\brev\lambda}}
\def\LambdaOp      {{\brev\Lambda{}'}}
\def\lambdao       {\lambdaO}
\def\lambdaob      {\bar{\lambdaO}}
\def\Lambdaob      {\bar{\LambdaO}}
\def\Lambdaobp     {\bar{\LambdaO}{}'}

\def\li            {\mbox{$\Lambda_{(i)}$}}
\def\lie           {Lie algebra}

\def\llb           {\mbox{\large[}}
\def\Llb           {\mbox{\Large[}}
\def\LLb           {\mbox{\Large(}}
\def\LO            {{\brev{L}}}
\def\lO            {{\brev{r}}}
\def\lrb           {\mbox{\large]}}
\def\Lrb           {\mbox{\Large]}}
\def\LRb           {\mbox{\Large)}}

\def\mh            {\hat m}
\def\Mid           {\!\mid\!}
\def\mlambda       {m_\lambda^{(\omega)}}
\def\mO            {\brev m}
\def\mod           {\ {\rm mod}\;}
\def\modinv        {modular invarian}
\def\mub           {\bar \mu}
\def\muf           {{\fcft\mu{1.1}{9.5}}}
\def\muO           {\brev{\mu}}
\def\muo           {\brev{\mu}}
\def\muob          {\bar{\muo}}
\def\muop          {{\brev{\mu}'}}

\def\mydollar      {$^{\pounds}$}
\def\natnum        {{\dl N}}

\def\none          { \{0\} }
\def\nud           {\nu}

\def\ocond         {linking condition}
\def\ofour         {\mbox{$\rho_4$}}
\def\olie          {orbit Lie algebra}
\def\om            {\omega}

\let\omchar=\ocha
\newcommand\omd[1] {{\dot\omega #1}}
\def\omD           {\dot\omega}
\newcommand\omdd[2] {{\dot\omega^{#1} #2}}
\def\omddj         {{\dot\omega^2j}}
\def\omdi          {{\dot\omega i}}
\def\omdj          {{\dot\omega j}}
\newcommand\omdm[1] {{\dot\omega^{-1} #1}}
\def\omdo          {{\dot\omega 0}}
\def\omj           {\mbox{$\sigma$}}
\def\omo           {\sigma_{n+1}^{}}
\def\omO           {\sigma_{n+1}}
\def\omT           {\mbox{$\omega^\star$}}
\newcommand\omt[1] {{\omega^\star #1}}
\newcommand\omtb   {\bar\omega^\star}

\def\omtalb        {{\omtb\alphab}}
\def\omtla         {{\omt\lambda}}
\def\omtLa         {{\omt\Lambda}}
\def\one           {\mbox{\small $1\!\!$}1}
\def\onedim        {one-dimen\-sional}
\def\onehalf       {\mbox{$\frac12$}}
\def\oneton        {1,2,...\,,n}
\def\onetonm       {1,2,...\,,n-1}
\def\onetonmr      {1,2,...\,,n-r}
\def\onetonp       {1,2,...\,,n+1}
\def\onetopmm      {1,2,...\,,\p-2}
\def\onetor        {1,2,...\,,r}
\def\otor          {0,1,...\,,r}
\def\oton          {0,1,...\,,n}
\def\otoNm         {0,1,...\,,N-1}
\let\p=\ell
\def\pbw           {Poin\-ca\-r\'e\hy Birk\-hoff\hy Witt}
\def\Pro           {\mbox{\sc p}^{}_{\!\omega}}
\def\PRo           {\tilde{\mbox{\sc p}}^{}_{\!\omega}}
\newcommand\Prod[2]{\mbox{$\prod_{#1}^{#2}$}}
\def\proj          {\mbox{\sc p}^{\star-1}_{\!\omega}}
\def\projm         {\mbox{\sc p}^{\star}_{\!\omega}}
\def\Prom          {\mbox{\sc p}^{-1}_{\!\omega}}
\def\prow          {\mbox{\sc p}_W}

\long\def\query#1{\hskip 0pt{\vadjust{\everypar={}\small\vtop to 0pt{\hbox{}%
     \vskip -13pt\rlap{\hbox to 50.0pc{\hfil{\vtop{\hsize=8pc\tolerance=6000%
     \hfuzz=.5pc\rightskip=0pt plus 3em\noindent#1}}}}\vss}}}}%
\newcommand\rank[1] {\mbox{rank}\,#1}

\def\rep           {representation}
\def\Rep           {Representation}
\newcommand\restr[1] {|\raisebox{-.5em}{$#1$}}
\newcommand\rhob   {\bar \rho}
\def\rhoO          {\brev{\rho}}
\def\rhoob         {\,\bar{\!\rhoO}}
\def\rhs           {right hand side}
\def\Sscya         {{S^{\mskip-9mu\raise 3pt\hbox{${\scriptstyle\circ}$}}}}

\newcommand\sect[1] {\section{#1}\setcounter{equation}{0}}
\newcommand\Sect[2]{\sect{#1}\label{s.#2} \ifnum\draftcontrol=1 \query{s.#2}\fi}
\def\sicu          {\mbox{$\sigma$}}

\def\slz           {\mbox{SL(2,\zett)}}
\newcommand\smallmatrix[1] {\mbox{\footnotesize $\left(\begin{array}#1
                   \end{array}\right)$}}
\def\SO            {\brev S}

\def\smat          {$S$-matrix}
\def\sO            {\brev s}
\newcommand\sumi[1] {{\displaystyle\sum_{#1\in\I}}}
\newcommand\sumI[1] {{\sum_{#1\in\I}}}
\newcommand\sumiO[1] {{\displaystyle\sum_{#1\in\IO}}}
\newcommand\sumIO[1] {{\sum_{#1\in\IO}}}
\newcommand\sumn[1] {\sum_{#1=1}^n}
\newcommand\sumne[1] {{\displaystyle\sum_{#1=1}^{N-1}}}
\newcommand\sumno[1] {{\displaystyle\sum_{#1=0}^{N-1}}}
\newcommand\sumnz[1] {{\displaystyle\sum_{#1=2}^{N-1}}}
\newcommand\sumre[1] {{\displaystyle\sum_{#1=1}^r}}
\newcommand\sumrE[1] {\sum_{#1=1}^r}
\newcommand\sumro[1] {{\displaystyle\sum_{#1=0}^r}}
\newcommand\sumrO[1] {\sum_{#1=0}^r}
\newcommand\sumto[1] {{\displaystyle\sum_{#1\in\brev I\setminus\{0\}}\!}}

\def\taujl         {[\vn,\vnp]}
\def\tauo          {{\tau_\omega}}
\let\tcha=\ocha
\def\Tcha          {Twining character}
\def\tchi          {\raisebox{.15em}{$\tilde\chi$}}
\def\tchil         {\tchi_\Lambda^{(\omega)}}
\def\ti            {{\brev I}}
\def\tildeR        {R^{\omega}_{}}
\def\tildER        {R^{\omega}}

\def\tria          {\mbox{$\rho_3$}}

\def\ttab          {\bar\theta}
\def\ttaob         {\bar{\brev\theta}}
\def\thetab        {\ttab}
\def\twtw          {^{(2)}}
\def\twtwtw        {^{(3)}}
\def\twodim        {two-di\-men\-si\-o\-nal}
\def\U             {{\sf U}}
\def\uaff          {untwisted affine Lie algebra}
\def\untw          {^{(1)}}
\def\uO            {\brev{u}}
\def\Vee           {{\scriptscriptstyle\vee}}
\newcommand\verma[1] {\Chi_\lambda^{(\omega)[#1]}}
\newcommand\version[1] {\ifnum\draftcontrol=1 \typeout{}\typeout{#1}\typeout{}
                   \vskip3mm \centerline{\fbox{\tt DRAFT -- #1 -- \today}}  
                   \vskip3mm \fi}
\def\vir           {\mbox{$ {\cal V}ir$}}
\def\vl            {\mbox{$v_\Lambda$}}
\def\vn            {\vec n}
\def\vnp           {\vec n'}
\def\vO            {\brev{v}}
\def\vmL           {{V_\Lambda}}

\def\vml           {\mbox{$V_\Lambda$}}
\def\vmlt          {\mbox{$V^\omega_\Lambda$}}
\newcommand\wf[5]  {w_#1 w_#2 w_#3 w_#4 w_#5}
\newcommand\wfs[6] {w_#1 w_#2 w_#3 w_#4 w_#5 w_#6}
\def\wh            {\hat w}
\def\Wh            {\hat W}
\def\Wnull         {\overline W}
\def\whi           {\hat w_i}
\def\Wl            {W_{(\lambda)}}
\def\whomi         {\hat w_{\omD i}}

\def\wi            {w_i}
\def\womi          {w_{\omD i}}
\def\wO            {\brev{w}}
\def\wOi           {\brev{w}_i}
\def\WO            {\brev{W}}
\def\wrt           {with respect to }
\def\wrtt          {with respect to the }

\def\wzwts         {WZW theories}

\def\zet           {{\dl Z}}

\def\zetpluso      {\mbox{${\zet}_{\geq 0}$}}
\def\zett          {\mbox{\small {\dl Z}}}

\global\def\draftcontrol{0}
\catcode`\@=12

\documentstyle[12pt,epsf]{article}

\begin{document}\version\versionno
%%% for draft versions, suppress in definitive version
%\draft
\let\dl=\bf

\begin{flushright}  {~} \\[-23 mm] {\sf hep-th/9506135} \\
{\sf NIKHEF 95-028} \\[1 mm]{\sf June 1995} \end{flushright} \vskip 2mm

%%%%%%%%%%%%%%%%%%%%%%%%%%%%%%%%%%%%%%%%%%%%%%%%%%%%%%%%%%%%%%%%%%%%%%%
\begin{center} \vskip 12mm

{\Large\bf FROM DYNKIN DIAGRAM SYMMETRIES} \vskip 0.3cm
{\Large\bf TO FIXED POINT STRUCTURES}
%{\Large\bf TWINING CHARACTERS AND}\vskip 0.3cm{\Large\bf ORBIT LIE ALGEBRAS}
\vskip 15mm

{\large J\"urgen Fuchs, \mydollar \ \, Bert Schellekens, \ Christoph Schweigert}
\\[5mm] {\small NIKHEF-H\,/\,FOM, Kruislaan 409}
\\      {\small NL -- 1098 SJ~~Amsterdam}
\end{center}
\vskip 16mm
\begin{quote}{\bf Abstract}.\\
Any automorphism of the Dynkin diagram of a symmetrizable \kma\ \g\ 
induces an automorphism of \g\ and a mapping $\tauo$ between highest 
weight modules of \g. For a large class of such Dynkin diagram
automorphisms, we can describe various aspects of these maps in terms 
of another \kma, the `orbit \lie' $\gO$. In particular, the generating
function for the trace of $\tauo$ over weight spaces, which we call the
`twining character' of \g\ (with respect to the automorphism), is equal 
to a character of $\gO$.
The orbit \lie s of untwisted affine \lie s turn out to be closely related 
to the fixed point theories that have been introduced in \cft.
Orbit \lie s and twining characters constitute a crucial step towards 
solving the fixed point resolution problem in \cft.
 \end{quote} %\vskip 5mm
\vfill {}\fline{} {\small \mydollar~~Heisenberg fellow} \newpage

\section{Introduction}

In this paper we associate algebraic structures to automorphisms of \dyd s
and study some of their interrelations. The class of \dyd s we consider are 
those of symmetrizable \kma s \cite{KAc3}. These are those \lie s which possess
both a Cartan matrix and a Killing form, which includes in particular the
simple, affine, and hyperbolic \kma s.

An automorphism of a \dyd\ is a permutation of its nodes which leaves
the diagram invariant. Any such map divides the set of nodes of the diagram 
into invariant subsets, called the orbits of the automorphism. We focus
our attention on two main types of orbits, namely those where each of
the nodes on an orbit is either connected by a single link to precisely 
one node on the same orbit or not linked to any other node on the same 
orbit. If all orbits of a given \dyd\ automorphism are of one of these
two types, we say that the automorphism satisfies the {\em \ocond}.
Except for the order $N$ automorphisms of the \aff s $A_{N-1}\untw$,
all diagram automorphisms of simple and affine \lie s belong to this class.

\subsection{Orbit \lie s}

In section \ref{s.olie} we show that for any automorphism satisfying the
\ocond\ one can define a `folded' \dyd\ (and associated Cartan matrix)
which is again the \dyd\ of a symmetrizable \kma. The folded \dyd\ has
one node for each orbit of the original diagram, and there is a definite
prescription for the number of links between any two nodes of the folded
diagram. If the \kma\ corresponding to the original \dyd\ is \g, we
denote the algebra corresponding to the folded \dyd\ by $\gO$ and call
it the {\em orbit \lie}. We show that the folding procedure preserves
the `type' of the \kma, where the type of a symmetrizable \kma\ can
be either `simple', `affine', `hyperbolic', or `non-hyperbolic indefinite'.
(However, `untwisted affine' and `twisted affine' are not separately 
preserved.)

Any automorphism of a \dyd\ (not necessarily satisfying the \ocond)
induces an outer automorphism of the associated \kma\ \g. This is described
in section \ref{s.lie}. For simple, affine and hyperbolic algebras the
induced automorphism is unique. In the case of simple \lie s, these outer
automorphisms are well known; they correspond to charge conjugation
(for $\g=A_n$ ($n>1$), $D_{2n+1}$ ($n>1$), and $E_6$), to the spinor
conjugation of $D_{2n}$ ($n>2$), and to the triality of $D_4$.
The induced outer automorphisms of \uaff s are either the
aforementioned ones (inherited from the simple horizontal subalgebra),
or certain automorphisms related to simple currents \cite{scya} of
WZW theories (i.e.\ \cfts\ for which the chiral symmetry algebra is
the semidirect sum of the \uaff\ and the Virasoro algebra), 
or combinations thereof.

\subsection{\Tcha s}

The automorphism of \g\ induces a natural map on the weight space of \g.
We can also employ the action on the algebra, in a less straightforward 
manner, to obtain an action, compatible with the action on the weight space, 
on the states of any \hwm\ of \g. We can therefore define a new type of
character-like quantities for these modules by inserting the generator
of the automorphism into the trace that defines the ordinary character.
We call the object constructed in this manner the {\em \tcha\/} of the
\hwm; its precise definition is presented in section \ref{s.omc}.

Trivially, the \tcha\ vanishes whenever the \hw\ is changed by the
automorphism. As a consequence, our interest is in those \hwm s whose 
\hw\ is not changed by the automorphism; we call these special 
modules the {\em fixed point modules\/} of the automorphism and refer
to their highest weights as {\em symmetric\/} \g-weights. For
fixed point modules, the \tcha\ receives a non-vanishing contribution 
from at least one state, namely the \hw\ state, but it is far from
obvious what happens for all the other states of the module. Note that
the weight of a state does not provide sufficient information for
answering this question. Rather, the action of the automorphism also
depends on the specific way in which the state is obtained from the
\hw\ state by applying step operators, as the automorphism acts 
non-trivially on these step operators. 

There is one interesting class of automorphisms for which
{\em only\/} the \hw\ state of a fixed point module contributes to the
\tcha. These are the order $N$ automorphisms of the \aff s $A_{N-1}\untw$.
In this particular case the Serre relations among the commutators of step 
operators conspire in such a way that all other states in the Verma module 
(and hence also in the irreducible module) cancel each others' contributions
to the \tcha. This statement will be proven in section \ref{s.ANN}
(following a route that does not rely on the explicit use of the Serre 
relations and hence avoids various technical complications).
Note that these automorphisms do {\em not\/} satisfy the \ocond.
Rather, all $N$ nodes of the \dyd\ of $A_{N-1}\untw$ lie on the same
orbit of this automorphism, and hence each node on the orbit is connected
to {\em two\/} other nodes on the orbit. Correspondingly, there is no
associated \olie\ (formally one obtains the `\lie' which has the 
$1\times1$ Cartan matrix $A=(0)\,$).

The main result of this paper, proved in section \ref{s.irr}, concerns
the fixed point modules of \dyd\ automorphisms which do satisfy the \ocond.
We prove that these modules are in one-to-one correspondence with the
\hwm s of the \olie\ $\gO$, and that the \tcha s of the fixed point
modules (both for Verma modules and for their irreducible quotients)
coincide with the ordinary characters of the \hwm s of $\gO$.
Note that we are not claiming that the \olie\ $\gO$ is embedded in
the original algebra \g. We can show, however, that the Weyl group
of $\gO$ is isomorphic to a subgroup of the Weyl group of \g. This
observation plays a key r\^ole in the proof, as it enables us to
employ constructions that are analogous to those used by Kac in his
proof of the Weyl\hy Kac character formula.

In the sections \ref{s.aff}, \ref{s.modtr} and \ref{s.curlb}, we specialize
to the case of untwisted affine \lie s and those automorphisms
which correspond to the action of simple currents. In section \ref{s.aff}
the action of such automorphisms is described in some detail, using the
realization of affine \lie s as centrally extended loop algebras. 
We find that for this special class of automorphisms the characters of $\gO$,
and hence also the \omchar s, have nice modular transformation properties. 
In section \ref{s.modtr} it is shown that the modification of the 
irreducible characters of $\gO$, and hence of the irreducible \tcha s
of \g, that is required in order to obtain these nice modular transformation
properties, differs from the modification of the irreducible characters of
\g\ only by an overall constant. Finally, in section \ref{s.curlb} we 
comment on those cases where the orbit \lie\ is one of the twisted \aff s
$\tilde B_n^{(2)}$ rather than an untwisted affine algebra. 

\subsection{Fixed point resolution}

Our main motivation for introducing and studying \tcha s stems from a
long-standing problem in conformal field theory, namely the
`resolution of fixed points'. \Tcha s and orbit Lie algebras constitute
important progress towards solving this problem. This will
not be discussed further in the present paper, except for the
following brief explanation of the relation between the two issues.

The fixed point resolution problem can be divided into two aspects.
The first aspect is the construction of representations of the modular group;
the second is the description of \rep\ spaces of the chiral symmetry
algebra whose characters transform in these \rep s of the modular group.
For theories with an extended chiral symmetry algebra, one tries to 
achieve the construction of representations of the modular group by starting
from the modular transformation matrices $S$ and $T$ of the original,
unextended chiral algebra. One then typically finds that certain irreducible 
modules appear in the spectrum more than once or appear only in reducible 
linear combinations; it follows in particular that the
original matrix $S$ does not contain enough information to derive
the matrix $S_{\rm ext}$ of the extended theory. If the extension of 
the chiral algebra is by simple currents (the corresponding modular
invariants are often referred to as `D-type invariants'),
these reducible modules originate from fixed points of these simple currents.

By requiring the characters of the extended theory to have the correct modular
transformation properties, one learns that the missing information is
supplied by another matrix $\fS$, which is defined only on the fixed point 
representations and together with a diagonal matrix $\fT$ again generates 
a representation of the modular group; $\fT$ is simply $T$ restricted
to the fixed points. By studying the spectrum of $\fT$ and comparing
it to known conformal field theories, conjectures regarding $\fS$
could be made for most, though not all, simple current invariants for
\uaff s. Indeed, it was found in \cite{scya5,scya6} that in all cases except
$B_{n}\untw$ and $C_{2n}\untw$ at even levels, $\fT$ is equal
up to an overall phase to the $T$-matrix of another untwisted affine
algebra.  One may call this the `fixed point algebra' (as we will see in a
moment, this is a more appropriate name than the term
`fixed point conformal field theory' that was chosen in \cite{scya5,scya6}).

The second aspect of the fixed point resolution problem is closely related
to the `field identification' in coset conformal field theories. From 
the point of view of the modular group,
this can be described in terms of  an `extension
of the chiral algebra by spin-zero currents'. As far as the matrix $S$ is
concerned we are then in exactly the same situation as  discussed above.
However, if the field identification currents have fixed points, then there 
is an additional problem: formally one either obtains a partition 
function with more than one vacuum state, or, if one normalizes it, a 
partition function with fractional multiplicities for the fixed point 
states. The solution to the latter problem
is that the various irreducible components of the reducible
module that is associated to the fixed point possess in fact different
characters. The difference of these characters must then transform like 
a character with respect to the new modular matrix $\fS$.

This implies that for field identification fixed points the characters of
the coset theory are not simply equal to the branching functions of the
embedding of affine \lie s, which are merely 
sums over the characters of the irreducible components. It may seem
that writing  down the correct irreducible characters
requires additional information that is not directly provided by the Lie
algebras \g\ and \h\ defining a coset theory ${\cal C}(\g/\h)$.
This additional information is contained in
the matrix $\fS$ and in the character modifications. 
 \futnot{The situation is complicated further by the fact that modifications
of the Virasoro-specialized characters appear to be needed only for coset
theories, but not for D-type modular invariants.}

As already mentioned, some of the diagram automorphisms introduced above 
are closely related to the action of simple currents. Simple currents act as
a permutation on the modules of the chiral symmetry algebra. More precisely,
their action is defined via the fusion rules of the \cft. On the other hand,
an action of simple currents on the Hilbert space of states of the theory 
could so far not be defined, nor was it required for the purpose for
which the simple currents were used, namely
the construction of modular invariants. In the special case of WZW models, 
simple currents act by permuting the integrable highest weight
modules of the underlying affine \lie. Since the action of some of the 
diagram automorphisms on highest weights is identical to this simple 
current action, and since the action of diagram automorphism is defined 
on {\it individual\/} states, the results of the present paper provide a 
natural definition of the simple current action on the entire Hilbert space.

In the application to field identification in coset models, this should 
enable us to prove that identified fields are really identical as modules 
of the chiral algebra. The action
on fixed point modules is more interesting still. In this case the
module is mapped to itself, and as the mapping has finite
order $N$, the module splits into invariant subspaces of eigenvectors 
with an $N$th root of unity as eigenvalue. These eigenspaces
are natural candidates for the irreducible modules. The \tcha s
are then natural ingredients for the character modifications.
Note, however, that although the \tcha s of \uaff s are non-trivial,
no character modifications (for the Virasoro-specialized characters)
are required for the D-type modular invariants of the WZW theories
which are associated to the \aff s. This already shows that more work 
will be needed to make all this precise; we plan to analyze this situation 
in detail in a separate publication. The effort will be worthwhile, 
however, since in this formalism we should be able to derive
the character modifications for a coset theory ${\cal C}(\g/\h)$ 
in terms of the \tcha s of the \kma s \g\ and \h\, rather than having to
introduce them as extraneous objects as was done in \cite{scya5,scya6}.

An obvious candidate for the fixed point resolution matrix $\fS$ is
the modular transformation matrix $\brev{S}$ of the \olie\ $\gO$. Indeed,
in \cite{scya5,scya6} the relation between the matrix $\fT$ and the 
spectrum of the WZW theory based on an affine algebra \g\
was proved by applying a folding procedure to
the weight space metric (the inverse of the symmetrized Cartan matrix)
of the horizontal Lie algebra \gb. For all simply-laced algebras and also
for $C_{2n+1}$ at even levels this folding procedure is
equivalent to the one discussed here, and consequently $\fS=\brev{S}$
in these cases. In other words, the fixed point algebra is equal to the
orbit Lie algebra defined here.

As remarked above, the folding discussed here does not necessarily
map untwisted to untwisted affine algebras. This turns out to be 
relevant for the remaining cases, i.e.\ for $B_{n+1}\untw$ and
$C_{2n}\untw$. Here a `fixed point conformal field theory' could only
be identified for odd levels $k=2p+1$, namely the WZW theory based on
$C_n\untw$ at level $p$ in both cases. For even levels $k=2p$, the 
fixed point spectra for $B_{n+1}\untw$ and $C_{2n}\untw$
were shown to differ by an overall constant, but they could not be identified
with any known conformal field theory, apart from a few special cases.
(These spectra were denoted as ${\cal B}_{n,p}$ in \cite{scya5,scya6}.
Meanwhile, the matrix $\fS$ has also been constructed in an indirect 
manner, using rank-level duality in $N=2$ supersymmetric coset models 
\cite{fuSc2}.)
A natural solution now suggests itself, namely that again the fixed point
algebra is equal to the orbit Lie algebra, just as in all other cases.
Applying our folding procedure, we find that the orbit Lie algebra is in fact 
a twisted algebra, namely $A_1\twtw$ (for $n=1$) or $\tilde B_{n}\twtw$.\,%
 \futnote{Here we use the notation of \cite{FUch}; in the notation of 
 \cite{KAc3}, these algebras are called $A_{2n}^{(2)}$.}
The fixed points of $B_{n+1}\untw$ and $C_{2n}\untw$ at level
\kv\ are in one-to-one correspondence with the representations of
$\tilde B_{n}\twtw$ at level \kv.

The modular transformations for characters of twisted algebras do not
always close within a given algebra. Rather, typically
the characters of one algebra are mapped to those of a different algebra.
In fact, $A_1\twtw$ and $\tilde B_{n}\twtw$ are precisely those twisted
\aff s whose characters possess well-defined modular transformations
among themselves, and just as for untwisted algebras, these transformations 
preserve the level of a module \cite{KAc3}. Remarkably, the modular
matrix $\brev{S}$ of $\tilde B_{n}\twtw$, respectively $A_1\twtw$,
appears to provide the correct fixed point resolution for even as well 
as odd levels. Indeed, $\brev{S}$ at level $k$ is related in the correct way
to the matrices $S$ of $C_{n}\untw$ at level $p$ (for $k=2p+1$), respectively
${\cal B}_{n,p}$ (for $k=2p$). At present we do not have a general proof that
these matrices resolve the fixed points correctly, but we have checked
it for algebras of low rank at low level.

\subsection{Organization}

Let us briefly summarize how this paper is organized. There are two main 
results, which concern the automorphisms of \dyd s satisfying
the \ocond\ and the order $N$ automorphisms of the affine $A_{N-1}\untw$
\dyd s, respectively. These two theorems are stated at the end of section 
\ref{s.omc}; the former is proven in section \ref{s.irr} (some details 
are deferred to appendix A), and the latter in section \ref{s.ANN}. 
In the earlier sections, various concepts which are necessary
for being able to formulate our theorems are introduced, such as the 
folding of Cartan matrices (section \ref{s.olie}), the induced 
automorphisms of \lie s and the concept of an \olie\ (section \ref{s.lie}),
and the maps induced on Verma and irreducible modules as well as the concept
of their \tcha s (section \ref{s.omc}). The remaining sections 
\ref{s.aff}, \ref{s.modtr} and \ref{s.curlb} contain further details
about the special case of affine \kma s, in particular about modular
transformation properties, which are relevant for applications in \cft.

\Sect{Folding Cartan matrices}{olie}

\subsection{\dyd\ automorphisms}

In this paper we consider symmetries of indecomposable symmetrizable Cartan 
matrices. A symmetrizable Cartan matrix is by definition a square matrix
$A=(\Ac ij)_{i,j\in I}^{}$, where $I\subset\zet$ is some finite index set, 
satisfying the properties $\Ac ij\in\zet$, $\Ac ii=2$, $\Ac ij\le 0$ for $i\ne j$,  
$\Ac ij=0$ iff $\Ac ji=0$, and that there is a non-singular diagonal matrix 
$D$ such that $DA$ is symmetric. To any symmetrizable Cartan matrix there is
associated a unique \lie\ \g\ with an invariant bilinear form 
$(\,\cdot\Mid\cdot\,)\!:\;\g\times\g\to\g$
(see \cite{KAc3} and section \ref{s.lie}). 
The \dyd\ of \g\ is defined as the graph with $|I|$ vertices which has 
coincidence matrix $2\cdot\one - A$, with \one\ the identity matrix. 
The \dyd\ is connected iff $A$ is indecomposable.

By an automorphism of the \dyd\ of \g\ (or of the associated Cartan 
matrix) we mean a bijective mapping $\omD\!:\ I\to I$ satisfying
  \be  \Ac\omdi\omdj=\Ac ij  \labl A
for all $i,j\in I$. We denote
by $N$ the order of $\omD$, i.e.\ the smallest positive integer such
that $\omD^N=\id$ ($N$ is finite since $I$ is finite), and by
  \be  N_i:= |\, \{i,\omdi,...\,,\omdd{N-1}i\} \,|  \labl{Ni}
the length of the $\omD$-orbit through $i$. Also, let
$\ti$ denote a set of representatives for the orbits of $\omD$. It will be
convenient to fix the choice of these representatives once and for all; for
definiteness we choose the smallest representatives of the orbits (for a
given labelling by $I\subset\zet$),
  \be  \IO:=\{i\ini \mid i\le\omdd ni {\rm \ \,for\ }1\le n\le N-1\}  
  \,. \labl{IO}
We will now show that a large class of automorphisms $\omD$ of symmetrizable
Cartan matrices can be used to `fold' the Cartan matrix $A$ such as to obtain
another matrix $\AO$ which is again a symmetrizable Cartan matrix. In 
particular, with the exception of the automorphism of order 
$N$ of $A_{N-1}^{(1)}$, all diagram automorphisms 
of all simple and affine \lie s belong to this class. 

\subsection{The folded Cartan matrix}

For any given automorphism $\omD$ of an indecomposable symmetrizable 
Cartan matrix $A$ and any $i\in I$ let us define the integer
  \be  s_i:= 3 - \frac{N_i}N  \sum_{l=0}^{N-1} \Ac{\omdd li}i 
  = 1- \sum_{l=1}^{N_i-1} \Ac{\omdd li}i  \,.  \labl{def.si}
In the following we restrict our attention to the class of those 
automorphisms $\omD$ which satisfy the relation
  \be  s_i \le2 \quad {\rm for\ all}\ i\in I  \,,  \labl{asum}
to which we will refer as the {\em \ocond}. As the last sum in \erf{def.si}
is non-positive and integer, this means that $s_i$ is either 1 or 2. Since 
each contribution to that sum is non-positive, in the case $s_i=1$ we have 
  \be  \Ac i{\omdd li}= 0  \labl 0
whenever $l\ne 0\mod N_i$, and accordingly the restriction of the \dyd\ 
of \g\ to the orbit of $i$ is isomorphic to the Dynkin diagram of the 
direct sum of $N_i$ copies of $A_1$. 
For $s_i = 2$, there is exactly one $m$, $1\le m\le N_i-1$, such that 
$\Ac{\omdd mi}i = -1$. In this situation we have $\omdd mi=\omdd {-m}i$
(otherwise $\Ac{\omdd {-m}i}i = \Ac i{\omdd {m}i}$ would be negative as well,
leading to a contradiction with the assumption \erf{asum}). This implies 
that in this case
$N_i$ and hence also $N$ are even; the restriction of the \dyd\ of \g\ to the
orbit of $i$ is then isomorphic to the Dynkin diagram of the direct sum
of $N_i/2$ copies of $A_2$. 

Next we introduce a $|\IO|\!\times\!|\IO|$-matrix $\AO$ that is obtained from $A$
by folding it in the sense of summing up the rows of $A$ that are related
by $\omD$ and multiplying them by $s_i$, and afterwards eliminating redundant 
columns; thus we define
  \be \AOc ij := s_i \frac{N_i}N  \sum_{l=0}^{N-1} \Ac{\omdd li}j \labl{AO}
for $i,j\inio$.
{}From the indecomposability of $A$ it is obvious that $\AO$ is indecomposable
as well.  We claim that $\AO$ is also again a symmetrizable Cartan  
matrix, i.e.\ that it satisfies the following five properties:
  \be  \begin{array}{ll}
  {\rm a)} & \AOc ii=2       \quad {\rm for\ all}\ i\inio\,,        \\[.4em]
  {\rm b)} & \AOc ij\in\zett \quad {\rm for\ all}\ i,j\inio \,,     \\[.4em]
  {\rm c)} & \AOc ij\le0 \quad {\rm for\ all}\ i,j\inio,\ i\ne j\,, \\[.4em]
  {\rm d)} & \AOc ij = 0 \,\Longleftrightarrow\, \AOc ji=0 \,,      \\[.4em]
  {\rm e)} & \mbox{there is a non-singular diagonal matrix $\DO$} \\
           & \mbox{such that $\BO:= \DO\AO$ is symmetric.}
  \end{array}\labl{a-e}

Let us prove the relations \erf{a-e} consecutively.
First, under the assumption that the \ocond\ \erf{asum} holds, we have 
  \be  \AOc ii = s_i\,(3-s_i) = 2 \,, \ee
which proves (\ref{a-e}a). The property (\ref{a-e}b) is fulfilled 
because in fact we only add up and multiply integers, as is made 
manifest by rewriting \erf{AO} as 
  \be  \AOc ij = s_i \sum_{l=0}^{N_i-1} \Ac{\omdd li}j  \,. \labl{s1}
Next we observe that if $i,j\in\IO$ are different, then $i,j\in I$ lie on
different orbits of $\omD$. As a consequence, $\omdd li \ne \omdd mj$ for all
$l,m$, and hence, as $A$ is a Cartan matrix, $\Ac{\omdd li}{\omdd mj}\le0$
for all $l,m$. Thus in particular the sum on the \rhs\ of \erf{s1}
is also smaller than zero, which proves (\ref{a-e}c).
Further, since all terms in the sum \erf{s1} are non-positive, $\AOc ij= 0$
implies that $\Ac{\omdd li}j = 0$ for all $l$. Because of \erf A this means
that also $\Ac i{\omdd lj} = 0$ for all $l$. Since $A$ is itself
a Cartan matrix, this in turn implies that also $\Ac{\omdd lj}i$ vanishes
for all $l$. Thus
      \be \AOc ji = \sum_{l=0}^{N_j-1} \Ac{\omdd lj}i = 0 \,, \labl{b.di}
and hence we obtain the property (\ref{a-e}d).

Finally, we know that there is a non-singular diagonal matrix $D=\diag(d_i)$
such that $B:=DA$ is symmetric. This matrix is unique
 \futnot{see \cite[\S 2.3]{KAc3}; indecomposability essential.}
up to scalar multiplication, and we can choose $d_i>0$ for all $i\ini$.
 \futnot{because $\Ac ij\le0$ for all $i\ne j$}
To verify (\ref{a-e}e), we first show that $d_i=d_{\omdi}$ for all $i\ini$.
To this end suppose that we are given a matrix $D$ which has the required
properties. Then we define the `orbit average' $\tilde D$ of $D$ as follows.
For any $l=0,1,\ldots, N-1$ we set $D_{(l)} := \diag(d_{\omdd li})$,
and then define $\tilde D:= \sum_{l=0}^{N-1} D_{(l)}$. The automorphism 
property of $\omega$ implies that $B_{(l)}:= D_{(l)}A$ satisfies
  \be  B_{(l)}^{i,j}= d_{\omdd li} \Ac ij = d_{\omdd li}
  \Ac{\omdd li}{\omdd lj} = B^{\omdd li,\omdd lj}  \, . \ee
This shows that $B_{(l)}$ is symmetric, and hence
$\tilde B:= \tilde D A = \sum_{l=0}^{N-1} B_{(l)}$ is symmetric as well. 
Thus $\tilde D$ possesses all the properties required for $D$, so that
by the uniqueness of $D$ it follows that $D\propto \tilde D$. This proves 
that $d_i=d_\omdi$ for all $i\ini$, as claimed. Next we define $\DO$ as
  \be  \DO = \diag(\dO_i)\,, \qquad \dO_i:=\Frac1{s_i}\, \Frac N{N_i}\, d_i \,. 
  \labl{dO}
Clearly, $\DO$ is a non-degenerate diagonal matrix with positive  
diagonal entries. Further, the entries of  $\BO := \DO\AO$ read
  \be  \BO^{i,j} = \frac N{s_i N_i}\, d_i\AOc ij = d_i \sum_{l=0}^{N-1}
  \Ac{\omdd li}j = \frac1N \sum_{l,l'=0}^{N-1} d_{\omdd li} \Ac{\omdd li}
  {\omdd{l'}j}   = \frac1N \sum_{l,l'=0}^{N-1} B^{\omdd li,\omdd{l'}j} \, .
  \labl{BO}
This shows that the matrix $\BO$ is symmetric, and hence
completes the proof of (\ref{a-e}e). As we will see below,
the formula \erf{BO} encountered in this proof is also interesting in its 
own right; it describes the relation between the invariant bilinear form
of \g\ and that of the \olie\ $\gO$ that will be defined in subsection 
\ref{ssolie}.

\subsection{Type conservation}

Symmetrizable Cartan matrices belong to one of the following three classes
(compare e.g.\ \cite[\S4.3]{KAc3}: they are either of finite, affine or 
indefinite {\em type}. We are now going to show that $\AO$ as obtained 
from $A$ by the prescription \erf{AO} is of the same type as $A$. 

If $A$ is symmetrizable and the bilinear form given by $B=DA$ is positive 
definite, then $A$ is said to be of {\em finite\/} (or simple) type. 
Now for any vector $\uO=(\uO_i)_{i\in\IO}$ we have by \erf{BO} the
relation 
  \be  \sumIO{i,j}\BO^{i,j}\uO_i\uO_j= \Frac N{N_iN_j}\sumIO{i,j}
  \sum_{l=0}^{N_i-1}\sum_{l'=0}^{N_j-1}B^{\omdd li,\omdd{l'}j}\uO_i\uO_j
  =N\sumI{i,j}B^{i,j}u_i u_j \,,  \ee
where $u_i=\uO_{\omdd mi}/N_i$ with $m$
chosen such that $\omdd mi\in\IO$. As a consequence, if $B$ is positive
definite, then so is $\BO$, and hence $\AO$ is of finite type as well. 

If $A$ is symmetrizable and the bilinear form $B$ is positive semidefinite 
and has exactly one eigenvector with eigenvalue zero, then $A$ is of 
{\em affine\/} type. The components of the left respectively right 
eigenvector of $A$ with eigenvalue zero,
  \be  \sumro i a_i \Ac ij = 0 = \sumro j \Ac ij \avj  \labl{aav}
are thus unique once the normalization of the eigenvector is specified
(in \erf{aav}, we set $I\equiv\{\otor\}$, which is the conventional 
labelling in the affine case).
Fixing the normalization in such a way that the minimal value (denoted by
$a_0$ and $\av0$, respectively) is equal to 1,
the components $a_i$ and $\avi$ are called the Coxeter labels and dual 
Coxeter labels of $A$, respectively. In the affine case \erf{BO} implies 
that $\BO$ is either positive definite or positive semidefinite. 

Now by \erf{aav} and the invariance \erf A of the Cartan matrix, the 
vector with $i$th component $a_\omdi^\Vee$ is also a right eigenvector 
with eigenvalue zero and hence is proportional to the vector of dual 
Coxeter labels, and an analogous statement holds for the Coxeter 
labels. The fact that $\omD$ has finite order (together with
the positivity of the (dual) Coxeter labels) then implies that
  \be  a_\omdi=a_i\,, \qquad  \av\omdi=\avi  \labl{aao}
for all $i\ini$. It follows that the vectors with entries 
  \be  \awO i:=\frac{s}{s_i}\,a_i \,,\qquad
  \avO i:=\frac{N_i}N\,\avi   \qquad {\rm for} \ i\in\ti \,, \labl{avo} 
with $s\equiv\max_{j\ini}\{s_j\}$, satisfy 
  \be  \sumiO i \awO i \AOc ij = \sumiO i \sum_{l=0}^{N-1}
  s\,\frac{N_i}N\, a_i \Ac{\omdd li}j = s\sumi i a_i \Ac ij =0   \labl{221}
and
 \futnote{The chosen normalization of $\avO i$ proves to be convenient for the
 treatment of the affine case. In the general case, with this specific
 normalization the coefficients $\avO i$ are, however, not necessarily integral.}
  \be  \sumiO j \AOc ij \avO j = \sumiO j \sum_{l=0}^{N-1}
  s_i\,\frac{N_iN_j}{N^2}\, \Ac{\omdd li}j\,\avj = s_i\,\frac{N_i}N\, \sumi j  
  \Ac ij\,\avj =0  \,. \labl{220}
In particular there is an eigenvector of $\BO$
with eigenvalue zero, i.e.\ $\BO$ is positive semidefinite rather than positive
definite. Also, the eigenspace of $\BO$ to the eigenvalue zero is \onedim,
since if $\vO$ with entries $\vO_i$, $i\in\ti$, is an eigenvector of $\BO$ to 
the eigenvalue zero, then the vector $v$ with entries $v_i= \frac1{N_i} 
\vO_{\dot\omega^l i},$ with $l$ such that $\omdd li\in\ti$, is an 
eigenvector of $B$ to the eigenvalue zero, which is, however, unique up to
normalization.

Finally, a symmetrizable Cartan matrix is said to be of {\em indefinite\/} 
type if it is neither of finite nor of affine type. Let us 
show that if $A$ is of indefinite type, then $\AO$ is also of indefinite 
type. By theorem 4.3 of \cite{KAc3}, $A$ is of indefinite type iff
\vspace{-.7em} \begin{enumerate}
\item there is a vector $u$ with strictly positive components,
such that $uA$ has strictly negative components, and
\vspace{-.7em}
\item the fact that a vector $v$ and $vA$ both have positive
components implies that $v$ is the zero vector.
\end{enumerate} \vspace{-.6em}
To show that the first condition is fulfilled for the folded Cartan matrix
$\AO$, assume that $u$ is a vector with strictly positive components for
which $uA$ has strictly negative components. Clearly, the vector $u'$ with 
$i$th component $u'_i=u_{\omD i}$ shares these properties of $u$, and hence
we can assume without loss of generality that $u_i = u_{\omD i}$. We then 
define $\uO_i:= u_i/s_i$; this is positive as well, and also obeys
  \be \sum_{i\in\ti}\uO_i\AOc ij = \sum_{i\in I}u_i \Ac ij < 0 
  \quad {\rm for\ all}\ j\in\IO \, . \ee
To show that the second condition is fulfilled, we assume that 
$\vO$ is a vector such that 
  \be \vO\AO \geq 0 \quad\mbox{and}\quad \vO\geq 0 \, . \labl{cond}
Define a vector $v$ by $v_i := s_i \vO_{\dot\omega^l i}$ for $i\in I$,
where $l$ is chosen such that $\omdd li\in\ti$. Then $v$ fulfills the 
conditions \erf{cond} with $\AO$ replaced by $A$. Since $A$ is by assumption
of indefinite type, $v$ and hence also $\vO$ have to vanish. Together,
these results imply that $\AO$ is of indefinite type as well.

We have thus shown that the matrix $\AO$ that is obtained by the folding
prescription \erf{AO} is always of the same type as the Cartan matrix $A$. 

A particularly interesting subclass among the Cartan matrices of indefinite 
type is given by the {\em hyperbolic\/} Cartan matrices. These are
characterized by the additional property that any indecomposable submatrix 
of the Cartan matrix $A$ that is obtained by deleting any row and the 
corresponding column of $A$ is of finite or affine type. 
Again one can show that if $A$ is hyperbolic then the same is true for
$\AO$. Namely, the pre-image (under the folding) of any proper subdiagram 
of the \dyd\ of $\AO$ is a subdiagram of the \dyd\ of $A$, which, as $A$ 
is assumed to be hyperbolic, is of finite or affine type.
But as we have just seen, these diagrams are mapped to diagrams of affine
or finite type, and hence the subdiagram of $\AO$ has to be of affine or 
finite type as well. This shows that also $\AO$ is hyperbolic. 

\subsection{Simple Cartan matrices}

In the next two subsections we will list all automorphisms of all simple
and affine \dyd s explicitly. The numbering of the nodes of the \dyd s 
is taken from \cite[p.\,43]{FUch}. Below we write \g\ and $\gO$ for the
\kma s which have Cartan matrix $A$ and $\AO$, respectively ($\gO$ will
be called the \olie\ associated to \g\ and $\om$, see subsection 
\ref{ssolie} below).

The non-trivial automorphisms of the \dyd s of simple \lie s are as  
follows. For $A_r$, $D_r$ and $E_6$ there is a reflection which we denote by  
\conj; it acts as $i\to r+1-i$ for $A_r$, as  
$r-1\leftrightarrow r$, $i\mapsto i$ else, for $D_r$, and as  
$1\leftrightarrow 5,\ 2\leftrightarrow 4,\ 3\mapsto3,\ 6\mapsto6$
for $E_6$. In addition, for $D_4$ there is the triality \tria, an
order three rotation which acts as $2\mapsto2,\;1\mapsto3\mapsto4\mapsto1$.

The algebras \gO\ for the cases when \g\ is simple are listed in  
table \erf{Ls}; this is in fact a well known list, as \gO\ plays an 
important role in the realization of the twisted affine \lie s as 
centrally extended twisted loop algebras \cite{KAc3}.
  \be   \begin{tabular}{|l|c|c|l|}
  \hline &&&\\[-.9em]
  \multicolumn{1}{|c|} {\g} &
  \multicolumn{1}{c|}  {$\om$} &
  \multicolumn{1}{c|}  {$N$} &
  \multicolumn{1}{c|}  {$\gO$}
  \\[1.2mm] \hline\hline &&&\\[-2.8mm]
   $A_{2n+1},\ {\scriptstyle n>0} $ & \conj &  2    & $B_{n+1} $      \\[1.9mm]
   $D_4      $       & \tria &  3    & $G_2 $          \\[1.9mm]
   $D_n ,\ {\scriptstyle n\geq 4} $ & \conj &  2    & $C_{n-1} $      \\[1.9mm]
   $E_6      $       & \conj &  2    & $F_4 $
          \\[.3em] \hline &&&\\[-.8em]
   $A_{2n}   $       & \conj &  2    & $C_n $
   \\[.4em] \hline \end{tabular} \labl{Ls}
In this table we have separated the $A_{2n}$ case from the others because 
in this case we have $s_n=2$ (and $s_i=1$ else), whereas in all other cases
all the $s_i$ are equal to 1.

\subsection{Affine Cartan matrices}

The relevant automorphisms $\omD$ for affine \lie s are the following.
For $\g=A_n\untw$, the automorphism group of the \dyd\ is the dihedral 
group ${\cal D}_{n+1}$ which is generated by the reflection $\conj\!:\  
i\mapsto n+1-i \;{\rm mod}\;n+1$ and the rotation $\omo\!:\ i\mapsto 
i+1\;{\rm mod}\;n+1$ which is of order $n+1$. Among the powers $\omO^{\,l}$, 
only those need to be considered for which $l$ is a divisor of $n+1$ so 
that the order is $N=(n+1)/l$.

For $\g=D_r\untw$ the automorphism group is generated by the `vector 
automorphism' \Jv, the `spinor automorphism' \Js\ and a conjugation \conj.
\Jv\ acts as $0\leftrightarrow1,\ r\leftrightarrow r-1$ and $i\mapsto i$ 
else, and hence is of order two; the map \conj\ acts as 
$r\leftrightarrow r-1$ and $i\mapsto i$ else. 
If $r$ is even, $\Js$ acts as $i\mapsto r-i$ and hence
has order two, while for odd $r$ the prescription $i\mapsto r-i$
only holds for $2\le i\le r-2$ and is supplemented by
$0\mapsto r\mapsto1\mapsto r-1\mapsto0$, so that \Js\ has order 4.
 \futnot{conjugate spinor automorphism: $\Jc=\Jv \circ\Js$}
If $r=4$, then the automorphism group is larger, namely the symmetric group
${\cal S}_4$; it contains as additional symmetries an order four rotation
$\ofour$, which acts as $ 0\mapsto1\mapsto3\mapsto4\mapsto0,\ 2\mapsto2$, and
an order three permutation $\rho_3$, acting like $1\mapsto3\mapsto4\mapsto1$,
$2\mapsto2$ and $0\mapsto0$. 

For the untwisted algebras $\g=B_r\untw$, $C_r\untw$ and $E_7\untw$  
and for the twisted algebras $\g=B_r\twtw$ and $C_r\twtw$, there is only a  
single non-trivial automorphism \conj\ which is a reflection.
For $\g=E_6\untw$ the automorphism group of the \dyd\ is the symmetric group
${\cal S}_3$; it is generated by the order three rotation $\sicu\!:\ 1\mapsto 5
\mapsto0\mapsto1,\ 2\mapsto4\mapsto6\mapsto2,\ 3\mapsto3$ and the reflection
$\conj\!:\ 1\mapsto 5,\ 2\mapsto4,\ 3\mapsto3,\ 6\mapsto6,\ 0\mapsto0$.
Finally, for $\g=E_8\untw,\,F_4\untw,\,G_2\untw$ and for the remaining twisted
algebras, there are no non-trivial \dyd\ automorphisms at all.

Let us remark that the notation has been chosen such that
for the untwisted affine algebras, \conj\ implements charge conjugation, 
while \omj\ corresponds, in \cft\ terms, to a simple current \cite{scya}.

We list $\gO$ for these automorphisms in table \erf{LS}.
In this table we again separate the cases where all the $s_i$ are 
equal to 1 from the others. Also, there is a single series of 
automorphisms which do not obey the \ocond\ \erf{asum}, namely for any
$N\ge2$ the automorphism of the \dyd\ of $A_{N-1}^{(1)}$ that 
has order $N$; this series is displayed in the last row of the table. In 
this case, which will be treated separately in section \ref{s.ANN},
there is only a single $s$, which takes the value zero, and the prescription 
\erf{AO} formally yields a one-by-one matrix with entry zero.

Also, in the table we use the notations of \cite[p.\,95]{FUch} for twisted 
affine algebras; the relation with the notation of \cite[p.\,55]{KAc3} is
\vskip.9em

\begin{tabular}{l|cccccc}
Notation of \cite{KAc3} 
& $A_2^{(2)}$           
& $A_{2n}^{(2)}$           
& $A_{2n-1}^{(2)}$         
& $D_{n+1}^{(2)}$          
& $E_6^{(2)}$              
& $D_4^{(3)}$              
 \\[1.5mm] \hline &\\[-2.2mm]
Notation of \cite{FUch}
& $A_1^{(2)}$
& $ \tilde B_n\twtw $
& $C_n^{(2)}$
& $B_n^{(2)}$
& $F_4^{(2)}$
& $G_2^{(3)}$
\end{tabular}

  \be   \begin{tabular}{|l|c|c|lll|}
  \hline &&&&&\\[-.9em]
  \multicolumn{1}{|c|} {\g} &
  \multicolumn{1}{c|}  {$\om$} &
  \multicolumn{1}{c|}  {$N$} &
  \multicolumn{3}{c|}  {$\gO$}
  \\[1.2mm] \hline\hline &&&&&\\[-2.8mm]
   $A\untw_n$      & $\!(\omo)^{(n+1)/N}\!$ & $N\!<\!n\!+\!1$
                   & \multicolumn{3}{c|} {$A\untw_{((n+1)/N)-1}$}  
\\[1.9mm]
   $A\untw_{2n+1}$ & \conj  & 2  &&& $ B\twtw_{n+1} $         \\[1.9mm]
   $B\untw_{n}$    & \jv    & 2  &&& $ \tilde B_{n-1}\twtw $  \\[1.9mm]
   $B\twtw_{2n}$   & \conj  & 2  &&& $ B\twtw_n $             \\[1.9mm]
   $C\untw_{2n}$   & \sicu  & 2  &&& $ \Bntilde $             \\[1.9mm]
   $C\untw_2$      & \sicu  & 2  &&& $ A\twtw_1 $             \\[1.9mm]
   $C\twtw_n$      & \conj  & 2  &&& $ C\untw_{n-1}$          \\[1.9mm]
   $D\untw_4$      & \ofour & 4  &&& $ A\twtw_1$              \\[1.9mm]
  %$D\untw_4$ &$(\ofour)^2$ & 2  &&& $ C\untw_2$              \\[1.9mm]
   $D\untw_4$      & \tria  & 3  &&& $ G\twtwtw_2$            \\[1.9mm]
   $D\untw_n$      & \jv    & 2  &&& $ C\untw_{n-2}$          \\[1.9mm]
   $D\untw_n$      & \conj  & 2  &&& $ C\twtw_{n-1}$          \\[1.9mm]
   $D\untw_{2n}$   & \js    & 2  &&& $ B\untw_n $             \\[1.9mm]
   $D\untw_{2n}$& \js \conj & 4  &&& $\tilde B_{n-1}\twtw $   \\[1.9mm]
   $E\untw_6$      & \sicu  & 3  &&& $ G\untw_2 $             \\[1.9mm]
   $E\untw_6$      & \conj  & 2  &&& $ F\twtw_4 $             \\[1.9mm]
   $E\untw_7$      & \sicu  & 2  &&& $ F\untw_4 $            %\\[1.9mm]
          \\[.3em] \hline &&&&&\\[-.8em]
   $A\untw_{2n}$   & \conj  & 2  &&& $ \Bntilde $             \\[1.9mm]
   $A\untw_{2n+1}$ & $\omO$\conj  & 2  &&& $ C\untw_{n} $     \\[1.9mm]
   $B\twtw_{2n+1}$ & \conj  & 2  &&& $ \Bntilde $             \\[1.9mm]
   $C\untw_{2n+1}$ & \sicu  & 2  &&& $ C\untw_n $             \\[1.9mm]
   $D\untw_{2n+1}$ & \js    & 4  &&& $ C\untw_{n-1}$          \\[1.9mm]
   $D\untw_{2n+1}$ & \js\conj   & 2  &&& $ C\twtw_{n}$       %\\[1.9mm]
          \\[.3em] \hline&&&&&\\[-.8em]
   $A\untw_n$      & $\omO$ &$n+1$&&&$ \none $
   \\[.4em] \hline \end{tabular} \labl{LS}

\pagebreak
\subsection{Hyperbolic Cartan matrices}

One can, of course, compile an analogous list for the hyperbolic \lie s as
well. However, the number of these algebras and their automorphisms 
(satisfying the \ocond) is rather large, and hence
we refrain from presenting this list here. Let us just
mention that the result that along with $\g$ also $\gO$ is a hyperbolic \lie\
may be easily verified case by case. As a by-product, this provides a check 
on the completeness of the list of hyperbolic \lie s that has been given in
the literature.\,%
 \futnote{In fact, the classification of hyperbolic \lie s presented in 
\cite{sacl} turns out to be not quite complete. We thank C.\ Saclioglu
for a correspondence on this issue.}

\binternal{
For example, applying to the hyperbolic Cartan matrix
\be A = \smallmatrix{{rrrrrr} 
                             2 & 0 &-1 & 0 & 0 & 0 \\[.2em]
                             0 & 2 &-1 & 0 & 0 & 0 \\[.2em]
                            -1 &-1 & 2 &-1 & 0 & 0 \\[.2em]
                             0 & 0 &-1 & 2 &-1 & 0 \\[.2em]
                             0 & 0 & 0 &-2 & 2 &-1 \\[.2em]
                             0 & 0 & 0 & 0 &-1 & 2 } \ee
(this is entry number 8 in the list of hyperbolic \lie s of rank 6 in \cite{sacl}) 
the involutive automorphism which interchanges the first and second node, yields
the rank 5 hyperbolic Cartan matrix
\be \AO = \smallmatrix{{rrrrr} 
                               2 & -2 & 0 & 0 & 0 \\[.2em]
                              -1 &  2 &-1 & 0 & 0 \\[.2em]
                               0 & -1 & 2 &-1 & 0 \\[.2em]
                               0 &  0 &-2 & 2 &-1 \\[.2em]
                               0 &  0 & 0 &-1 & 2 }\,. \ee
The algebra defined by the latter Cartan matrix
was overlooked in the classification of hyperbolic 
\lie s presented in \cite{sacl}. The `stretched version' of this algebra 
(and its dual Lie algebra) with rank 6 does appear (as entry number 9)
in the list of \cite{sacl}.
}\einternal

\Sect{\lie\ automorphisms}{lie}

In this section we show that any automorphism $\omD$ of finite order of the
Cartan matrix $A$ induces an automorphism of the same order of the \kma\ 
\g\ which has Cartan matrix $A$. To this end we first sketch how \g\ can 
be constructed from the Cartan matrix \cite{KAc3}. Then we show how 
$\omD$ induces an automorphism $\omega$ of \g\ and investigate to
what is extent this automorphism is unique. 

\subsection{Symmetrizable \kma s}

To any symmetrizable Cartan matrix $A$ there is associated a \lie, denoted 
by $\g\equiv\g(A)$ and called a symmetrizable \kma, which is unique up to
isomorphism \cite[Prop.\,1.1]{KAc3}. \g\ is constructed from $A$ as follows.
Denote by $n$ the dimension and by $r$ the rank of the matrix $A$.
We introduce a complex vector space $\go$ of complex dimension $2n-r$.
Next we choose $n$ linearly independent elements $H^i$ of $\go$ and
$n$ linearly independent functionals $\als i\in\go^\star$ (called the simple
roots of \g), such that $\als i(H^j)=\Ac ij$ for $i,j=1,2,...\,,n$. 
This choice is unique up to isomorphism. 

The \kma\ \g\ is then the Lie algebra that is generated freely by the 
elements of $\go$ and $2n$ further elements $\EE i\pm\equiv 
E^{\pm\alpha^{(i)}}$, with $i\ini\equiv\{1,2,...\,,n\}$, modulo the relations
  \be \begin{array}{l}
  [x,y] =0  \qquad\mbox{for all}\ x,y\in\go \,, \\[2.5 mm]
  [x,\EE i\pm]= \pm \als i(x)\,\EE i\pm \quad\mbox{for all}\ x\in\go \,, \\[2.5mm]
  [\EE i+,\EE j-]=\delta_{ij}\, \HH j{} \,, \\[2.7 mm]
  {({\rm ad}^{}_{\EE i\pm})}^{1-\Ac ji}_{}\EE j\pm =0 \quad{\rm for}\ i\ne j 
  \,, \end{array}\labl C
where the map ${\rm ad}^{}_x$ is defined by ${\rm ad}^{}_x(y):=[x,y]$.
Thus the subspace $\go$ is an abelian subalgebra of \g; it is 
called the \csa\ of \g. Also, \g\ has a triangular decomposition
  \be  \g = \g_+ \oplus \go \oplus \g_- \,,  \labl{+0-}
where $\g_\pm$ are subalgebras and generated freely by the $\EE i\pm$ 
modulo the relations in the last line of \erf C, which are known as the 
Serre relations.

The algebra $\gd:=[\g,\g]$\, is called the derived algebra of \g. It 
contains all central elements and has a triangular decomposition
  \be  \gd\equiv[\g,\g] = \g_+ \oplus \goh \oplus \g_- \,, \ee
where $\goh$ is the span of the elements $H^i$, $i=1,2,...\,,n$. Thus 
$\goh\subseteq\go$ is the \csa\ of \gd, and the {\em derivations}, i.e.\ 
the generators of a complement of $\gd$ in \g, span a complement of 
$\goh$ in $\go$.
We will also denote by $\gK$ the common kernel of all the simple
roots $\als i$ (and hence of all roots, since any root $\alpha$ is a 
linear combination of the $\alpha^{(i)}$). $\gK$ is a subspace of $\goh$. 

By definition, the non-degenerate bilinear form of \g\ satisfies
  \be  (H^i\mid x) = d_i\,\als i(x) \quad {\rm for\ all}\ x\in\go  \ee
with $d_i$ as defined after \erf{b.di}, and hence in particular
  \be  (H^i\mid H^j) = d_i\,\Ac ij=\GG ij \quad {\rm for}\ i=\oneton \,.  
  \labl{BP}

\subsection{Induced outer automorphisms}\label{s.ioa}

We are now in a position to construct an automorphism $\omega$ of \g\ using
any symmetry $\omD$ of the \dyd\ of \g. We start by defining $\omega$ on the
generators $E^i_\pm$ of $\g_\pm$:
  \be   \omega(E^i_\pm) :=  E^{\omd i}_\pm       \,.  \labl{oei}
Because of \erf A this mapping preserves the Serre relations, and hence it
provides automorphisms of both $\g_+$ and $\g_-$. Further, the automorphism
property of $\omega$ implies that it has to act on the $H^i$ as
  \be \omega(H^i)= \omega([E^i_+,E^i_-]) = [ E^{\omd i}_+, E^{\omd i}_-]
  = H^{\omd i}  \, . \ee
This way we have constructed a unique automorphism of the derived algebra 
\gd\ of \g.  This automorphism has the same order $N$ as the automorphism 
$\omD$ of the \dyd. To show how $\omega$ acts on the rest of the \csa\ of \g,
i.e.\ on the derivations, requires a bit more work. To this end it is helpful to 
work with a special basis of $\go$. 

Since $\alpha^{(i)}(H^j) =\Ac ij =\Ac{\omD i}{\omD j} = \alpha^{(\omD i)}
(H^{\omD j})$ and since the $H^j$ span $\goh$, for all $x\in\goh$ we have
  \be \alpha^{(\omd i)}(\omega(x)) = \alpha^{(i)}(x) \,.\ee
Hence the subspace $\gK$ is mapped by $\omega$ bijectively to itself. We can 
therefore diagonalize $\omega$ on $\gK$ and choose a basis of $n-r$ 
eigenvectors $K^a$, $a=\onetonmr$, such that 
  \be  \omega(K^a) = \zeta^{n_a} K^a         \, ,                \ee
where 
  \be  \zeta:=\exp\LLb\frac{2\pi\ii}N\LRb  \labl{zeta}
is a primitive $N$th root of unity. We can extend the basis of $\gK$ to a basis
of $\goh$ by adding further eigenvectors of $\omega$ on $\goh$, which we 
denote by $\JJ p$, $p=\onetor$. We write
  \be \omega(\JJ p) = \zeta^{m_p} \JJ p \,, \ee
and denote the span of all \JJ p by $\gH$. Clearly, the
restriction of the invariant bilinear form to $\gH$ is non-degenerate. 
Moreover, we find that 
  \be (K^a\mid x) =0 \quad\mbox{for all}\ x\in\goh \, .\ee
This holds because writing $K^a=\sumn j \K^a_j H^j$ we have 
  \be 0 = \alpha^{(i)}(K^a) = \sumn j \K^a_j \alpha^{(i)}(H^j) 
  = \sumn j \Ac ij \K^a_j \ee
for all $i=\oneton$, so that the invariant bilinear form obeys
  \be (H^i\mid K^a) = \sum_j \K^a_j\,(H^i\mid H^j) = \sum_j d_i\Ac ij \K^a_j 
  = 0 \ee
for all $i=\oneton$. 

Using the fact that the invariant form is non-degenerate on $\go$, we 
conclude that there are $n-r$ unique elements $D^a$ of $\go$ such that
  \be \begin{array}{lll}
  (D^a\Mid K^b) = \delta^{ab}& \mbox{for all}& 1\leq a,b \leq n-r\,,  \\[.3em]
  (D^a\Mid D^b)  = 0   &  \mbox{for all}  &   1\leq a,b \leq n-r \,,  \\[.3em]
  (D^a\Mid \JJ p)= 0   &  \mbox{for all}  &   1\leq a \leq n-r,\;
                                             1\leq p\leq r \,.
  \end{array}\labl{315}
The elements $D^a$ are linearly independent and span a complement $\gD$ of 
$\goh$ in $\go$. 

We can now study the action of $\omega$ on the derivations 
$D^a$. To this end we first show that $\omega(D^a)$ is again an element of 
the \csa\ $\go$.  Namely, let us start from the most general ansatz $\omega(D^a) 
=h+\sum_{\alpha,\ell} \labda_{\alpha,\ell} E^{\alpha,\ell}$, where $h\in\go$
and the elements $E^{\alpha,\ell}$ are generators of the root space for the root
$\alpha$ (thus the number of possible values of the index $\ell$ equals the
(finite) dimension of this root space).
Assume now that $\labda_{\alpha,\ell}\ne0$ for some root $\alpha>0$
(the argument for $\alpha$ a negative root is completely parallel) and
some $\ell$.  The step operator $E^{-\alpha,\ell}$ is an element of $\g_-$, 
and hence so is $\om^{-1}(E^{-\alpha,\ell})$. Because of $D^a\in\go$ and
the fact that $\omega$ is an automorphism of \g, this implies that
$[D^a,\om^{-1}(E^{-\alpha,\ell})]$ and
$[\om(D^a),E^{-\alpha,\ell}]= \om([D^a,\om^{-1}(E^{-\alpha,\ell})])$ 
are elements of $\g_-$, too. On the other hand, we have
$[E^{\alpha,\ell},E^{-\beta,\ell'}]=\delta_{\alpha,\beta}\delta_{\ell,\ell'}
h^{\alpha,\ell}$ with non-vanishing $h^{\alpha,\ell}\in\go$, so that
by inserting the ansatz we made above we find that the element
$[\om(D^a),E^{-\alpha,\ell}]$ of \g\ 
has a component $\labda_{\alpha,\ell}h^{\alpha,\ell}$ in $\go$. This is a
contradiction, and hence the assumption that $\labda_{\alpha,\ell}\ne0$ is
wrong. Thus we conclude that $\omega(D^a)\in\go$. 
 \futnot{also:
  $ 0= \omega([H^j, D^a]) = [H^{\omD j}, \omega(D^a)] = \sum_{\alpha,\ell} 
  \labda_{\alpha,\ell}\,\alpha(H^{\omD j}) E^{\alpha,\ell}$ for any $j\in I$.
Since the $E^{\alpha,\ell}$ are linearly independent, this implies 
that $\labda_{\alpha,\ell} \alpha(H^{\omD j}) = 0$ for all $j\in I$. 
Now since the roots are non-zero functionals, 
for any real root $\alpha$ there is a choice of $j$ such that 
$\alpha(H^{\omD j}) \neq 0$. It follows that $\labda_{\alpha,\ell} =0$ 
for all real roots $\alpha$ and all $\ell$.\\
But this argument does not work for imaginary roots}

We can therefore make the general ansatz 
  \be \omega(D^a)  = \sum_{b=1}^{n-r} ( U^a_b D^b + \zeta^{n_b} V^a_b K^b)
  + \sum_{p=1}^{r} W^a_p\JJ p  \,. \labl{uvw}
Here in the second term we have introduced an explicit phase factor,
which will simplify the discussion below.
We now impose the condition that $\omega$ preserves the invariant bilinear
form, i.e.\ require that $(\omega(D^a)\Mid \omega(\JJ p))=0=
(\omega(D^a)\Mid \omega(D^b))$ and $(\omega(D^a)\Mid \omega(K^b)) =\delta^{ab}$
for all $p=\onetor,\ a,b=\onetonmr$.
Inserting the ansatz \erf{uvw}, the first of these conditions reads
  \be 0 = (\omega(D^a)\mid \omega(\JJ p)) = \zeta^{m_p}\, (\sum_{q=1}^{r} 
  W^a_q \JJ q\mid\JJ p)  \ee
for all $p=\onetor$. As the metric on $\g_J^{}$ is non-degenerate, this 
implies that $W^a_q$ vanishes. The second requirement then amounts to
  \be \delta^{ab}=(\omega(D^a)\mid \omega(K^b))= \sum_{c=1}^{n-r} U^a_c 
  \zeta^{n_b} (D^c\mid K^b) =  U^a_b \zeta^{n_b} . \ee
Thus the ansatz \erf{uvw} for $\omega(D^a)$ gets reduced to  
  \be \omega(D^a)  = \zeta^{-n_a} D^a + \sum_{b=1}^{n-r} V^a_b \zeta^{n_b} K^b 
  \,.\labl{ans2}
The last requirement then constrains the matrix $V$; we obtain
  \be 0 = (\omega(D^a)\mid \omega(D^b)) = V^a_b + V^b_a \, , \ee
i.e.\ $V$ has to be an antisymmetric matrix. 

To summarize, we have shown that the only freedom we are left with 
consists in adding terms proportional to central elements to $\omega(D^a)$,
and that this freedom is parametrized in terms of an antisymmetric 
$(n-r)\times(n-r)$ matrix. In the particularly interesting cases
where $\g$ is simple, affine, or hyperbolic, 
there is thus no freedom left at all; in the simple and hyperbolic 
cases there are no derivations, while in the affine case 
no term with the central element $K$ appears (the only antisymmetric 
one-by-one matrix is zero) so that just $\om(D)=D$.

We can restrict the freedom in $\om(D)$ even more by imposing the requirement
that $\omega$ has order
$N$ also on the derivations $D^a$. The relation \erf{ans2} implies that 
  \be \omega^l (D^a)  = \zeta^{-ln_a} D^a + \sum_{b=1}^{n-r} V^a_b 
  (\sum_{t=1}^l \zeta^{t n_b - (l-t) n_a})  K^b \, . \ee
It follows that $\omega$ has order $N$ if and only if $V^a_b$ vanishes 
whenever $n_a=-n_b\;{\rm mod}\,N$. It is also clear that these constraints 
always possess the trivial solution $V\equiv 0$. 

{}From the invariance of the bilinear form on $\go$ it follows that 
$\omega$ as defined above is in fact an automorphism of \g. 
The only identity that still has to be shown to this end is that
$\alpha^{(i)}(D^a)$ coincides with $\alpha^{(\omD i)}(\om(D^a))$;
this follows by
  \be \alpha^{(i)}(D^a) = \frac1{d_i}\,(H^i\Mid D^a) = \frac1{d_{\omD i}}\,
  (H^{\omD i}\Mid \om(D^a)) = \alpha^{(\omD i)}(\om(D^a)) \,.\ee

{}From now on we will assume that $V$ has been chosen such that $\omega$ 
has in fact order $N$ on all of $\g$. We will refer to such an automorphism 
which respects the triangular decomposition as a {\em strictly 
outer automorphism\/} or as a {\em diagram automorphism\/} of \g. The first 
of these terms is appropriate because any such automorphism is 
indeed outer, as can be seen e.g.\ by the fact that (compare section 
\ref{s.omc} below) it induces a non-trivial map on the representation ring 
of \g, whereas inner automorphisms do not change the isomorphism class of
a \rep.

\subsection{Orbit \lie s} \label{ssolie}

We denote by \gO\ the symmetrizable \kma\ that has $\AO$ as its Cartan
matrix and call \gO\ the {\em orbit Lie algebra\/} that is associated
to the \dyd\ automorphism $\omD$, respectively to the automorphism
$\omega$ of \g\ that is induced by $\omD$. We would like to stress that 
\gO\ is {\em not\/} constructed as a subalgebra of \g; in particular it 
need not be isomorphic to the subalgebra
of \g\ that consists of those elements which are mapped to themselves by
$\omega$. There does exist, however, a subalgebra $\mathaccent"017F{\;\g}$
of \g\ (to be described elsewhere) which is pointwise fixed
under $\om$ and whose Cartan matrix is closely related to the Cartan matrix 
$\AO$ of \gO; namely, the transpose of the Cartan matrix of
$\mathaccent"017F{\;\g}$ is equal to the matrix $(A^t)\,\brev{}$ that one
obtains when applying our folding procedure to the transpose $A^t$ of 
the Cartan matrix $A$ of \g.

Later on, we will use this orbit \lie\ to describe aspects of the action that
$\omega$ induces on irreducible highest weight modules of \g. To this end,
we need to set up some relations between \g\ and $\gO$. In preparation
for these considerations, we first show that
there is a close relation between the \csa\ $\gOo$ of the orbit \lie\ and 
the eigenspace $\geins$ of $\omega$ to the eigenvalue $\zeta^0=1$ in $\go$.
This relation is described by a map $\Pro$ which is defined as follows.
First consider the subalgebra $\goheins:= \geins \cap \goh$ of $\geins$.
The elements of $\goheins$ are those elements
$h=\sum_{i=1}^n v_i H^i$ of $\goh$ which are fixed under $\omega$, 
$\omega(h)=h$, which implies that $v_i = v_{\omD^l i}$ for all $l$. To any
such $h$ we associate the element 
  \be  \Pro(h) := \sum_{i\in\ti} \vO_i \HO^i  \labl{Pro}
of $\gOoh$, where
  \be  \vO_i := N_i v_i \labl{vNv}
for all $i\in\ti$. It is obvious that the map $\Pro$ is an 
isomorphism between $\goheins$ and $\gOoh$. 

Moreover, the invariant bilinear forms on $\goheins$ and $\gOoh$
 \futnot{As it can always be read off the notation to which algebra an 
element belongs, it is safe to employ the same symbol for both of these
forms.}
 satisfy
  \be (h\mid h') = \frac 1N\, (\Pro(h)\mid \Pro(h')) \,  \labl{normrel}
for all $h,h'\in \goheins$. To prove this relation, it is sufficient to check it
on a basis of $\goheins$. As a basis, we choose
  \be h^i := \frac 1N \sum_{l=0}^{N-1} H^{\omD^l i}=
  \frac 1{N_i} \sum_{l=0}^{N_i-1} H^{\omD^l i} \ee
for $i\in\ti$. Then we have $\Pro(h^i)=\HO^i$, and we can use \erf{BP} to find
  \be (h^i\mid h^j) = \frac1{N^2}\sum_{l,l'=0}^{N-1} \GG{\omD^l i}{\omD^{l'}j}
  = \frac1N\, \GGO ij = \frac1N\,(\Pro(h^i)\mid \Pro(h^j)) \, . \ee

Next we show that $\Pro$ yields a one-to-one correspondence between central 
elements in $\goheins$ and central elements of $\gO$. First note that
$K= \sum_{i=1}^n \kappa_i H^i$ is central iff 
  \be  \sum_{i\in I}\Ac ji \kappa_i = 0 \quad\mbox{for all}\quad j\in I \,. 
  \labl{central}
Now \erf{central} implies that
  \be  \sum_{i\in\ti} \AOc ji N_i\kappa_i = N_j\, s_j\,\sum_{i\in I}
  \Ac ji \kappa_i = 0 \quad\mbox{for all}\quad j\in\ti \, ,\ee
and hence also the element $\Pro(K)=\sum_{i\in\ti}\KOa_i\HO^i$ of $\gO$,
with $\KOa_i:= N_i\kappa_i$, is central. Conversely, 
with $\brev K$ also the pre-image $\Prom(\brev K)$ is central. 
This result shows in particular that the dimension of $\gkeins$
is precisely $|\IO|-\lO$, where $\lO$ is the rank of $\AO$. 

Finally we can continue the range of definition of $\Pro$ to all of $\goeins$
such that \erf{normrel} is still valid: we use again the basis of eigenvectors
of $\omega$ introduced in subsection \ref{s.ioa}. Given the derivation $D^a$, 
consider the projection $\Pro(K^a) \in\gO$ of the corresponding central 
element $K^a\in\g$. Since the bilinear form on $\gOo$ is non-degenerate, 
for each $D^a\in\gdeins$
we can define $\Pro(D^a)$ to be the unique derivation in $\gOo$ for which
  \be  (\Pro(D^a)\mid \Pro(D^b)) = 0 \,, \qquad 
  (\Pro(D^a)\mid \Pro(K^b)) = N \delta^{ab} \, \labl{331}
for all $K^b\in\gkeins$, and
  \be  (\Pro(D^a)\mid \Pro(x)) = 0  \qquad\mbox{for all}\ x\in \g_J^{(0)}
  \, \labl{332}
(the factor of $N$ in the second of the conditions \erf{331} ensures 
that the relation \erf{normrel} between the invariant bilinear forms on
$\goheins$ and $\gOoh$ extends to all of $\goeins$ and $\gOo$).
This completes the definition of $\Pro$. 

We can use the action of $\om$ to define a dual action, denoted 
by $\omT$, on the space $\go^\star$ that is dual to $\go$, i.e.\ on the 
weight space of \g, namely as
  \be  (\omT\beta)(x) := \beta(\omega^{-1}x) \labl{333}
for all $\beta\in\go^\star$ and all $x\in\go$.
The natural correspondence between $\go^{(0)}$ and the \csa\ $\gOo$ of \gO\
implies a corresponding relation for the dual spaces, the weight spaces.
We therefore have a bijective map
  \be  \projm:\quad \gOo^\star\to \gstar{0}  \labl{projm}
between the weights of $\gO$ and the weights $\lambda\in\gstar0$, i.e.\
those weights of \g\ that are fixed under $\omT$, $\omtla=\lambda$. 
We will refer to the elements of $\gstar0$ as {\em symmetric\/} \g-weights.
For brevity, we will also often denote the pre-image $\proj(\lambda)\in\
\gOo^\star$ of $\lambda\in\gstar{0}$ by $\lambdaO$.

By duality, the invariant bilinear form on $\go^{(0)}$ defines an
invariant bilinear form on $\gstar{0}$, and analogously for $\gO$.
The relation \erf{normrel} between the restriction of the invariant
bilinear form on $\go^{(0)}$ and the bilinear form on $\gOo$ therefore
implies an analogous relation between the bilinear form on symmetric weights 
$\lambda\in \gstar{0}$ and the one on $\gO$-weights:\,\,%
\futnote{As for the equations \erf{nrw} and \erf{normrel}, the following remark 
is in order. For an arbitrary symmetrizable \kma\ there is no canonical 
normalization of the invariant bilinear symmetric 
form. On the other hand, in \erf{nrw} and \erf{normrel} the relative
normalization of these forms on $\g$ and $\gO$ 
has been fixed in a convenient way. This can be
in conflict with the conventional normalization as soon as $\g$, and along with
$g$ also $\gO$, is simple or affine.\\ 
For simple \lie s and affine \lie s other than $\tilde B^{(2)}_n$ one usually 
fixes the normalization by requiring that the long roots have length squared 2 
\cite[(6.4.2)]{KAc3}, while for $\tilde B^{(2)}_n$ one normalizes the
bilinear form such that the roots have length squared $1, 2$ or $4$. 
If one sticks to this normalization, then the factor $N$ in equation 
\erf{normrel} and \erf{nrw} must be replaced by a different factor
$N'$ in the following cases: for $\g=A_{2n}$, for $\g=A^{(1)}_{2n}$ 
with the order two automorphism $\gamma$, and for $\g=B^{(2)}_{2n+1}$ one has
$N'=2N=4$, while for the order two automorphism of $C_n^{(2)}$
one has $N'=N/2=1$, and for the order four automorphism $\js\conj$ of 
$D^{(1)}_{2n+1}$ one needs $N'=2N=8$.}
  \be  (\lambda\mid\mu) = N\cdot (\proj(\lambda)\mid \proj(\mu)) 
  \equiv N\cdot(\lambdaO\mid\muO)\, . \labl{nrw}

\Sect{\Tcha s}{omc}

\subsection{The map $\tauo$}

Let $V$ be a vector space and
  \be  R:\quad \g \to {\sl End}(V) \ee
a \rep\ of a \lie\ \g\ by endomorphisms $R(x)\!:\, V\to V$. Any 
automorphism $\om$ of \g\ induces in a natural manner a map on the 
\g-module $(V,R)$. Namely, via
  \be  \tildeR(x) := R(\om(x))  \labl T
for all $x\in\g$, the action of $\om$
provides another \rep\ $\tildeR$ of \g. 
This is again a representation of \g\ in
${\sl End}(V)$. To describe the structure of the module $(V,\tildeR)$ in
more detail, we first note that the construction does not change $V$ as a 
vector space. However, this identity between
vector spaces in general does {\em not\/} extend to an isomorphism
of \g-modules, i.e.\ in general the map does change the 
(isomorphism class of the) module.
 \futnot{In other words, $R(\g)$ and $\tildeR(\g)$ describe two, generically
inequivalent, embeddings of \g\ into the algebra {\sl End}($V$). It is only
after applying to $(V,R)$ and $(V,\tildeR)$ the forgetful functor from the
category of \g-modules to the category of vector spaces
that $(V,R)$ and $(V,\tildeR)$ become identical objects.}

Here we are interested in the case where 
$\om$ is a strictly outer automorphism and where the module is a
\hwm.  If the \hwm\ with \hw\ $\Lambda$ is a Verma module, we denote it by 
$(V,R_\Lambda)$, while if it is the irreducible quotient of $(V,R_\Lambda)$, 
we write $(\hil,R_\Lambda)$.
 \futnot{Note that in both cases we write $R_\Lambda$.}
A natural basis of a \hwm\ consists of eigenvectors of the action of
the \csa\ $\go\subset\g$. Both for Verma and irreducible modules, the 
eigenspaces $\Wl\subset V$ of weight $\lambda$ \wrtt action of $\tildeR
(\go)$ coincide with the eigenspaces \wrtt original action $R(\go)$.

Further, recall that the action of $\omega$
preserves the triangular decomposition \erf{+0-} of \g,
i.e.\ not only maps the \csa\ to the \csa, but also the generators  
for positive (negative) roots to generators for positive (negative) roots.
As a consequence, $(V,\tildER_\Lambda)$ is again a Verma module. 
Moreover, since $\omega$ maps $\g_+$ to $\g_+$, the sets of primitive
singular vectors, i.e.\ those vectors which are
annihilated by the enveloping algebra $\U(\g_+)$, of $(V,R_\Lambda)$ and
$(V,\tildER_\Lambda)$ coincide. 
Now an irreducible\ \hwm\ \hill\ has a single primitive singular vector,
namely its \hwv, and hence the previous observation implies that
$(\hil,\tildER_\Lambda)$ is again an \ihwm. 

\binternal{
The primitive singular vectors of 
$(V,R_\Lambda)$ are, besides the \hwv\ \vl, the vectors
  \be  v_i:=(R_\Lambda(E^i_-))^{\Lambda^i+1}\cdot\vl \ee
for $i\in I$. $v_i$ is thus mapped on $(R_\omtLa(E^\omdm i_-))^{\Lambda^i+1}
\cdot v_\omtLa$. But this is precisely the primitive null vector
  \be  \tilde v_{\omdm i}=(R_\omtLa(E^\omdm i_-))^{(\omtLa)^\omdm i+1}\cdot 
  v_\omtLa  \ee
of $(V,\tildER_\Lambda)$.
}\einternal

To obtain a more detailed description of the relation between 
$(V,R_\Lambda)$ and $(V,\tildER_\Lambda)$ (respectively 
$(\hil,R_\Lambda)$ and $(\hil,\tildER_\Lambda)$) as \g-modules, we note 
that the \hwv\ in both modules is the same element of the underlying 
vector space $V$ (respectively $\hil$). 
However, as an element of the module, its associated weight has to be 
transformed by the map \omT\ defined by \erf{333}, so that in fact
the \hwv\ $v^{}_{\rm h.w.}\in V$ has highest weight $\Lambda$ in 
$(V,R_\Lambda)$, but highest weight $\omT\Lambda$ in $(V,\tildER_\Lambda)$.
We thus conclude that as a module,
$(V,R_\Lambda)$ is isomorphic to the abstract Verma module \vml\ (and hence
$(\hil, R_\Lambda)$ is isomorphic to the irreducible quotient $\hil_\Lambda$,
the irreducible highest weight module with highest weight $\Lambda$), while
$(V,\tildER_\Lambda)$ and $(\hil,\tildER_\Lambda)$ are isomorphic to 
$V_\omtLa$ and $\hil_\omtLa$, respectively:
  \be  \begin{array}{ll} (V, R_\Lambda)\cong \vml \, , &
  (V,\tildER_\Lambda) \cong V_\omtLa \,, \\[.85em]
  (\hil, R_\Lambda)\cong \hil_\Lambda\, ,\quad &
  (\hil,\tildER_\Lambda) \cong \hil_\omtLa \, .  \end{array}\ee

Via these isomorphisms, one and the same element $v$ of the vector space $V$ 
(respectively $\hil$) is identified with an element $v'$ of \vml\
and another element $v''$ of $V_\omtLa$ (respectively of $\hil_\Lambda$ and 
$\hil_\omtLa$). In other words, the automorphism $\om$ induces maps 
$\tauo\!:\,\vml\to\vmlt$ and $\tauo\!:\,\hill\to\hillt$ acting as $v'\mapsto
v''$ (for simplicity we use the same symbol for the map on the Verma module and
its restriction to the irreducible quotient). By definition, this map $\tauo$
thus satisfies  
  \be  \tauo(R_\Lambda(x)\cdot v)= R_\omtLa(\omega(x))\cdot\tauo(v) \,
  \labl{C'}
for all $x\in\g$ and all $v\in\vml$ (respectively $\hill$), i.e.\ the 
diagrams
%  CORR:  \omega  at `other side'
  \be  \begin{array}{rcl} 
  \begin{array}{rcl} \vml &  
  \stackrel{R_\Lambda(x)}{\mbox{---------}\!\!\!\longrightarrow}
  & \vml \\[2 mm]
  {\scriptstyle \tauo}\,\downarrow\ && \ \downarrow\,{\scriptstyle \tauo}
  \\[2 mm]  V_\omtLa \! & \stackrel{R_\omtLa(\omega(x))}
  {\mbox{---------}\!\!\!\longrightarrow} & \! V_\omtLa  \end{array}
               & \quad {\rm and}\quad &
  \begin{array}{rcl} \hill&  
  \stackrel{R_\Lambda(x)}{\mbox{---------}\!\!\!\longrightarrow}
  & \hill\\[2 mm]
  {\scriptstyle \tauo}\,\downarrow\ && \ \downarrow\,{\scriptstyle \tauo}
  \\[2 mm] \hillo \!\!\!  & \stackrel{R_\omtLa(\omega(x))}
  {\mbox{---------}\!\!\!\longrightarrow} & \!\!\hillo \end{array} 
  \end{array}\ee
commute.  As \erf{C'} generalizes the defining property of an intertwining 
map, we will refer to the relation \erf{C'} as the {\em $\omega$-twining 
property\/} of $\tauo$. Also note that for any weight $\lambda$ of the 
module, the action of $\tauo$ restricts to an action
  \be  \tauo\restr{\scriptstyle\Wl}:\quad \Wl\to W_{(\omtla)}  \ee
on the (\findim) weight space $\Wl$.

\subsection{\Tcha s}

Of particular interest in applications, e.g.\ in \cft, are those
irreducible highest weight \rep s for which $\omT\Lambda=\Lambda$; 
in the physics literature they are known as `fixed points' of the diagram
automorphism \cite{scya6}. While $\tauo$ is generically
a map between two different \irmod s, in this case it is an endomorphism 
of a single \irmod.\,%
 \futnote{As a consequence, we will always be dealing with a definite \rep\ $R$,
and correspondingly often simplify notation by writing $x$ in place of $R(x)$.}
In this situation the following definition makes sense. 
For any strictly outer automorphism $\omega$ of \g\
let us define the {\em \atcha s\/}, or, briefly, {\em \tcha s\/} $\Chil$ of 
a Verma module \vml\ and $\chil$ of its irreducible quotient \hill, 
as follows. They are (formal) functions on the \csa\ $\go$, defined
analogously to ordinary characters, but
with an additional insertion of the map $\tauo$ in the trace. Thus in the case of
Verma modules the \tcha\ $\Chil$ reads 
\be \begin{array}{llcll} {}\\[-.8em]
\Chil: \\[-2.2em]
       && \go  & \to    & \complex \,,\\[.7em]
       && \Chil(h)&:=   & {\rm tr}_\vmL^{} \tauo\, \eE^{2\pi\ii R_\Lambda(h)} \,,
       \end{array}\labl{Chil}
and analogously the \tcha\ $\chil$ of the irreducible module is given by
\be \begin{array}{llcll} {}\\[-.8em]
\chil: \\[-2.2em]
       && \go  & \to    & \complex \,,\\[.7em]
       && \chil(h)&:=   & {\rm tr}_\hilL^{} \tauo\, \eE^{2\pi\ii R_\Lambda(h)} \, . 
       \end{array}\labl{chil}
These \tcha s are majorized by the ordinary characters, and hence in
particular they are convergent wherever the ordinary characters converge.
Note that generically some contributions to the \tcha s have non-zero
phase, so that
instead of using the term character one might prefer to call these objects
character-valued indices. However, by the identifications \erf{1a} and
\erf{1b} below, it follows that the expansion coefficients of the \tcha s
are still non-negative integers.
 \futnot{By a suitable re-interpretation of the symbol `tr' one can formally 
extend the definitions \erf{Chil} and \erf{chil} to \hwm s
with non-symmetric \hw s. These formal objects vanish identically.} 

The \omchar\ can be interpreted as the generating functional of the trace of the
map $\tauo$ restricted to the various weight spaces. Taking the trace separately
on each weight space and extending the definition of the weights as functionals
on $\go$ to formal exponentials, $\eE^{2\pi\ii\lambda}(h):=
\exp(2\pi\ii\lambda(h))$, we can rewrite the \omchar\ $\Chil$ in the form
  \be \Chil = \sum_{\lambda\leq\Lambda} \mlambda \eE^{2\pi\ii\lambda} \, , 
  \labl{weide} 
and analogously for the \omchar\ $\chil$ of the irreducible module. 
Here $\mlambda$ denotes the trace of the restriction of $\tauo$ to the 
(\findim) weight space $\Wl$ of weight $\lambda$, 
and we write $\lambda \leq \Lambda$ iff $\Lambda-\lambda$ is a 
%% CORRok: `positive' replaced by 
non-negative linear combination of simple roots. 
Because of the trace operation, the coefficient $\mlambda$ can be
different from zero only for $\lambda\in\gstar0$, i.e.\ only if $\lambda$
is a symmetric weight. Hence we can restrict the sum in \erf{weide} to 
symmetric weights.

Combining the cyclic invariance of the trace and the $\om$-twining property
\erf{C'} of $\tauo$, we also learn that
%  CORR:  \om  instead  \om^{-1}
  \be \begin{array}{ll}
  \chil(h)\!& = {\rm tr}_\hilL^{} \tauo \eE^{2\pi\ii R_\Lambda(h)}   
    = {\rm tr}_\hilL^{} \eE^{2\pi\ii R_\Lambda(\om(h))} \tauo \\[.5em]
  & = {\rm tr}_\hilL^{} \tauo \eE^{2\pi\ii R_\Lambda(\om(h))} 
    = \chil(\om(h)) \,   \end{array}\ee
for the character of the irreducible module, and an analogous result
holds for the character of the Verma module. 

\subsection{Eigenspace decompositions}

In the discussion in subsection \ref{ssolie}, the eigenspace $\go^{(0)}$
in $\go$ to the eigenvalue $\zeta^0$ of $\om$ played an important r\^ole.
Similarly, when analyzing the properties of \tcha s, it proves to be 
convenient to decompose elements of $\go$ into their components in 
all eigenspaces $\go^{(l)}$ (to the eigenvalue $\zeta^l$) of $\om$.
Also, the map $\omT$ on the weight space
has the same order $N$ as $\omega$, and hence
we can decompose the weight space into eigenspaces $\gstar j$
of $\omega^\star$ to the eigenvalue $\zeta^j$,
  \be  \go^\star = \bigoplus_{j=0}^{N-1} \gstar j    \, .\labl{goj}
The elements of the subspaces of $\gstar{j}$ can be characterized by the fact 
that for any $l$ different from $-j\bmod N$, their restriction on $\go^{(l)}$ 
vanishes. To see this, consider arbitrary elements $\beta\in\gstar{j}$ and 
$x\in \go^{(l)}$. Then we have
  \be \beta(x) = (\omega^\star\beta)(\omega x) = \zeta^{j+l}\beta(x) 
  \,, \labl{j+l}
which shows that $\beta(x)$ has to vanish whenever $j+l\neq 0\bmod N$. 
Conversely, if an element $\beta$ of $\go^\star$ vanishes on all
elements of $\go$ except for those of $\go^{(l)}$, then by 
decomposing any element $h\in\go$ into its components in the various
eigenspaces $\go^{(j)}$ according to $h=\sum_j h^{(j)}$ with $\om(h^{(j)})=
\zeta^jh^{(j)}$, we find that
  \be (\omega^\star \beta)(h) = \beta(\omega^{-1} h) = \zeta^{-l} 
  \beta(h^{(l)}) = \zeta^{-l}\beta(h) \,, \ee
and hence that $\beta\in 
%% CORRok: ` \gstar{-l}$' replaced by 
\gstar{-l\mod N}$. 

Consider now the \omchar\ in the formulation \erf{weide}, i.e.
$\Chil(h) = \sum_{\lambda\leq\Lambda} \mlambda \eE^{2\pi\ii\lambda}(h)$, 
and decompose $h\in\go$ into its components $h^{(j)}$ as above.
As only symmetric weights contribute in \erf{weide}, the relation \erf{j+l}
can be employed in a similar manner as above to conclude that
  \be  \begin{array}{ll} 
  \Chil(h) \!\!& % \equiv\Chil(\sum_{j=1}^{N-1} h^{(j)})
  = \sum_{\lambda\leq\Lambda} \mlambda \,\eE^{2\pi\ii\lambda}(\sum_j h^{(j)}) 
  \\[.4em] &
  = \dsum_{\lambda\leq\Lambda} \mlambda \,\exp\llb 2\pi\ii\sum_{j=0}^{N-1}
    \lambda(h^{(j)}) \lrb
  = \dsum_{\lambda\leq\Lambda} \mlambda \,\exp\llb 2\pi\ii\,\lambda(h^{(0)}) 
    \lrb \,.  \end{array} \ee
Thus we have
  \be  \Chil(h)=\Chil(h^{(0)}) \,,  \labl{chilh0}
and analogously for the \omchar\ $\chil$ of the irreducible module.
In other words, the \omchar s depend on $h\in\go$ non-trivially only through
its component in the subspace $\go^{(0)}$ of the \csa\ $\go$ that consists 
of fixed points of $\omega$. Correspondingly, from now on we will consider
the \tcha s just as functions on $\go^{(0)}$.

\subsection{The main theorems}

We are now in a position to state the main result of this paper.
Recall that there is a natural mapping $\Pro$ \erf{Pro} from
$\go^{(0)}$ to $\gOo$, which induces a corresponding dual map 
$\projm$ \erf{projm} between the respective weight spaces. 
Let $\omD$ satisfy the \ocond\ \erf{asum}, and let $\Lambda$
be a symmetric \g-weight. Then we have 

\begin{quote}
{\bf Theorem 1:}\\[.3em]
a) The \omchar\ $\Chil$ of the Verma module of \g\ with highest weight $\Lambda$
coincides with the ordinary character of the Verma module with highest weight
$\proj(\Lambda)$ of the \olie\ $\gO$ in the sense that
\end{quote}
  \be \Chil(h) = \ChiO_{\proj(\Lambda)}(\Pro(h)) \, . \labl{1a}
\begin{quote}
b) The \omchar\ $\chil$ of the irreducible \g-module with dominant 
integral highest weight $\Lambda$ coincides with the ordinary character 
of the irreducible module with highest weight $\proj(\Lambda)$ of the 
\olie\ $\gO$ in the sense that
\end{quote}
  \be \chil(h) = \chiO_{\proj(\Lambda)}(\Pro(h)) \, . \labl{1b}
 As already mentioned, the \ocond\ \erf{asum} is in fact satisfied for
all diagram automorphisms of all affine and simple \lie s with the
exception of the order $N$ automorphisms of $A_{N-1}^{(1)}$. In these
exceptional cases for any value of the level there is only a single highest 
weight $\Lambda$ on which $\proj$ is defined. These cases can still be 
treated with our methods; they are covered by
 
\begin{quote}
{\bf Theorem 2:}\\[.3em]
In the case of $\g=A_{N-1}^{(1)}$ and the outer automorphism of order $N$,
the coefficients $\mlambda$ in the expansion \erf{weide} for
the \omchar\ of both the irreducible and Verma modules obey
\end{quote}
  \be  \mlambda=0 \quad{\rm for}\ \lambda\ne\Lambda \, , \ee
\begin{quote}
i.e.\ except for the contribution from the highest weight vector, all 
contributions cancel against each other.  
\end{quote}
The theorems 1 and 2 will be proven in section \ref{s.irr} and section 
\ref{s.ANN}, respectively. In section \ref{s.ANN} we will also present
the explicit expression for the \omchar\ for $A_{N-1}^{(1)}$ with respect 
to the order $N$ automorphism.

\Sect{The \omchar\ and the Weyl group $\WO$}{irr}

Our proof of Theorem 1 proceeds in several rather distinct steps which
are inspired by Kac' proof of the Weyl\hy Kac character
formula (see e.g.\ \cite[pp.\,152,\,172]{KAc3}). An additional crucial 
ingredient of the proof consists in the identification of a natural 
action of $\WO$, the Weyl group of the orbit \lie\ $\gO$, on the \omchar s. 
\futnot{compare file kac.tex for sketch of Kac's proof.}

\subsection{The action of the Weyl group $\WO$}\label{acwo}

We have seen that in the description \erf{weide} of the
\tcha\ only weights lying in $\gstar0$ contribute, and that this part of 
the weight space of $\g$ is isomorphic to $\gOo^\star$ via the map $\projm$. 
Hence we can employ $\projm$ to push
the action of $\WO$ on $\gOo^\star$ to an action of $\WO$ on $\gstar{0}$. 

To describe the Weyl groups explicitly, we denote by $\wi$ the fundamental
reflections which generate the Weyl group $W$ of \g, i.e.\ the reflections 
of the weight space of $\g$ \wrt the hyperplanes perpendicular to the 
simple roots $\alpha^{(i)}$, and analogously by $\wOi$ the fundamental 
reflections for $\gO$. Now for any fundamental reflection $\wOi$ of $\WO$ 
we can find an element of the Weyl group of \g, to be denoted by $\whi$, 
which acts on $\gstar 0$ precisely like $\wOi$ acts on $\gOo^\star$, i.e.\ 
which satisfies $\proj(\whi(\lambda)) = \wOi(\proj(\lambda))$ for all 
$\lambda\in\gstar0$.  We will denote the mapping which maps $\wOi$ to
$\whi$ by $\prow$,
  \be  \prow:\quad \wOi\mapsto\whi\,,\qquad (\prow(\wOi))(\lambda):=
  \whi(\lambda)\equiv \projm(\wOi(\proj(\lambda)))   \, . \labl{push}
Moreover, we will see that $\whi$ commutes with $\omega^\star$.

Let us first deal with those fundamental reflections $\wOi$ for which 
the integer $s_i$ defined in \erf{def.si} is $s_i=1$. In this case define 
  \be \whi := \prod_{l=0}^{N_i-1} w_{\omD^l i}  \, . \labl{whi}
Note that because of $s_i=1$ we have $\Ac i{\omD^li}=0$
whenever $i\neq\omD^li$, so that $w_{\omD^l i}$ and $w_{\omD^{l'} i}$
commute, and hence the product in \erf{whi} is well-defined.
The fact that $w_{\omD^l i}$ and $w_{\omD^{l'} i}$
commute also ensures that $\whi^2=\id$, and that $\whi$ commutes
with the induced automorphism $\omega^\star$. This implies in particular 
that the action of $\whi$ respects the
orbits of $\omega^\star$. For $s_i=1$ we also have $(\alpha^{(i)}\mid 
\alpha^{(\omD^l i)})=0$, so that the action of $\whi$ on \g-weights 
$\lambda$ reads
  \be \whi(\lambda) = \lambda - \sum_{l=0}^{N_i-1} 
  (\lambda\mid \alpha^{(\omD^l i)^\Vee})\, \alpha^{(\omD^l i)}  \,. \labl{w1}

Let us now describe how $\whi$ acts on the positive roots of $\g$. We have
  \be \whi(\alpha^{(\omD^l i)}) = w_{\omD^l i} (\alpha^{(\omD^l i)}) 
  = - \alpha^{(\omD^l i)} \, , \labl{wii1}
while $\whi$ maps any positive root which is not on the $\omega^\star$-orbit
of $\alpha^{(i)}$ to a positive root which is also not on that orbit. 
This can be seen as follows: let $\beta=\sum_{j=1}^n n_j \alpha^{(j)}$ be a
positive root which is not on the orbit of $\alpha^{(i)}$. Then there is some
index $j$, which is not on the orbit of $i$, $j\neq\omD^l i$, for which $n_j$
is strictly positive. Since the only effect of $\whi$ on $\beta$ is to add 
terms proportional to the $\alpha^{(\omD^l i)}$,
  \be \whi(\beta) = \sum_{j=1}^n n_j \alpha^{(j)} + \sum_{l=1}^{N_i-1} 
  \xi_l\, \alpha^{(\omD^l i)}  \, , \ee
the expansion of $\whi(\beta)$ in terms of simple roots still contains the 
term $n_j  \alpha^{(j)}$. Since $\whi(\beta)$ is again a root of \g, and 
since one coefficient is positive, it is again a positive root of \g. 
Moreover, since $n_j\neq 0$ it is clear that it cannot be on the orbit of 
$\alpha^{(i)}$. 

To deal with the case $s_i=2$ we first recall that in this case $N_i$ is even
and that the restriction of the \dyd\ of \g\ to this orbit is the \dyd\ of
$N_i/2$ copies of the simple \lie\ $A_2$. As a consequence, in the sequel 
we can in fact restrict ourselves to the case $N_i=2$. Otherwise we first 
treat the automorphism $\omega^{N_i/2}$, which has order two and possesses
$N_i/2$ orbits each of which corresponds to the \dyd\ of $A_2$. On the 
%% CORRok: `orbits' replace by
set of orbits of $\omega^{N_i/2}$,
the automorphism $\omega$ induces an automorphism $\om'$
of order $N_i/2$; all orbits \wrt this automorphism $\om'$ have $s'_j=1$. 

For $N_i=2$ and $s_i=2$ we define 
  \be \whi := \wi\,\womi\,\wi \, .\labl{whi2}
Clearly, $\whi$ has order 2, $\whi^2=\id$. Since $\Ac i{\omD i}= -1$, 
we also have $(\wi\womi)^3=\id$ and hence 
  \be  \whomi =\womi\wi\womi =\wi\womi\wi = \whi \,. \ee
This implies that again $\whi$ and $\omega^\star$ commute. 
The action of $\whi$ on the roots of our main interest reads
  \be \whi(\alpha^{(i)}) = - \alpha^{(\omD i)} \,,
  \qquad \whi(\alpha^{(\omD i)}) = - \alpha^{(i)}\,, \qquad 
  \whi(\alpha^{(i)}+ \alpha^{(\omD i)} ) = -(\alpha^{(i)} + 
  \alpha^{(\omD i)}) \, ,\labl{wii2}
while any other positive root is again mapped on a positive root different
from $\alpha^{(i)},\ \alpha^{(\omD i)}$ and $\alpha^{(i)}+ \alpha^{(\omD i)}$. 
This can be checked explicitly by using arguments which are completely
parallel to those used in the case $s_i=1$.

Finally, we again compute the action of $\whi$ on weights in $\gstar 0$.
For any such \g-weight we have $(\lambda\Mid\alpha^{(i) \Vee}) =
(\lambda\Mid\alpha^{(\omD i) \Vee}) =: l$, and hence
  \be \begin{array}{ll} \whi(\lambda) \!\!\! &= \wi\womi\wi(\lambda) 
               = \wi\womi(\lambda - l \alpha^{(i)}) \\[.6em]
              &= \wi(\lambda-l \alpha^{(i)} - 2l \alpha^{(\omD i)}) 
               = \lambda - 2l\cdot ( \alpha^{(i)} + \alpha^{(\omD i)}) \, .
\end{array}\labl{w2}

We can summarize the formul\ae\ \erf{w1} and \erf{w2} by
  \be \whi(\lambda) = \lambda - s_i\cdot\sum_{l=0}^{N_i-1} 
  (\lambda\mid \alpha^{(\omD^l i)^\Vee})\, \alpha^{(\omD^l i)}  \, . \labl{w12}
Let us check that the prescription \erf{w12} indeed describes the
mapping $\prow$ defined by \erf{push}.
Knowing how $\Pro$ acts on $\go$, it is straightforward to determine how
$\projm$ acts on $\gOo^\star$. Let us first compute the action of $\projm$ 
on the simple coroots $\alphaO^{(i)^\Vee} := d_i \, \alphaO^{(i)}$. We
observe that the invariant bilinear form on the \olie\ $\gO$ identifies
$\gOo$ with its weight space $\gOo^\star$ in such a way that
$\alphaO^{(i)^\Vee}$ corresponds to $\HO^i$.
Also, since $\omega$ leaves the bilinear form invariant and
$(\Pro h\Mid \Pro h') = N (h\Mid h')$, the identification of $\gOo$ with 
$\gOo^\star$ corresponds to identifying the maps $\Pro$ and $\proj$
up to a rescaling by $N$. As a consequence, the dualization of the identity
$\Pro( \sum_{l=1}^{N_i} H^{\omD^l i})= N_i \HO^i$ reads
  \be \projm(\alphaO^{(i)^\Vee}) = \frac N{N_i} \sum_{l=0}^{N_i-1}
  \alpha^{(\omD^l i)^\Vee} \, .\ee
Using the relation \erf{dO} between $d_i$ and $\dO_i$ and the fact that 
$\projm$ is a linear map, we can also compute the action on the simple roots,
  \be  \projm(\alphaO^{(i)}) = \Frac1{\dO_i} \projm(\alphaO^{(i)^\Vee})
  = s_i\cdot \sum_{l=0}^{N_i-1} \alpha^{(\omD^l i)} \, .\labl{pai}
With these results, the formula \erf{push} for $\whi(\lambda)$ becomes 
  \be \begin{array}{ll}  \whi(\lambda) \equiv \projm(\wO_i(\lambdaO)) \!\!&
  = \projm(\lambdaO - (\lambdaO\Mid \alphaO^{(i)^\Vee})\, \alphaO^{(i)})
  = \lambda - \Frac1N (\lambda\Mid \projm \alphaO^{(i)^\Vee})\cdot
  s_i \dsum_{l=0}^{N_i-1} \alpha^{(\omD^l i)}  \\[3mm]
  &= \lambda - \Frac1N \, \dsum_{l=0}^{N-1} \Frac N{N_i} \,
  (\lambda\Mid \alpha^{(\omD^l i)^\Vee} ) \, s_i \dsum_{l=0}^{N_i-1}
  \alpha^{(\omD^l i)} \\{}\\[-.8em]
  &= \lambda - s_i \cdot \dsum_{l=0}^{N_i-1}
  (\lambda\Mid \alpha^{(\omD^l i)^\Vee})\, \alpha^{(\omD^l i)}  \, . 
  \end{array} \ee
Thus $\whi(\lambda)$ as defined in \erf{push} coincides with \erf{w12},
as promised. In short, both for $s_i=1$ and for $s_i=2$ we have shown that 
we can represent the generators of the Weyl group $\WO$ by elements of $W$ 
which commute with $\omega^\star$.

Of particular interest is the case where the \g-weight on which $\whi$ acts
is a Weyl vector of \g, i.e.\ an element $\rho$ of $\g^\star$ which obeys 
$\rho(H^i)=1$ for all $i\in I$. In this case \erf{w12} reads
  \be \whi(\rho) = \rho - s_i \sum_{l=0}^{N_i-1} \alpha^{(\omD^l i)} 
  \,.\labl{wref}

\subsection{$\Wh$ as a subgroup of $W$} \label{s.ww}

We can now define $\Wh$ as the subgroup of $W$ that is 
%% CORRok: `spanned' replaced by
generated
by the elements $\wh_i$ of $W$ that are defined by \erf{w12}.
In this section we show that $\Wh$ is in fact isomorphic to $\WO$, the 
Weyl group of the orbit \lie\ $\gO$, or in other words,
that the map $\prow$ which maps $\wO_i$ to $\wh_i$ as defined in \erf{push}
extends to an isomorphism of the groups $\WO$ and $\Wh$. The proof involves 
a few lengthy calculations which will be described in detail
in appendix \ref{s.wwiso}. 

First recall that the Weyl group $\WO$ can be described as a Coxeter group,
namely as the group that is freely generated by the generators $\wO_i$
modulo the relations
  \be \begin{array}{ll}
  (\wO_i)^2 = \eins &\mbox{for all}\quad i\in\IO\,, \\[1.9mm]
  (\wO_i\wO_j)^{\mO_{ij}} = \eins & \mbox{for all} \quad i,j\in\IO,\ i\neq j \,.
  \end{array} \labl{cox}
The integers $\mO_{ij}$ take the specific values
$\mO_{ij}= 2,3,4,6$ for $\AOc ij\AOc ji = 0,1,2,3$, while for $\AOc ij\AOc ji
\geq 4$ one puts $\mO_{ij}=\infty$ (and
uses the convention that $x^\infty = \eins$ for all $x$). 

We have to show that the generators $\wh_i$ obey exactly the same relations. 
Above we have already seen that the $\wh_i$ square to the identity; thus,
denoting by $\mh_{ij}$ the order of $\wh_i\wh_j$ in $W$, it remains to be
shown that $\mh_{ij}=\mO_{ij}$. To see this, we first prove that 
$\mO_{ij}$ is a divisor of $\mh_{ij}$ (and hence a fortiori
$\mh_{ij}\geq \mO_{ij}$). Namely, 
assume that $(\wh_i\wh_j)^{\mh_{ij}}\in W$ is the identity element of $W$;
then in particular it acts as the identity on the subspace $\gstar{0}$
of $\go^\star$. Hence by construction also
$(\wO_i\wO_j)^{\mh_{ij}}\in\WO$ acts as the identity on the
weight space of $\gO$; this means that it is the identity element of $\WO$,
which in turn by \erf{cox} implies that $\mh_{ij}$ must a divisible by 
$\mO_{ij}$.

The inequality $\mh_{ij}\geq \mO_{ij}$ automatically proves our assertion
for $\AOc ij\AOc ji\geq 4$. 
In the remaining cases, one can show in a case by case study that in fact
already $(\wh_i\wh_j)^{\mO_{ij}} = \eins$, which 
then concludes the proof of the isomorphism property of $\prow$. These
calculations are straightforward, but somewhat lengthy, and accordingly
we present them in appendix \ref{s.wwiso}. 

For later convenience, we also introduce the homomorphism $\epsh$ from $\Wh$ 
to $\zet_2=\{\pm1\}$ that is induced by the sign function $\epsO$ on $\WO$,
  \be \epsh(\wh):=\epsO(\prow^{-1}(\wh)) \, .        \labl{newsign}
Note that $\epsh$ is typically different from the sign function that
$\Wh$ inherits as a subgroup from the sign function $\eps$ of $W$. 

\subsection{The action of $\Wh$ on the \omchar\ of the Verma module}\label{omver}

We now consider the action of $\Wh$ on the \omchar s. As
$\Wh$ is a subgroup of $W$, its action on the \omchar\ is defined in the same
way as the action of $W$ on the ordinary characters is, i.e.\ via the action
\erf{w12} on \g-weights. In this subsection we show that the function
  \be  \CHI :=\eE^{-\rho-\Lambda} \Chil \,,  \labl{chidef}
with $\Chil$ the \omchar\ of the Verma module with highest weight $\Lambda$,
is odd under the action \erf{w12} of $\Wh$, i.e.\ 
  \be  \wh(\CHI)=\epsh(\wh) \,\CHI\, .  \labl{wco}
Note that here the sign function $\epsh$ on $\Wh$ defined in \erf{newsign}
appears, rather than the sign function $\eps$ of $W$. We also remark that, 
since the only dependence of the \omchar\ $\Chil$ of the Verma module on the 
specific highest weight $\Lambda$ is by a multiplicative factor of 
$\eE^\Lambda$, the quantity $\CHI$ is independent of the choice of $\Lambda$. 

It is sufficient to check \erf{wco} for the fundamental reflections $\whi$ which
generate $\Wh$. Thus in the sequel we consider a reflection $\whi$ with fixed
$i\in\IO$, for which we have to show that
  \be  \whi(\CHI)=-\CHI \,.  \labl{wico}
To prove this, we make use of the \pbw\ theorem. To this end we must first
choose a basis \bminus\ of $\g_-$, including some enumeration of the elements 
of \bminus. Let us first deal with the case $s_i=1$. 
In this case we choose as the first $N_i$ elements of \bminus\
the step operators $E^{\omD^l i}_- $ for $l=0,1,...\,, N_i-1$ 
(the root spaces corresponding to simple roots are \onedim\
so that this prescription makes sense),
and then the step operators associated to all other negative roots in an
arbitrary ordering. The \pbw\ theorem then asserts that the set of all products
  \be   \begin{array}{ll}
  \Enm = \Ene \cdot \Emz \,, \quad&
  \Ene:=(E_{}^{-\alpha^{(i)}})^{n_0}_{}\,
  (E_{}^{-\alpha^{(\omdi)}})^{n_1}_{}\, \ldots
  (E_{}^{-\alpha^{(\omD^{N_i-1}i)}})\raisebox{.7em}{$\scriptstyle n_{N_i-1}$}\,, 
  \\[.7em]&
  \Emz:= (E_{}^{-\beta_1})^{m_1}_{} (E_{}^{-\beta_2})^{m_2}_{} \ldots \,,
  \end{array}\labl{pwform}
forms a basis of the universal enveloping algebra $\U(\g_-)$; 
 \futnote{The automorphism $\om$ of \g\ extends to an automorphism of the 
universal enveloping algebra $\U(\g)$ by simply defining
$\om(xx')=\om(x)\om(x')$ for all $x,x'\in\g$ as well as $\om(\bfe)=\bfe$.}
 here the exponents $n_i$ and $m_i$ can take all values in the non-negative 
integers in such a way that only finitely many of them are different from 
zero. As the elements $\Enm$ are linearly independent, commutator
terms that arise when reshuffling the products of generators of $\g_-$ can
never give rise to a non-zero contribution to the \omchar\ $\CHI$; 
furthermore, an element $v^{(\vec n,\vec m)}=\Enm\cdot v_\Lambda^{}$
of the Verma module can contribute to $\CHI$ only
if $n_0 = n_1 =\ldots=n_{N_i-1}^{}=:n$. 

The \pbw\ theorem also implies that the contributions to $\CHI$ stemming 
from the products $\Ene$ and $\Emz$ in \erf{pwform} factorize, so that
we can investigate their transformation properties under $\whi$
separately. First, 
$\whi$ commutes with $\omT$ and maps any $\omT$-orbit of negative roots 
to some other orbit of negative roots, i.e.\ only permutes the orbits that
contribute to the second factor.
Thus, by the fact that commutator terms are irrelevant for the trace,
the contribution to $\CHI$ coming from operators of the 
type $\Emz$ is invariant under $\whi$. On the other hand, the 
contribution $(\CHI)_1$ of operators of the type $\Ene$
to $\CHI=\eE^{-\rho-\Lambda}\Chil$ can be computed explicitly as
  \be  (\CHI)_1 = \eE^{-\rho}\sum_{n=0}^\infty \exp[{n(-\alpha^{(i)}
  -\alpha^{(\omD i)} - \ldots -\alpha^{(\omD^{N_i -1}i)})}]
  = \frac{\eE^{-\rho}}{1-\exp[{-\sum_{l=0}^{N_i-1} \alpha^{(\omD^l i)}}]} \, . 
  \ee
Acting on this expression with $\whi$, with the help of \erf{wii1} and 
\erf{wref} we obtain
  \be  \whi((\CHI)_1) =\frac{\exp[{-\rho + \sum_{l=0}^{N_i-1} 
  \alpha^{(\omD^l i)}}]} {1-\exp[{\sum_{l=0}^{N_i-1} \alpha^{(\omD^l i)}}]}
  = - (\CHI)_1 \,. \ee
Combining the two factors of the product \erf{pwform}, we thus
arrive at the desired result \erf{wico}.

Next consider the case $s_i=2$. In this case we choose a different basis \bminus\ 
of $\g_-$ in order to obtain a decomposition analogous to \erf{pwform}. 
As the first three elements of \bminus\ we take the step operators 
$E_{}^{-\alpha^{(i)}}$, $E_{}^{-\alpha^{(\omdi)}}$ and $E_{}^{-\alpha^{(i)}
-\alpha^{(\omD i)}}$, 
and then again the step operators corresponding to all other negative roots
in an arbitrary ordering. A basis of $\U(\g_-)$ is then given by \erf{pwform}
with the first factor $\Ene$ replaced by 
  \be  \Ete:=(E_-^i)^{n_0}_{} (E_-^{\omD i})^{n_1}_{} 
  (E_{}^{-\alpha^{(i)}-\alpha^{(\omD i)}})^{n'}_{}  \, . \ee
The same type of arguments as in the previous case then shows that the
contribution to $\CHI$ from operators of the type $\Emz$ again
transforms trivially under $\whi$. Further, in order to have a contribution 
$(\CHI)_1$ from operators of the type $\Ete$, we need again $n_0 = n_1
=:n$. Now the transformation properties
  $\omega(E_-^i) = E_-^{\omD i}$ and $\omega( E_-^{\omD i}) = E_-^i$ imply
that 
  \be \omega( E_{}^{-\alpha^{(i)}-\alpha^{(\omD i)}}) 
  = \omega([E^i_-,E^{\omD i}_-])
  = [E^{\omD i}_-, E^i_-] = - E_{}^{-\alpha^{(i)}-\alpha^{(\omD i)}} \, . 
  \labl{minus}
This allows us to compute the contribution $(\CHI)_1$ to the \omchar\ as 
  \be \begin{array}{l}
  (\CHI)_1 = \eE^{-\rho}\dsum_{n=0}^\infty \Llb \eE^{n(-\alpha^{(i)})} \cdot
    \eE^{n(-\alpha^{(\omD i)} )} \Lrb \cdot
  \dsum_{n'=0}^\infty (-1)^{n'} \eE^{n'(-\alpha^{(i)}-\alpha^{(\omD i)})} 
  \\{}\\[-.6em] \hsp{12.1}
  = \eE^{-\rho} (1-\eE^{- \alpha^{(i)}-\alpha^{(\omD i)}})^{-1} 
             (1+\eE^{- \alpha^{(i)}-\alpha^{(\omD i)}})^{-1}  
  = \eE^{-\rho} (1-\eE^{- 2 \alpha^{(i)}-2 \alpha^{(\omD i)}})^{-1} \, .
  \end{array} \ee
Using the transformation properties \erf{wii2} and \erf{wref}
(note the additional factor of $s_i=2$ in the transformation law \erf{wref}
of $\rho$) we find that this contribution to the \omchar\ $\CHI$ changes 
sign under the action of $\whi$. Hence again we obtain \erf{wico}.
This completes the proof of \erf{wico}, and hence of \erf{wco}. 

\subsection{The action of $\Wh$ on the irreducible \omchar\ $\chIl$}

In this subsection we show that the \omchar\ $\chil$ of an \ihwm\ with
dominant integral \hw\ $\Lambda$ is even under the action of $\Wh$, i.e.\ 
  \be  \wh(\chil)=\chil \,. \labl{wcc}
Again, it is sufficient to check this for all generators $\whi$ of $\Wh$.
Thus for all $i\in\IO$
we have to show that any weight $\lambda\in\gstar{0}$ contributes in the same
way to the \omchar\ as the weight $\whi(\lambda)$. 

Let us first deal with the case $s_i =1$. Then the subalgebra $\g_i$ of \g\
that is spanned by the generators $E_\pm^{\omdd li}$ and $H^{\omdd li}$,
  \be  \g_i := \langle E_\pm^{\omdd li}, H^{\omdd li}\mid
  l=0,1,\ldots,N_i-1 \rangle \,, \ee
is isomorphic to a direct sum of $N_i$ copies of $A_1$ algebras,
  \be  \g_i \cong \underbrace{A_1 \oplus A_1 \oplus \ldots
  \oplus A_1 }_{N_i\ {\rm summands}} \, . \ee
Now consider the decomposition
  \be  \calh_\Lambda = \bigoplus \calh_{(L_k)} \,  \ee
of the \irmod\ $\calh_\Lambda$ of \g\ into \irmod s $\calh_{(L_k)}$ 
of $\g_i$. As the \hw\ $\Lambda$ is dominant integral, each of the
modules $\calh_{(L_k)}$ has dominant integral \hw, i.e.\ for any value
of $k$ each of the $N_i$ numbers $L_k$, $k=1,2,...\,,N_i$, is a 
non-negative integer. Also, according to the representation theory of $A_1$, 
any weight of such a module \wrt the subalgebra $\g_i$ is then a sequence 
of $N_i$ integers $\ell_k$, $k=1,2,...\,,N_i$, and all these weights
are non-degenerate. 

A module $\calh_{(L_k)}$ of $\g_i$ can of course only contribute to the 
\omchar\ if it is mapped onto itself by $\tauo$. For the rest of the 
discussion we will assume that the module under consideration fulfills
this condition (otherwise no state of the module contributes to the
trace, so that the contribution is trivially symmetric under $\Wh$). 
Now to the \omchar\ only those states can 
contribute for which $\ell_1=\ell_2=\ldots=\ell_{N_i}$. 
Thus we have to show that the unique state $v$ in 
%% CORRok: ` $\calh_{(\ell_k)}$ ' replaced by
$\calh_{(L_k)}$ 
with $\ell_1=\ell_2=\ldots=\ell_{N_i} =: l>0$ contributes precisely with the 
same phase to $\chil$ as the unique state $v'$ with 
$\ell'_1=\ell'_2=\ldots=\ell'_{N_i} =: - l$. 
Now $v'$ can be obtained by acting on $v$ as
  \be v' = (E_-^i E_-^{\omD i} \ldots E_-^{\omD^{N_i-1}i})^l\, v \, . \ee
We now combine the identity $\omega(E_-^i E_-^{\omD i} \ldots 
E_-^{\omD^{N_i-1}i}) = E_-^i E_-^{\omD i} \ldots E_-^{\omD^{N_i-1}i}$ 
and the $\om$-twining property \erf{C'} of the map $\tauo$ 
to find that the eigenvalue equation $\tauo(v)= \zeta^k v$ implies
  \be \tauo(v')= \tauo((E_-^i E_-^{\omD i} \ldots E_-^{\omD^{N_i-1}i})^l v)
  = \omega( (E_-^i E_-^{\omD i} \ldots E_-^{\omD^{N_i-1}i})^l ) \tauo(v)
  = \zeta^k(v') \,. \ee
Thus $v$ and $v'$ contribute the same phase $\zeta^k$, which
proves our claim \erf{wcc} in the case $s_i=1$. 

To deal with the case $s_i=2$ we can again assume that $N_i=2$. In this case
we define $\g_i$ as the subalgebra
  \be  \g_i := \langle E_{}^{\pm\alpha^{(i)}\pm\alpha^{(\omD i)}} , 
  H^i+H^{\omD i}\rangle \ee
of \g, which is isomorphic to $A_1$. The automorphism $\omega$ acts on 
$\g_i$ as (compare \erf{minus})
  \be \omega( E_{}^{\pm\alpha^{(i)}\pm\alpha^{(\omD i)}}) 
  = -  E_{}^{\pm\alpha^{(i)}\pm\alpha^{(\omD i)}}  , \qquad
  \omega( H^i+H^{\omD i}) = H^i+H^{\omD i} \, .  \labl{oddom}
Again we decompose the \irmod\ $\calh_{\Lambda}$ of \g\ into \irmod s 
$\calh_{\ell}$ of $\g_i$, for which again the weights are non-degenerate. 
Only states which have the same eigenvalue for $H^i$ and $H^{\omD i}$
can contribute to the \omchar. Thus we have to show that the unique state 
$v$ with $H^i v = H^{\omD i} v = l v$ ($l>0$)
contributes the same phase as the unique state $v'$ obeying
$H^i v' = H^{\omD i} v' =-l v'$. Now we have
  \be  v' = ( E_{}^{-\alpha^{(i)} - \alpha^{(\omD i)}})^{2l} v \,, \ee
where the factor of two arises because the $H^i$- and $H^\omdi$-eigenvalues
are added up. Thus only even powers of the step operator 
$E^{-\alpha^{(i)}-\alpha^{(\omD i)}}$ occur. Because of \erf{oddom} 
the vectors $v$ and $v'$ therefore contribute with the same phase (which for
$N_i=2$ is a sign), and hence the claim \erf{wcc} is again proven.

\subsection{The linear relation between irreducible and Verma characters}

In this subsection we show that the \omchar\ $\chil$ of the \irmod\ \hill\
can be written as an (infinite) linear combination, with complex 
coefficients, of the \omchar s of certain
Verma modules. We first need to introduce some notation; as in \erf{weide},
for two weights $\lambda, \mu \in \go^\star$ we write $\mu\leq \lambda$ iff
the difference $\lambda -\mu$ is a
non-negative linear combination of the simple roots of \g, i.e.\ iff
  \be \lambda-\mu = \sum_{i\in I} n_i \, \alpha^{(i)} \qquad\mbox{with}\qquad
  n_i\in \zetpluso \, . \labl{lmm}
To any such pair of weights we associate a non-negative integer, the depth, by
  \be \dpth_\lambda(\mu):= \sum_{i\in I} n_i \, . \ee

Let us assume that $\lambda$ is a symmetric weight, $\lambda\in\gstar0$. 
We claim that for any such $\lambda$  we can find complex numbers
$\tilde c_{\lambda \mu}$ with $\tilde c_{\lambda \lambda} =1$ such that
  \be  \CHil = \sum_{\mu\leq\lambda} \tilde c_{\lambda \mu}\, 
  \chi_\mu^{(\omega)} \, , \labl{551}
where $\CHil$ denotes the \omchar\ of the Verma module with highest weight
$\lambda$ and $\chii_\mu^{(\omega)}$ the \omchar\ of the irreducible module
with highest weight $\mu$. Note that the weights $\lambda,\mu$ need not be 
dominant integral. (Also, for non-symmetric weights $\CHil$ vanishes, so 
that the assertion is trivially true.)

To prove \erf{551} we define inductively a sequence $\verma n$ of (finite) 
linear combinations of \omchar s of irreducible modules,
  \be \verma n:= \sum_{\scriptstyle\mu\leq\lambda \ \ \atop \scriptstyle 
  \dpth_\lambda(\mu)\leq n}  \!\!\!
  \tilde c^{[n]}_{\lambda\mu}\, \chii_\mu^{(\omega)} \, , \ee
such that the coefficient of $\eE^\mu$ in $ \CHil - \verma n$ vanishes
for any $\mu$ with $ \dpth_\lambda(\mu)\leq n$ and that 
  \be \tilde c_{\lambda\mu}^{[n]} = \tilde c_{\lambda\mu}^{[\dpth_\lambda(\mu)]} 
  \qquad \mbox{for all }\ n\geq \dpth_\lambda(\mu)\, . \labl{552}
At depth zero, there is a single state we have to take into account, namely
just the highest weight vector, and hence we define $\tilde c_{\lambda\mu}^{[0]}
:= \delta_{\lambda,\mu}$. Next, suppose that we have already defined $\verma n$
for some value of $n>0$. Then the difference $\CHil - \verma n$ is of the form
  \be \CHil - \verma n = \sum_{\scriptstyle\mu\leq\lambda \atop
  \scriptstyle\dpth_\lambda(\mu)= n+1\ } \!\!\!\!\!\! d^{[n]}_\mu\, \eE^\mu 
  + \!\!\!\sum_{\scriptstyle\mu\leq\lambda \atop \scriptstyle
  \dpth_\lambda(\mu)>n+1\ } \!\!\!\!\!\! d^{[n]}_\mu  \, \eE^\mu \, , \ee
where $d^{[n]}_\mu$ are some complex numbers. We then define
  \be \verma{n+1} := \verma n + \!\!\!\sum_{\scriptstyle\mu\leq\lambda 
  \atop \scriptstyle\dpth_\lambda(\mu)= n+1\ } \!\!\!\!\!\! 
  d_\mu^{[n]} \,  \chii_\mu^{(\omega)} \,. \ee
This way we only add terms proportional to $\eE^\nu$ with 
$\dpth_\lambda(\nu) \geq n+1$ to $\verma n$, so that
$\tilde c_{\lambda\mu}^{[n+1]} =  \tilde c_{\lambda\mu}^{[n]} 
  = \tilde c_{\lambda\mu}^{[\dpth_\lambda(\mu)]}$
for $\dpth_\lambda(\mu) \leq n$. Moreover, all terms proportional to
$\eE^\mu$ with $\dpth_\lambda(\mu) = n+1$ are removed from the difference
$\CHil - \verma n$, because any irreducible \omchar\ contributes at 
$\dpth(\mu)$ only through the highest weight vector, while all other 
contributions are at higher depths. This shows that the quantities 
$\verma n$ possess the properties stated above. From \erf{552} and the 
properties of the $\tilde c_{\lambda \mu}^{[n]}$ described above 
it then follows immediately that \erf{551} holds, with the coefficients 
$\tilde c_{\lambda \mu}$ given by 
  \be  \tilde c_{\lambda\mu}:= \tilde c_{\lambda\mu}^{[\dpth_\lambda(\mu)]}
  \labl{553}

The weights $\mu$ which give a non-vanishing contribution to the sum
\erf{551} have to obey further requirements in addition to $\mu\le\lambda$.
First note that the \omchar\ $\chi_\mu^{(\omega)}$ vanishes unless 
$\mu\in\gstar 0$. Writing $\lambda - \mu$ as in \erf{lmm},
the fact that both $\lambda$ and $\mu$ are fixed under  
$\omT$ implies that $n_{\omdi}=n_i$ for all $i\ini$, so that
  \be  \lambdaO - \muO = \sum_{i\in\IO} n_i \alphaO^{(i)} \, , \ee
with $\lambdaO = \proj(\lambda),\ \muO = \proj(\mu)$, and hence
$\muO\le\lambdaO$. 

Next, we consider the generalized second order Casimir 
operator of \g, defined as \cite{KAc3}
  \be \ctwo := 2\, (\rho\Mid H) + \sum_{I=1}^{2n-r}  (u^I\Mid u_I) + 
  2 \sum_{\alpha>0} \sum_\ell E^{\alpha,\ell}_- E^{\alpha,\ell}_+ . \ee
Here $\{u^I\}$ and $\{u_I\}$ denote any two dual bases of $\go$, the sum 
over $\ell$ in the last 
term takes care of the possible (finite) degeneracies of roots,
and in the first term we implicitly identify $\go$ with its dual space 
$\go^\star$ with the help of the invariant bilinear form. Finally,
the \g-weight $\rho$ is a Weyl vector of \g, i.e.\ a weight which obeys 
$\rho(H^i)=1$ for all $i\in I$ (if the determinant of the Cartan matrix 
$A$ vanishes, then this element is not unique; in this case we make some 
arbitrary, but definite choice for $\rho$). For any  Weyl vector of \g,
the projected $\gO$-weight $\rhoO=\proj(\rho)$ is a Weyl vector of $\gO$. 
With the above results, we can then relate the eigenvalues of the 
generalized second order Casimir operators of \g\ and $\gO$.
The operator $\ctwo$ has the constant value 
  \be  C_2(\lambda) = (\lambda+2\rho\Mid\lambda)
  = |\lambda +\rho|^2 -|\rho|^2   \ee
(with $|\mu|^2\equiv(\mu\Mid\mu)$\,) on $V_{\lambda}$; taking into account 
the relation \erf{nrw} between the invariant bilinear forms
of \g\ and \gO, it therefore follows that
  \be  N\cdot |\lambdaO+\rhoO|^2 = |\lambda+\rho|^2 = |\mu+\rho|^2
  = N\cdot |\muO+\rhoO|^2  \,  \ee
for all weights $\mu$ which appear in \erf{551}.

In summary, for any \g-weight $\Lambda$ all weights appearing in the 
decomposition of the \tcha\ of the Verma module $V_\Lambda$ that is 
analogous to \erf{551} are contained in the subset 
  \be  \BB(\Lambda):= \{\, \lambda=\projm(\lambdaO) \mid \lambdaO 
  \le\LambdaO, \ |\lambdaO+\rhoO|^2 = |\LambdaO+\rhoO|^2 \,\}    \ee
of the weight space of \g.
Also, we can assume that the elements of $\BB(\Lambda)$ are  indexed by
the positive integers, $\BB(\Lambda)= \{\lambda_i\Mid i\in\natnum\}$,
in such a way that $\lambdaO_j \le \lambdaO_i$ implies that
$i$ is smaller than $j$. Applying the formula \erf{551} to all elements 
of $\BB(\Lambda)$, it then follows that for all $\lambda_i$ we have
  \be  \Chili = \sum_{\lambda_j\in \BB(\Lambda)}
  \tilde c_{ij}\, \chilj  \labl{lincom3}
with complex coefficients $\tilde c_{ij}$.

Moreover, by construction we have $\tilde c_{ii} =1$, and $\tilde c_{ij}$ 
can be non-zero only if $\lambdaO_j\le\lambdaO_i$, which due to the chosen 
ordering in $\BB(\Lambda)$ implies that $i\leq j$. 
Hence the (infinite) matrix $\tilde c=(\tilde c_{ij})$
is upper triangular so that it can be inverted; its inverse $c=(c_{ij})$ 
is upper triangular as well and obeys $c_{ii}=1$. This shows that the 
following kind of inverse of the formula \erf{551} holds: the \omchar\ 
$\chil$ of the \irmod\ with \hw\ $\Lambda$ can be written as an (infinite) 
linear combination \futnot{not finite hence not linear combination}
  \be  \chil = \sum_{\lambda\in \BB(\Lambda)} c^{}_{\lambda} \,
  \CHil  \labl{lincom}
of the \omchar s of Verma modules with highest weights in $\BB(\Lambda)$, 
where the $c_\lambda$ are complex numbers such that $c_\Lambda=1$.

\subsection{The character formula}

We are now in a position to complete the proof of Theorem 1. 
Assume that the \hw\ $\Lambda$ is dominant integral, and let us
write the linear relation \erf{lincom} as
  \be  \chil = \sum_{\lambda\in \BB(\Lambda)} c^{}_{\lambda} \, 
  \eE^{\rho+\lambda} \cdot \eE^{-\rho-\lambda} \CHil
  = \LLb\!\! \sum_{\lambda\in \BB(\Lambda)} c^{}_{\lambda} \, \eE^{\lambda+\rho}
  \LRb \cdot \CHI \,, \labl{lc}
where we used the fact that $\CHI\equiv\eE^{-\rho-\lambda} \CHil$ is 
independent of $\lambda$.
The results of the two preceding subsections show that $\CHI$ is odd under 
the action of $\Wh$ (\wrt the sign function $\epsh$ inherited from $\WO$),
while the left hand side of \erf{lc}
is even under $\Wh$. This implies that the sum in brackets on the \rhs\ must 
be odd under $\Wh$, which means that $c_{\lambda}=\epsh(\wh)c_{\mu}$
whenever there is an element $\wh\in\Wh$ such that $\wh(\lambda+\rho)=\mu+\rho$. 
Thus for all $\wh\in\Wh$ we have
  \be  c_\lambda= \epsh(\wh)\, c_{\wh(\lambda+\rho)-\rho}    \ee
with $\epsh(\wh)$ as defined in \erf{newsign}.

Moreover, as $\Lambda$ (and hence also $\LambdaO=\proj(\Lambda)$) is 
dominant integral, with any weight $\lambda$ the weight system of the 
irreducible module already contains the full $\Wh$-orbit of $\lambda$.
As a consequence, we actually need to know $c_{\lambda}$ only for a single
element of each orbit of the action of $\Wh$; moreover, only weights in 
$\gstar0$ contribute. Now $\proj$ intertwines the action of $\WO$ on 
$\gOo^\star$ and the 
action of $\Wh$ on $\gstar 0$; any orbit of the $\WO$-action contains a 
unique representative in the fundamental Weyl chamber of $\gO$. Since the 
only weight $\lambdaO$ in the fundamental Weyl chamber of $\gO$ with 
$\lambdaO\leq\LambdaO= \proj(\Lambda)$ for which $\lambdaO+\rhoO$ has the
same length as $\LambdaO+\rhoO$ is the \hw\ $\LambdaO$ itself, we learn 
that $\BB(\Lambda)$ contains only a single orbit of $\Wh$, namely that of
the \hw\ $\Lambda$. Together with $c_\Lambda=1$
this implies that \erf{lc} can be rewritten as
  \be  \chil = ( \dsum_{\wh\in\Wh} \epsh(\wh)\, \eE^{\wh(\Lambda+\rho)})
              \cdot \CHI  \, . \labl{fast}

We can now use the fact that any symmetrizable \kma\ \g\ possesses the 
trivial one-dimensional \irmod\ with highest weight $\Lambda=0$. 
This weight is obviously a symmetric weight; also, by definition, $\tauo$ 
leaves the highest 
weight vector fixed, and hence in this special case the \omchar\ is constant, 
$ \chii^{(\omega)}_0 = 1$. Evaluating \erf{fast} for this case we find
  \be 1 = \chii^{(\omega)}_0 = 
  ( \sum_{\wh\in\Wh} \epsh(\wh)\, \eE^{\wh(\rho)}) \cdot \CHI  \, . \ee
This allows us to read off the explicit expression 
  \be \CHI \equiv ( \sum_{\wh\in\Wh} \epsh(\wh)\, \eE^{\wh(\rho)})^{-1} \, \ee
for $\CHI$. When inserted into \erf{chidef} and \erf{fast}, this yields 
the explicit expressions 
  \be \Chil = \eE^{\Lambda+\rho} \CHI = \eE^{\Lambda+\rho} \,
  ( \sum_{\wh\in\Wh} \epsh(\wh) \,\eE^{\wh(\rho)})^{-1}  \labl{vres}
for the \omchar\ of the Verma module, and
  \be \chil = \frac{ \sum_{\wh\in\Wh} \epsh(\wh)\,  \eE^{\wh(\Lambda+\rho)}}
  { \sum_{\wh\in\Wh} \epsh(\wh) \, \eE^{\wh(\rho)}}    \labl{umf} 
for the \tcha\ of the irreducible module.

We now observe that for any $h\in\go^{(0)}$ we have
  \be  (\wh(\Lambda+\rho))(h) = ([\prow(\wO)](\Lambda+\rho))(h)
  = (\projm \wO \proj(\Lambda+\rho))(h) = (\wO (\LambdaO +\rhoO))(\Pro h)  \ee
with $\LambdaO= \proj(\Lambda)$. When combined with \erf{umf}, this implies
that 
  \be \chil(h) = \frac{ \sum_{\wO\in\WO} \epsO(\wO)\,  
  \eE^{(\wO(\LambdaO+\rhoO))(\Pro h)}}
  { \sum_{\wO\in\WO} \epsO(\wO) \, \eE^{(\wO(\rhoO))(\Pro h) }} \ee
for all $h\in\go^{(0)}$.
By the usual Weyl\hy Kac character formula for the integrable highest 
weight module with highest weight $\LambdaO$, this means that
  \be  \chil(h) = \chii_\LambdaO (\Pro h) \, .  \ee

This completes the proof of part b) of theorem 1. Analogously, part a)
of theorem 1 follows by comparing \erf{vres} with the formula for the
Verma module characters of $\gO$ (since $\CHI$ is independent of 
$\Lambda$, this result holds for arbitrary
highest weights, not just for dominant integral ones).

\Sect{Simple current automorphisms of untwisted affine \lie s}{aff}

\subsection{Centrally extended loop algebras}

Let us now specialize to the case where \g\ is an \uaff, which is relevant
for applications in \cft. Among the diagram automorphisms of the untwisted 
affine \lie s, there is a particularly interesting subclass which 
corresponds to the action of simple currents in the WZW models of \cft. 
{}From now on we will also restrict to these specific diagram automorphisms; 
abstractly, they can be characterized as the elements of the
unique maximal abelian normal subgroup ${\cal Z}(\g)$ of the
group $\Gamma(\g)$ of diagram automorphisms; this abelian subgroup
is isomorphic to the center of the universal covering Lie group 
that has the horizontal subalgebra $\gb\subset\g$ as its \lie.
Also, the remark at the end of section \ref{s.lie} shows that in this situation
the equations \erf{normrel} and \erf{nrw} are valid in the conventional
normalization of the invariant symmetric bilinear form. 

In the affine case, the rank $r$ of the $n\times n$
Cartan matrix $A$ is $n-1$. Hence the space $\gD$ of derivations
is \onedim. However, one usually does not choose the 
derivation $D$ in the way we did in the general case, i.e.\ such that 
$\omega(D)=D$, but rather in a way which is suggested by the realization of 
affine \lie s in terms of centrally extended loop algebras. We will denote
the latter derivation by $L_0$. 

In the description of untwisted \aff s via loop algebras, a basis of 
generators of \g\ is given by $\HH im$ and $\EE\alphab m$ together with the 
canonical central  
element $K$ and the derivation $L_0$. Here $m$ takes values in \zet, $i$
takes values in the index set $\Ib$ that corresponds to the \hsa\ \gb\ of \g,
and $\alphab$ is a root of \gb. The \hsa\ is a simple \lie; for 
$\g=X_r^{(1)}$ it is given by $\gb=X_r$. 
 \futnot{ By a slight abuse of terminology, in other papers often both \g\ 
and its derived algebra \gd\ are called the (untwisted)
affine \lie\ associated to \gb. This is justified because the  
extension by $D$ is unique up to isomorphism (although of course $D$ itself
is not unique).}
The rank of \gb\ is equal to the rank $r=n-1$ of $A$, so that it is natural to
write the index set $I$ as
 \futnote{Note, however, that by construction the index set $\ti$ is then
generically {\em not\/} the set $\{0,1,...\,,\rank\gOb\}$.}
  \be  \I:= \Ib\cup\{0\}=\{0,1,2,...\,,r\} \,.  \labl{Ib}
In this basis the Lie brackets of \g\ read
  \be \begin{array}{l}
  [\HH im,\HH jn]=m\,\GGb ij\,\delta_{m+n,0}\,K \,, \\ [2.2 mm]
  [\HH im,\EE\alphab n]=\alphab^i \EE\alphab{m+n} \,,\\ [2.2 mm]
  [\EE\alphab m,\EE{\alphab'}n]=e_{\alphab,\alphab'}\EE{\alphab+\alphab'}{m+n} 
      % \qquad {\rm for\ } \alphab+\alphab'\,\mbox{ a \gb-root}
  \,, \\ \mbox{ }\\[-2.1 mm]
  [\EE\alphab m,\EE{-\alphab}n]= (\alphab^\Vee,\HH{}{m+n}) + m\delta_{m+n,0}K 
  \,, \\ \mbox{ }\\[-2.1 mm]
  [L_0,\HH im]=-m\,\HH im\,, \qquad [L_0,\EE\alphab m]=-m\,\EE\alphab m\,,
  \end{array}\labl L
together with $[\,\cdot,K]=0$. It is implicit in \erf L that 
$e_{\alphab,\alphab'}=0$ if $\alphab+\alphab'$ is not a \gb-root. 
Further, $\bar B$ is the symmetrized Cartan matrix of \gb,
   \be  \GGb ij = (\bar\alv i,\bar\alv j) = \frac2{(\bar\als i,\bar\als i)}
   \,\Acb ij  \labl{GG}
for $i,j\inib$,
with $\alphab^{(i)}$ the $i$th simple root and $\alphab^\Vee\equiv2\alphab/
(\alphab,\alphab)$, and we implicitly identify the Cartan subalgebra of 
\gb\ with the weight space inasfar as we use the notation
  $  (\lambdab,\HH{}m) = \sumrE i \lambdab_i \HH im \equiv
  \sumrE{i,j} \Gb ij \lambdab^i \HH jm  $,
with $\bar G$ the inverse of $\bar B$, for any \gb-weight $\lambdab$. 
The relation between the inner product $(\cdot\,,\cdot)$ on \gb\ 
and the invariant bilinear form $(\cdot\Mid\cdot)$ on \g\ is
  \be  (h\mid h') = (\bar h\,,\bar h') + \xi\eta' + \xi'\eta \ee
for $h=\bar h+\xi K-\eta L_0$ with $\bar h\in\bar\g_\circ$. 
  \futnot{$\bar h=\sumrE i h_i H^i$}
Note that the normalization of $\bar G$, or
equivalently the normalization of $(\cdot\,,\cdot)$ 
is arbitrary; we fix this freedom such that the 
highest \gb-root $\ttab$ has length squared 2. Then in particular the 
{\em level\/} $\kvl$, defined for any vector of \g-weight $\lambda$ by 
$\kvl=2k_\lambda/(\ttab,\ttab)$, is equal to the  
eigenvalue $k_\lambda$ of the canonical central element $K$.

The step operators associated to the simple roots $\als i$ with $i\inib$ 
are given by
  $  \EE i\pm \equiv\EE{\pm\alpha^{(i)}}{} = \EE{\pm\alphab^{(i)}}0$  %\labl{Ei}
and the corresponding Cartan subalgebra elements by $\HH i{}\equiv
[\EE i+,\EE i-]=\HH i0$, while
the step operators associated to the zeroth simple root $\als0$ read
  $  \EE 0\pm = \EE{\mp\ttab}{\pm1}$,   %\labl{E0}
and the corresponding \csa\ element is $\HH0{}\equiv[\EE0+,\EE0-]=\HH00$. 
More generally, we introduce the elements $\HH0n$ as the linear combinations
  \be  \HH0n = K\,\delta_{n,0} - \sumre j \avj \HH jn  \,. \labl{H0}
The level of a weight $\lambda$ is then related by
  \be  \kvl= \sum_{i=0}^r a_i^\Vee \lambda^i  \labl{kvl}
to its Dynkin components $\lambda^i$.
Further, for $i\inib$ the Coxeter and dual Coxeter labels  coincide
with the expansion coefficients of the highest \gb-root $\theta$ in the basis
of simple roots and the basis of simple coroots, respectively, so that
\erf{H0} may be rewritten as
  $    \HH0n = K\,\delta_{n,0} - (\ttab,\HH{}n)$.  %\labl{H00}
We also note that according to \erf{kvl} the component $\lambda^0$ of a
weight is redundant if only weights at a fixed level are considered.

\binternal{
As the level of the adjoint \rep\ of \g\ is zero, in the case of roots 
$\beta$ the zeroth component satisfies
  \be  \beta^0 = -\sumre j \avj\beta^j   \,.  \labl{ao}
This is one ingredient in the verification of the fact that the components 
of the (horizontal projections of the) simple roots $\alphab^{(i)}$ in 
the Dynkin basis are the rows of the Cartan matrix. Namely,
for the simple roots $\alphab^{(i)}$ with $i\inib$, which satisfy
$(\alphab^{(i)})^j = \Acb ij=\Ac ij$ for all $j\inib$, \erf{ao} together
the second of the eigenvalue equations \erf{aav} implies
  \be  (\alphab^{(i)})^0 = -\sumre j \avj \Ac ij = \av0 \Ac i0 = \Ac i0 \,.
  \labl{ali0}
Similarly, for the zeroth simple root $\alphab^{(0)}=-\ttab$  one has
  \be  (\alphab^{(0)})^j = -\sumre i a_i (\alphab^{(i)})^j
  = - \sumre i a_i \Ac ij = a_0 \Ac0j = \Ac0j  \quad\; {\rm for\ all\ }  
  j\inib\,, \ee
and hence
  \be  (\alphab^{(0)})^0 = -\sumre j \avj \Ac0j = \Ac00 =2  \,.  \labl{al00}
}\einternal

\subsection{The derived algebra}

We will again first describe how the automorphism $\omega$ acts on the 
derived algebra $\gh$ of \g\ and later study the action on the derivation.
On the generators $\HH im$ and $\EE\alphab m$, which together with $K$
span $\gh$, the diagram automorphism $\om$ acts as
  \be  \om(\HH in) = \HH{\omd i}n \,, \qquad 
  \om(\EE\alphab n) = \etaa\,\EE{\omtb\alphab}{n+\ela} \,, \labl{om}
while it leaves the canonical central element $K$ fixed,
  \be  \om(K) = K  \,. \labl{omK}
Here we use the following notation.
In \erf{om} the prefactors $\etaa$ are signs which are +1 for the simple 
roots and can be deduced for all other roots by writing the step operator 
for a non-simple root as a (multiple) commutator of step operators for 
simple roots and then using the automorphism property of $\om$ to extend 
the action \erf{oei} of $\omega$ on the step operators for simple roots.
(We have already encountered an example with $\etaa=-1$ when we
calculated $\omega(E^{\alpha^{(i)} + \alpha^{(\omD i)}})$ in subsection
\ref{omver}, cf.\ equation \erf{minus}.)
Also, the index $i$ in principle only takes values in the unextended
index set $\Ib$, and for $i=\omdm0$ the identity \erf{H0}
is implicit on the right hand side (note that $\omdm0\in\Ib$ for all
$\om\in{\cal Z}(\g)$). However, owing to this same identity
and the invariance \erf{omK} of $K$, the relations \erf{om}
are still valid if one allows for
$i$ to lie in the extended index set $\I$, i.e.\ including $i=0$.
Further, in \erf{om} we introduced for any \gb-root $\alphab$ the number
$\ela$ defined by 
  \be  \ela:= (\alphab,\lab{\omdd{-1}0}) \,, \labl{ela}
  \futnot{$\ela= \alpha^{}_{\omdd{-1}0}$}
where we denote by $\lab i$, $i=\onetor$, the {\em horizontal 
fundamental weights}, i.e.\ the fundamental weights of the \hsa\ \gb.
Finally, we introduced a map $\omtb$ on the weight space of $\gb$; it is 
defined by the following action on the Dynkin components $\lambdab^j$ 
of the weight $\lambdab$:
  \be (\omtb \lambdab)^j = \lambdab^{\omD^{-1} j} \quad{\rm for}\ j \neq 
  \omD 0\,, \labl{omtla}
while
  \be (\omtb \lambdab)^{\omD 0} = \lambda^0 \equiv \kvl- \sum_{j=1}^r a_j^\Vee 
  \lambdab^j\,. \ee
Hence
  \be  \omtb \lambdab = \kvl\Lambdab_{(\omD 0)} + \sum_{{\scriptstyle 
  j=1\atop \scriptstyle j\neq\omD0}}^r \lambdab^{\omD^{-1}j} \Lambdab_{(j)} 
  - (\sum_{j=1}^r a_j^\Vee \lambdab^j)\, \Lambdab_{(\omD 0)}\,.  \ee
Note that $\omtb$ is an affine mapping on the weight space of \gb. 

As the components of the simple \gb-roots $\alphab^{(i)}$ in the Dynkin
basis are just the rows of the Cartan matrix of \gb, the definition of $\omtb$
implies in particular that
  \be  \omtb{\alphab^{(i)}} = \alphab^{(\omdi)} 
  \ \ {\rm for}\ i\neq\omD^{-1}0\,, \qquad \omtb(\alphab^{(\omD^{-1}0)})
  =-\thetab \,. \labl{oai}
By making use of the Serre relations and the invariance property
of the Cartan matrix, it then follows that $\omtb \alphab$ is a \gb-root  
whenever $\alphab$ is (and hence the notation $\EE{\omtb\alphab}{n+\ela}$  
introduced in \erf{om} indeed makes sense).

The action of $\omtb$ on the roots of \gb\ can be described more concretely 
as follows. We write any root $\betab$ of \gb\ as a linear combination of 
simple coroots as $\betab = \sumrE i \beta_i \alphab^{(i)^\Vee}$, and
for simplicity set $\beta_0=0$ for all roots $\betab$ as well as 
$\alphab^{(0)^\Vee} = -\thetab$. With these conventions we have
  \be \ela = (\alphab,\Lambdab_{(\omD^{-1}0)}) = \alpha_{\omD^{-1}0}^{} \, ,\ee
while the action of $\omtb$ on the roots is
  \be \omtb \alphab = \sum_{j=0}^r \alpha_j \alphab^{(\omD j)^\Vee} = 
  -\ela \thetab + \sum_{j=1}^r \alpha_{\omD^{-1}j}^{}
  \alphab^{(j)^\Vee} \, .\labl{da1}
We can use this result to derive a relation which we will need in subsection
\ref{ssvir}; since for untwisted affine \lie s we have $a_0^\Vee 
= a_{\omD 0}^\Vee =1$, so that $(\Lambdab_{(\omD 0)},\thetab) 
= a_{\omD 0}^\Vee =1$, the relation \erf{da1} implies
  \be (\Lambdab_{(\omD 0)}, \omtb\alphab) = -\ela +\alpha_0 = - \ela \, .
  \labl{da2}

\binternal{
Let us mention some properties of the numbers $\ela$ that were defined in
\erf{ela}. First, we have $|\ela|\leq\el_{\bar\theta}=1$, hence
  \be  \ela\in \{1,0,-1\}  \labl{ela1}
for all roots $\alpha$ of the \hsa, as well as $\ela\geq0$ for positive and
$\ela\leq0$ for negative \gb-roots. Next, the definition immediately
shows that $\el_{-\alphab}=-\ela$ and
  \be  \el_{\alphab+\alphab'} = \ela+\el_{\alphab'}    \labl{el+}
if $\alphab+\alphab'$ is a root. Furthermore, we have
  \be  \el_\omtalb = - \ela+ \alpha^{}_{\omdd{-2}0}  \,,  \labl{l-2}
as follows immediately from \erf{da1}.
}\einternal

\subsection{The derivation}\label{ssder}

Let us now describe how $\omega$ acts on $L_0$.
The derivation $L_0$ is defined as the unique element of the \csa\ of \g\ 
which has the commutation relations
  \be [L_0, E^i_\pm] = \mp \delta_{i,0}\, E^0_\pm  \ee
and satisfies $(L_0\Mid L_0)=0$.
Then the automorphism property of $\om$ demands that
$[\omega(L_0), E^i_\pm] = \mp \delta_{i,\omD0} E^{\omD0}_\pm$; this
fixes $\omega(L_0)$ up to a term proportional to the central element $K$,
$\omega(L_0) = L_0 - (\Lambdab_{(\omD 0)},H) + c K.$ Indeed, using 
$(\Lambdab_{(\omD 0)},\alphab^{(i)}) = \delta_{\omD 0,i}$ for $i=\onetor$,
and $(\Lambda_{(\omD 0)},-\thetab) = -a^\Vee_{\omD 0} = -1$, we find that
$ [\omega(L_0), E^i_\pm] 
 \futnot{$=\mp\delta_{i,0} E^0_\pm \pm\delta_{i,0}E^0_\pm \mp\delta_{\omD 0}^i 
 E^i_\pm$}
= \mp \delta_{\omD 0,i} E^i_\pm$. The constant of proportionality $c$ can 
be obtained from the requirement that the invariant bilinear form is 
$\om$-invariant, which implies that $(\omega(L_0)\Mid \omega(L_0))$ vanishes,
  \be \begin{array}{ll} 0 \!\!&
  = (\omega(L_0)\mid \omega(L_0)) = ( L_0 - (\Lambdab_{(\omD 0)},H) + c K
  \mid L_0 - (\Lambdab_{(\omD 0)},H) + c K) \\[2.3mm]
  &= -2c + (\Lambda_{(\omD 0)}, \Lambda_{(\omD 0)}) \, .\end{array} \ee
Hence we obtain
  \be \omega(L_0) = L_0 - (\Lambdab_{(\omD 0)},H) + \onehalf \, 
  (\Lambdab_{(\omD 0)},\Lambdab_{(\omD 0)}) \,  K \, . \labl{618}

With this result we can now make the relation between the derivation $D$ 
that we used in the general setting and the derivation $L_0$ explicit. 
To this end recall that $D$ is uniquely characterized by the relations 
\erf{315} for $(D\Mid\cdot\,)$. To make the definition concrete,  
we choose a basis of eigenvectors of $\omega$ for the \csa\ 
$\gh_\circ$ of the derived algebra. Thus we introduce vectors 
  \be  h_m^i:= \frac1{N_i}\sum_{l=1}^{N_i} \zeta_i^{-ml} H^{\omD^l i} \, ,\ee
where $i$ takes values in $\IO$, $m=1,2,...\,, N_i$, and $\zeta_i
=\zeta^{N/N_i}$, with $\zeta$ as defined in \erf{zeta}, is a primitive  
$N_i$th root of unity. As a basis of $\gho$ we now choose all $h_m^i$,
except for $i=0=m$, together with the central element $K$. 
Rewriting the conditions \erf{315} in terms of the generators
$H^i$, we then find that $D$ is characterized by
  \be (D\Mid D) = 0\,,\qquad (D\Mid H^i) = 0 \ \ \mbox{for}\ i\neq \omD^l0\,,
  \qquad (D\Mid H^{\omD^l 0}) = \frac 1N \ \ \mbox{for all}\ l\, . \ee
\futnot{If non-simple current automorphisms are considered, $N$ has to
be replaced by $N_0$.} 
This fixes $D$ to
  \be  D = - L_0 + \frac1N \sum_{l=1}^{N-1}(\Lambdab_{(\omD^l 0)},H) 
  - \frac{\coo}{2N^2} \,K \, \labl{dlo}
with
  \be \coo := \sum_{l,l'=1}^{N-1} (\Lambdab_{(\omD^l 0)}, 
  \Lambdab_{(\omD^{l'} 0)})
  \equiv \sum_{l,l'=1}^{N-1} \Gb{\omD^l 0}{\omD^{l'} 0} \, . \labl{coo}

Let us also determine the relation between the derivation $\DO=\Pro(D)$ 
defined by \erf{331} and \erf{332} and the element $\LO_0$ 
of the orbit \lie\ $\gO$. {}From $\Pro(h_0^i)\eq\HO^i$ we learn that $\DO$
is characterized by
  \be (\DO\Mid \DO) = 0\,,\qquad (\DO\Mid \HO^i) = 0 \ \ \mbox{for}\ i\in\IO
  %%\setminus \{0\}\,,\qquad (\DO\Mid \KO) = N \, . \ee
  \setminus \{0\}\,,\qquad (\DO\Mid\Pro(K)) = N \, . \ee
  %%This fixes $\DO$ to $\DO=-N\LO_0$. 
Now \erf{avo} implies that $\Pro(K)\eq N\KO$,
   so that these properties fix $\DO$ to $\DO\eq{-}\LO_0$. 
Together with $\DO=\Pro(D)$ and \erf{dlo}, this relation shows that 
  %%\be  \Pro(L_0) = N \LO_0 +\frac1N\, \Pro(\sum_{l=1}^{N-1} 
  %%\Lambdab_{(\omD^l0)},H) -\frac{\coo}{2N^2}\, \KO \, . \ee
  $$ \PRo(L_0) = \LO_0 +\frac1N\, \PRo(\sum_{l=1}^{N-1} 
  \Lambdab_{(\omD^l0)},H) -\frac{\coo}{2N}\, \KO \, . $$ 
Here the notation $\PRo(h)$ stands for the value of the projection $\Pro$
on the symmetric part of $h$; for the middle term on the \rhs\ the 
symmetric part is explicitly
$(\sum_{l=1}^{N-1}(\Lambdab_{(\omD^l0)},H))^{(0)}_{}/N = \coo K/N^2$,
\futnot{JF: I get this by explicit calculation. the calculation is presented
below (separate page); Bert and Christoph, please check it carefully.}
    so that when using again $\Pro(K)\eq N\KO$ the latter equation reduces to
    \be  \PRo(L_0) = \LO_0 + \frac{\coo}{2N}\, \KO \, . \ee
\binternal{
The explicit calculation goes as follows. Define
  \be  X:=\llb\sumne l(\Lambdab_{(\omD^l0)},H)\lrb^{(0)}_{} \,. \ee
By the definition of `$^{(0)}$', i.e. of the symmetric part of an element of
the CSA, this is
  \be  X = \Frac1N\, \sumno m \LLb\, \sumne l\, (\Lambdab_{(\omD^l0)},
  \om^m(H))\LRb \,. \ee
Writing out the \gb-inner product, this becomes
  \be  X = \Frac1N\, \sumno m \; \sumne l \sumre i\, \Gb{\omD^l 0}i\,
  \om^m(H^i) \,. \ee
Because of $\Gb j0=0=\Gb0j$ we can extend the summations over the indices $l$
and $i$ to $l=0$ and $i=0$:
  \be  X = \Frac1N\, \sumno{l,m}\,\sumro i\, \Gb{\omD^l 0}i\, \om^m(H^i) \,. \ee
The action of $\om$ on $H^i$ is just by $i\mapsto\omD i$, hence
  \be  X = \Frac1N\, \sumno{l,m}\,\sumro i\,\Gb{\omD^l 0}i\,H^{\omD^m i} \,. \ee
Next we redefine, for each term in the $m$-summation, $i$ to $\omD^{-m}i$:
  \be  X = \Frac1N\,\sumno{l,m}\,\sumro i\,\Gb{\omD^l0}{\omD^{-m}i}\,H^i \,. \ee

So far everything is trivial. Now we use the identity \erf{6x} with $j=0$, i.e.
  \be N a_i^\Vee\, \Gb{\omD^l 0}{\omD^l 0} = a_i^\Vee \sumno m
  \Gb{\omD^l 0}{\omD^m 0} + \sumno m \Gb{\omD^l 0}{\omD^m i} \,, \ee
which implies
  \be \sumno{l,m} \Gb{\omD^l 0}{\omD^{-m} i} = a_i^\Vee\, \llb\, N\,\sumno l
  \Gb{\omD^l 0}{\omD^l 0} - \sumno{l,m} \Gb{\omD^l 0}{\omD^m 0} \lrb \,, \ee
or, together with \erf{6y}
  \be \sumno{l,m} \Gb{\omD^l 0}{\omD^{-m} i} = a_i^\Vee\cdot (N\cdot
  \Frac2N\,\coo -\coo) = a_i^\Vee\cdot\coo \,.  \ee
Inserting this into the formula for $X$ yields
  \be  X = \Frac1N\,\coo \sumro i a_i^\Vee\, H^i \,,  \ee
and by definition of $K$ this is precisely
  \be  X = \Frac1N\,\coo \, K \,.  \ee
}\einternal

\subsection{The action of $\omT$}

{}From the action of $\omega$ we can also derive the action of its dual map 
$\omT$. First, % we show that $\omega^\star(\alpha^{(i)})= \alpha^{(\omD i)}$. 
from 
  \be  (\omega^\star(\alpha^{(i)}))(H^j) = \alpha^{(i)}(H^{\omD^{-1} j})
  = \Ac i{\omD^{-1}j}=\Ac{\omD^i}j= \alpha^{(\omD i)}(H^j) \ee
it follows that $\omT(\alpha^{(i)}) = \alpha^{(\omD i)} + \xi_i \delta$,
where $\xi_i$ is some number and $\delta$ is the specific element
  \be \delta := \sum_{i=1}^r a_i \alpha^{(i)} \,, \labl{delta}
with the $a_i$ the Coxeter labels of \g, of the weight space $\g^\star$. 
(Note that the imaginary roots of \g\ are precisely the integral multiples 
of $\delta$.) Also, applying $\omega$ to the relation 
$[D,E_+^i] = \alpha^{(i)}(D) E_+^i$, we obtain
  \be \alpha^{(i)}(D) E_+^{\omD i} = \omega([D, E_+^i]) = [D, E_+^{\omD i}] 
  = \alpha^{(\omD i)}(D)  E_+^{\omD i} \,, \ee
which shows that $\alpha^{(i)}(D) = \alpha^{(\omD i)}(D)$.
To determine the constants $\xi_i$, we now apply $\omT(\alpha^{(i)})$ to
$D$. By the results just obtained, this yields
  \be  (\omT(\alpha^{(i)}))(D) = \alpha^{(\omD i)}(D) +\xi_i \,\delta(D) \,; \ee
on the other hand, from the definition of $\omT$ we obtain
  \be  (\omT(\alpha^{(i)}))(D)= \alpha^{(i)}(\omega^{-1}D) =\alpha^{(i)}(D)\,.
  \ee
Thus $\xi_i \delta(D) = 0$, which because of $\delta(D) \neq 0$ means
that $\xi_i=0$. Hence we have
  \be  \omT(\alpha^{(i)}) = \alpha^{(\omD i)}  \ee
for all $i\in I$. This implies in particular that
  \be  \omT(\delta) = \sum_i a_i\,\omT(\alpha^{(i)}) = \delta \,. \labl{dd}

Analogously, one derives how the fundamental weights 
$\Lambda_{(i)}\in\go^\star$, defined by
  \be \Lambda_{(i)}(H^j)=\delta_i^j \qquad\mbox{and}\qquad
  \Lambda_{(i)}(L_0) = 0 \,, \ee
transform under $\omT$. We find 
  \be \omT(\Lambda_{(i)}) = \Lambda_{(\omD i)} + \llb \Gb{\omD^{-1}0}i 
  - \onehalf a_i^\Vee\, (\Lambdab_{(\omdm0)}, \Lambdab_{(\omdm0)}) \lrb\, 
  \delta \, .  \labl{tcheck}
       \futnot{\def\ttauv{\mbox{$\tilde\tau_\mu^{}$}}
\def\sigv{\mbox{$\omega_\mu$}}\def\sigV{{\omega_\mu}}\def\v{\bar\mu}
More generally, the automorphism \sigv\ induces an associated
bijection of the weight space of the affine algebra \g. This map
$\ttauv$ acts on affine weights $\lambda\equiv(\hat\lambda;\,h_\lambda)$
(with $\hat\lambda$ a weight of \gd, and $h_\lambda$ the $L_0$-eigenvalue) 
as $\ttauv:\ (\hat\lambda;\, h_\lambda) \mapsto (\hat\lambda+\kv\v;\, 
h_\lambda+(\v,\lambdab)+\onehalf\,k\,(\v,\v)\,).$
This generalizes the action of elements of the abelian
normal subgroup of translations of the Weyl group of \g.
In the special case that $\mu$ is an element of the coroot lattice  
$L^\Vee$ of \gb, $\ttauv$ is indeed just an element of this translation 
subgroup of the Weyl group. Also note that the formula given already 
determines the automorphism \sigv\ of \g\ uniquely up to phases \Cite{gmow}.}

\binternal{
To check that this is the right relation, we apply both sides of the
relation to a basis of $\go$, for which we choose the $H^i$ and $L_0$. We have
  \be  (\omT(\Lambda_{(i)}))(H^j) = \Lambda_{(i)}(H^{\omD^{-1}j}) =
  \delta_i^{\omD^{-1}j} = \delta_{\omD i}^j  \,, \ee
which is manifestly the same result that one obtains by applying the 
right hand side  of \erf{tcheck}
to $H^j$. To check the relation on $L_0$, we use the relations
  \be \Lambda_{(i)}(L_0) = 0\,, \qquad \delta(L_0) = -1\,,
  \qquad\mbox{and}\qquad \Lambda_{(i)}(K) =a_i^\Vee \, . \ee
They show that application of the right hand side of \erf{tcheck} to $L_0$ 
gives $ -\Gb{\omD^{-1}0}i + \onehalf a_i^\Vee\, (\Lambda_{(\omdm0)}, 
\Lambda_{(\omdm0)})$, while \erf{618} shows that 
  \be  \begin{array}{ll}  (\omT(\Lambda_{(i)}))(L_0) \!\!\!&
  = \Lambda_{(i)}(\omega^{-1} L_0)
  =  \Lambda_{(i)}(L_0 - (\Lambdab_{(\omD^{-1} 0)},H) 
  +\onehalf (\Lambdab_{(\omdm0)}, \Lambdab_{(\omdm0)}) K)  \\[.8em] &
  = -\Gb{\omD^{-1}0}i + \onehalf (\Lambdab_{(\omD 0)}, \Lambdab_{(\omD 0)}) \,
  a_i^\Vee \,.  \end{array}\ee
}\einternal

Together with the element $\delta$ \erf{delta}, the fundamental 
weights \li\ form
a basis of the weight space $\g^\star$. Another basis of $\g^\star$ is 
given by $\delta$, $\kappa := \Lambda_{(0)}$ and by the 
horizontal fundamental weights 
  \be  \Lambdab_{(i)} = \Lambda_{(i)} - a_i^\Vee \Lambda_{(0)}  \labl{hfw}
with $i=\onetor$. The relation between
the components of a weight $\lambda$ in the two bases is
   \be  \sum_{i=0}^r \lambda^i \Lambda_{(i)} + n_\lambda \delta = \lambda 
   =\sum_{i=1}^r \lambdab^i \Lambdab_{(i)} +k^\Vee_\lambda \kappa + n_\lambda 
   \delta  \,, \ee
i.e.\ $\lambdab^i=\lambda^i$\, for $i=\onetor$\, and
$\kvl= \sum_{i=0}^r a_i^\Vee \lambda^i$. 
We also set $\Lambdab_{(0)} \equiv 0$, which will be convenient in some
calculations.

The horizontal fundamental weights \erf{hfw} transform under $\omT$ as
  \be\begin{array}{ll}
  \omT(\Lambdab_{(i)}) \!\!&= \Lambda_{(\omD i)} -a_i^\Vee \Lambda_{(0)}
  + a_i^\Vee (\Lambda_{(0)} -\Lambda_{(\omD 0)} ) + \Gb{\omD^{-1}0}i \,\delta 
  \\[2mm]
  &= \Lambdab_{(\omD i)} - a_i^\Vee \Lambdab_{(\omD 0)} + \Gb{\omD^{-1}0}i
  \,\delta\, .  \end{array}\ee
Using the relations
  \be (\delta\mid\Lambdab_{(i)}) =\Lambdab_{(i)}(K) = 0 \,,\qquad
  (\delta\mid\delta) = \delta(K) = 0 \,, \ee
and the fact that along with $\omega$ also $\omT$ is an isometry, we then 
find that the metric $\bar G$ on the horizontal weight space obeys
  \be \begin{array}{ll} \Gb ij \!\!& \equiv
  (\Lambdab_{(i)}\mid \Lambdab_{(j)}) = (\omega^\star \Lambdab_{(i)}\mid 
  \omega^\star \Lambdab_{(j)}) \\[2mm]
  &= \Gb{\omD i}{\omD j} -a_i^\Vee \Gb{\omD 0}{\omD j} -a_j^\Vee \Gb{\omD 0}
  {\omD i} + a_i^\Vee a_j^\Vee \Gb{\omD 0}{\omD 0} \, . \end{array} \ee
Applying the analogous relation for the automorphism $\omega^m$, we see that 
  \be  \begin{array}{ll} \sum_{l,l'=0}^{N-1} \Gb{\omD^l i} {\omD^{l'} j} =\!\!&
  \sum_{l,l'=0}^{N-1} \Gb{\omD^{l+m} i} {\omD^{l'+m} j}
  -N a_i^\Vee \sum_{l=0}^{N-1} \Gb{\omD^m 0}{\omD^{l+m} j} \\[.5em]
  & -N a_j^\Vee \sum_{l=0}^{N-1} \Gb{\omD^m 0}{\omD^{l+m} i}
  + N^2 a_i^\Vee a_j^\Vee \Gb{\omD^m 0}{\omD^m 0} \,, \end{array} \ee
and hence
  \be N a_i^\Vee a_j^\Vee \Gb{\omD^m 0}{\omD^m 0} =
  a_i^\Vee  \sum_{l=0}^{N-1} \Gb{\omD^m 0}{\omD^l j}
  +a_j^\Vee  \sum_{l=0}^{N-1} \Gb{\omD^m 0}{\omD^l i} \, . \labl{6x}

Define now $X_i := \sum_{l,l'=0}^{N-1} \Gb{\omD^l 0} {\omD^{l'} i}$.
Summing equation \erf{6x} over $m$, we obtain the system 
  \be  a_i^\Vee X_j + a_j^\Vee X_i  = N \sum_{m=1}^{N-1} \Gb{\omD^m 0} {\omD^m 0}
  \,a_i^\Vee a_j^\Vee =: \xi \, a_i^\Vee a_j^\Vee \,  \labl{6s}
of linear equations for the $X_i$.
These equations have the unique solution $X_i = \onehalf
\,\xi\, a_i^\Vee$. With $a_0^\Vee=1$, it then follows that
  \be  \coo \equiv \sum_{l,l'=0}^{N-1} \Gb{\omD^l 0} {\omD^{l'} 0}
  = X_0 = \frac N2 \sum_{m=1}^{N-1} \Gb{\omD^m 0} {\omD^m 0} \,  \labl{6y}
and
  \be  X_i=\sum_{l,l'=0}^{N-1} \Gb{\omD^l 0} {\omD^{l'} i} = a_i^\Vee\,X_0
  = a_i^\Vee\, \coo \labl{vstr}
with $\coo$ as in \erf{coo}. 
 \futnot{this is a remarkable relation}

\subsection{The action of $\projm$}

It is straightforward to determine the action of $\projm$ on 
$\gOo^\star$ from the action of $\Pro$ on $\go$. We only need to
observe that the invariant bilinear form identifies 
$\gOo$ with the weight space $\gOo^\star$ in such a way that 
$\alphaO^{(i)}$ corresponds to $\HO^i/\dO_i$, and that
since $\omega$ leaves the bilinear form invariant and
$(\Pro h\Mid \Pro h') = N (h\Mid h')$, the identification of
$\gOo$ with $\gOo^\star$ corresponds to identifying $\Pro$ and $\proj$
up to a rescaling by $N$. Then in particular for the simple roots we have
  \be \projm(\alphaO^{(i)}) = s_i\cdot \sum_{l=0}^{N_i-1} 
  \alpha^{(\omD^l i)} \,, \ee
as already deduced for the general case in \erf{pai}.

For untwisted affine \lie s the general relation \erf{avo} between the Coxeter 
labels and dual Coxeter labels of \g\ and of the orbit algebra \gO\ can be
made more concrete: because of the normalizations $\awO 0=a_0=1$ and 
$\avO 0=(N_0/N)\av0=\av0=1$, the relations \erf{220} and \erf{221} tell us 
that the numbers defined by \erf{avo} are precisely the conventionally 
normalized Coxeter and dual Coxeter 
labels of \gO, respectively (in particular for all untwisted affine algebras
they are integral, which for $\avO i$ is not manifest in \erf{avo}).
For the generator $\deltaO :=\sum_{i\in\IO} \aO_i \alphaO^{(i)}$ whose 
integral multiples are the imaginary $\gO$-roots, this implies
  \be \projm(\deltaO)= \delta \, . \labl{dtrans}

We can now also determine how the fundamental weights 
$\LambdaO_{(i)}\in\go^\star$ are mapped by $\projm$. As can be
checked by considering the action on $L_0$ and on the $h_0^i$, we have   
  \be \projm(\LambdaO_{(i)}) = \sum_{l=0}^{N_i-1} \Lambda_{(\omD^l i)} +
  \xi a_i^\Vee N_i\, \delta  \,,   \ee
where $\xi:= (1-2N)\coo/2N^3$ is a constant that depends only on $\omega$. 

Using the relation $a_i^\Vee=\frac N{N_i} \aO_i^\Vee$ we compute the action
of $\projm$ on the horizontal fundamental weights:
  \be\begin{array}{ll}
  \projm(\Lambdaob_{(i)}) \!\!&= \dsum_{l=0}^{N_i-1} \Lambda_{(\omD^l i)} 
  + \xi a_i^\Vee N_i\, \delta - \aO_i^\Vee ( \dsum_{l=0}^{N-1} \Lambda_{(\omD^l 0)}
  + \xi a_0^\Vee N \delta)  \\{}\\[-3mm]
  &= \dsum_{l=0}^{N_i-1} \Lambdab_{(\omD^l i)} -\drac{N_i}N a_i^\Vee
  \dsum_{l=0}^{N-1} \Lambda_{(\omD^l 0)}
  = \drac{N_i}N (\dsum_{l=0}^{N-1} \Lambda_{(\omD^l i)} - a_i^\Vee
  \Lambda_{(\omD^l 0)}) \, . \end{array}\ee
With the definition \erf{coo} of $\coo$ and the identity \erf{vstr},
this yields the relation 
  \be  \overline{\GO}_{ij} = (\Lambdaob_{(i)}\Mid \Lambdaob_{(j)}) 
  = \frac1N\, (\projm\Lambdaob_{(i)}\Mid \projm\Lambdaob_{(j)})
  = \frac{N_i N_j}{N^3} \,\Llb \sum_{l,l'=0}^{N-1} \Gb{\omD^l i}{\omD^{l'}j}
      - a_i^\Vee a_j^\Vee \coo \Lrb    \labl{Go2}
between the metrics of the horizontal part of the weight spaces of \g\ and
$\gO$.

\subsection{The Virasoro algebra}\label{ssvir}

It is natural to consider the extension
of the affine algebra \g\ to the semi-direct sum of $\g$ and the
Virasoro algebra \vir. The \lie\ \vir\ is spanned by generators $C$ and
$L_n$, $n\in\zet$; $C$ is a central element, and $L_0$ is the derivation 
of \g\ described in subsection \ref{ssder}. The Virasoro algebra 
has Lie brackets
  \be  [L_m,L_n]= (m-n)\,L_{m+n} +\mbox{$\frac1{12}$}\,(m^3-m)\,C \,,
  \labl{vir}
and its semi-direct sum with \g\ is defined by 
  \be  [L_m,\HH in]=-n\,\HH i{m+n}\,, \qquad [L_m,\EE\alphab n]=-n\,
  \EE\alphab{m+n}  \labl{virg}
and $[C,\cdot\,]=0=[K,\cdot\,]$.
It is in fact possible to extend $\omega$ to this semi-direct sum, namely via
  \be  \om(L_m)= L_m- ( \lab{\omd0}, H_m) +  
  \onehalf\,(\lab{\omd0},\lab{\omd0}) \,\delta_{m,0}\, K  \labl{omvir}
and $\om(C)=C$. It is readily checked (using in particular the identity
\erf{da2}) that $\om$ defined by \erf{om} and \erf{omvir}
is an automorphism of the Virasoro algebra and of its semi-direct sum
with \g. Note that, just as the extension from the derived algebra $\gd$
to all of \g, the extension of $\omega$ to the semi-direct sum 
$\g\oplus \vir$ is unique.

\binternal{
That \erf{omvir} defines an automorphism of the Virasoro algebra follows by
  \be  \begin{array}{l}  [\om(L_m),\om(L_n)]= [L_m,L_n]
  - [(\lab{\omd0},H_m), L_n] + [(\lab{\omd0},H_n), L_m]
  + [(\lab{\omd0},H_m),(\lab{\omd0},H_n)] \\[2.7mm] \hsp{9.8}
  = (m-n)\,L_{m+n} +\frac1{12}\,(m^3-m)\,C - (m-n)\,(\lab{\omd0},H_{m+n})
  + m\,(\lab{\omd0},\lab{\omd0}) \,\delta_{m+n,0}\, K  \\[2.7mm] \hsp{9.8}
  = (m-n)\,\om(L_{m+n}) +\frac1{12}\,(m^3-m)\,C \,.  \end{array} \ee
That $\om$ defined by \erf{om} and \erf{omvir}
is also an automorphism of the semi-direct sum of \vir\ with \g\ is seen as
follows. First, for $i=\onetor$ we have
  \be  [\om(L_m),\om(\HH in)]= [L_m-(\lab{\omd0},H_m), \HH\omdi n] 
  = -n\, \HH\omdi{m+n}-m\delta_{m+n,0}\,(\lab\omdo)^\omdi_{}\,K \,,   \labl{oloh}
which agrees with $\om([L_m,\HH in])= -n\,\HH\omdi{m+n}$ because of
$(\lab i)^j=\delta_{ij}$; second,
  \be  [\om(L_m),\om(\EE\alphab n)]= [L_m-(\lab{\omd0},H_m), \EE\omtalb{n+\ela}] 
  = -\,(n+\ela+(\lab\omdo,\omtalb))\,\EE\omtalb{m+n+\ela} \,,  \labl{oloe}
which coincides with $\om([L_m,\EE\alphab n])=-n\,\EE\omtalb{m+n+\ela}$ as a
consequence of the identity \erf{da2}.
}\einternal

\medskip
A symmetric weight satisfies by definition $\omtla=\lambda$,
which because of \erf{tcheck} implies in particular that
  \be  \lambda^i=\lambda^{\omdd li}  \labl{osll}
for all $i=\otor$ and all $l$.
This identity is certainly a necessary condition for $\omtla=\lambda$, but
in fact it is also sufficient. Namely,
for any \g-weight $\lambda = \sum_{i=0}^r \lambda^i \Lambda_{(i)} 
+ n_\lambda \delta$ with $\lambda^i=\lambda^{\omD i}$ one has 
  \be  \kvl \equiv\sumrO i \avi\lambda^i=\sum_{i\in\IO} \avi N_i\lambda^i
  \,. \labl{kvli}
Furthermore, by an argument analogous to the derivation of \erf{vstr} from 
\erf{6s}, it can be deduced from the set of equations \erf{6x} that the 
metric on the weight space of \gb\ satisfies the identities
  \be  \sumno m \Gb\omdo{\omdd mi} = \half N\,\avi\, \Gb\omdo\omdo \ee
for all $i\ini$ which are not of the form $i=\omdd n0$ for some $n$, and
  \be  \sumnz m \Gb\omdo{\omdd m0} = \half (N-2)\, \Gb\omdo\omdo \ee
(which is of course non-trivial only for $N>2$).

\binternal{
Proof: Define $Y_i := \sum_{l=0}^{N-1} \Gb{\omD0}{\omD^li}$.
Equation \erf{6x} implies the system 
  \be  a_i^\Vee Y_j + a_j^\Vee Y_i  = N \Gb{\omD 0} {\omD 0}
  \,a_i^\Vee a_j^\Vee =: \xi' \, a_i^\Vee a_j^\Vee \, \ee 
of linear equations for the $Y_i$. This has the same homogeneous part as the 
system \erf{6s} of equations for the quantities $X_i$. Therefore, there is
again a unique solution $Y_i = \onehalf \,\xi'\, a_i^\Vee$, which shows that 
  \be  Y_i=\sum_{l=0}^{N-1} \Gb{\omD 0} {\omD^{l} i} = a_i^\Vee\,Y_0
  = \onehalf N a_i^\Vee\,  \Gb{\omD 0} {\omD 0}  \, . \ee}
\einternal

Combining these identities with \erf{kvli}, one finds that for any symmetric
\g-weight one has $\sumrE i \Gb\omdo i\lambda^i=\onehalf\,
%% CORRok: `\kv' replaced by
\kvl \Gb\omdo\omdo$, or what  is the same,
  \be  (\lab\omdo,\lambdab) = \half\kvl\,(\lab\omdo,\lab\omdo) \,.  
  \labl{llk}
Now according to \erf{dd} and \erf{tcheck} the weight 
$\lambda=\sum_{i=0}^r \lambda^i \Lambda_{(i)} + n_\lambda \delta$ 
is mapped by $\omega^\star$ to 
  \be \begin{array}{ll}
  \omega^\star(\lambda) \!\!&= \sum_{i=0}^r \lambda^i \Lambda_{(\omD i)}
  + ( n_\lambda + \sum_{i=0}^r [ \Gb{\omD^{-1}0 }i -\frac12a_i^\Vee 
  (\Lambdab_{(\omD^{-1}0)}, \Lambdab_{(\omD^{-1}0)})] \lambda^i )\,\delta 
  \\[3mm] &= \sum_{i=0}^r \lambda^i \Lambda_{(\omD i)} + n_\lambda\delta+
  ( \sum_{i=0}^r \Gb{\omD^{-1}0 }i \lambda^i - \onehalf (\Lambdab_
  {(\omD^{-1}0)}, \Lambdab_{(\omD^{-1}0)}) k^\Vee_\lambda )\, \delta  \, . 
  \end{array}\ee
The relation \erf{llk} thus shows that $\omtla=\sum_{i=0}^r \lambda^i 
\Lambda_{(\omD i)}+n_\lambda\delta=\lambda$ if \erf{osll} holds. Thus 
\erf{osll} is a sufficient condition for $\lambda$ being a symmetric weight. 

It follows in particular that the pre-image $\proj(\lambda)$ of a symmetric 
weight $\lambda$ is the unique weight of \gO\ that is obtained by 
restricting to components $\lambda^i$ with $i$ in the index set $\IO$,
and with $\breve n_\lambdao=n_\lambda$, i.e.
  \be  \proj: \quad \lambda\mapsto\lambdaO=\proj(\lambda)\,, \quad
  \lambdaO^i=\lambda^i\,\ {\rm for}\ i\inio,\ \breve n_\lambdao=n_\lambda 
  \,. \ee

\futnot{With \erf{llk},
the relation \erf{omvir} shows that for all vectors of \g-modules whose
weight stays fixed under $\omT$, the action of $\om(L_0)$ coincides with
the action of $L_0$. (Another way to obtain this result is to use the 
Sugawara formula for the Virasoro generators.)}

\Sect{The order $N$ automorphism of $A_{N-1}^{(1)}$}{ANN}

We would like to be able to treat all diagram automorphisms of all affine
\lie s. Except for the automorphism of order $N$ of $\g=A_{N-1}^{(1)}$ 
which rotates the \dyd, all of these are already covered by theorem 1. 
The remaining exceptional case is the subject of theorem 2, which
we prove in the present section.

For the automorphism $\om$ of order $N$ of $A_{N-1}^{(1)}$, 
the symmetric weights $\Lambda$ obey $\Lambda^i={\rm const}=1/\kvL$
for $i=\otoNm$, so that the level $\kvL$ of any dominant integral 
symmetric weight is divisible by $N$.  Furthermore, the subspace 
$\go^{(0)}$ of $\go$ that stays fixed under $\omega$ is \twodim; it is 
spanned by the two elements $K=h_0 = \sum_{i=1}^{N-1} H^i$ and
$D = -\frac1N \sum_{l=0}^{N-1} \omega^l(L_0)+\xi K$, with $\xi$ some number
which can be deduced from \erf{dlo}. Now only symmetric weights 
$\lambda\in \gstar0$ contribute to the \omchar; for these we have
  \be \lambda(D) = -\frac1N \sum_{l=0}^{N-1} ((\omega^\star)^l \lambda)(L_0) + 
  \lambda(\xi K) = \lambda(-L_0 +\xi K) \, .  \ee
This implies that the \omchar\ of the Verma module obeys  
  \be \Chil(t K + \tau L_0) = \Chil( (t+\xi\tau)K-\tau D) \,, \ee
and an analogous formula holds for the irreducible \tcha\ $\chil$. 
As $K$ acts as a constant $k_\Lambda$ on any highest weight module, the 
dependence of the \tcha\ on the variable $t$ is only via an exponential factor,
  \be  \Chil(t,\tau) \equiv \Chil(t K + \tau L_0) = \eE^{2\pi\ii tk_\Lambda} 
  \cdot {\rm tr}_{V_\Lambda} \tauo \eE^{2\pi\ii \tau R(L_0)} \, . \labl{73}

In the rest of this section we will show that the only non-vanishing
contribution to the trace in \erf{73} comes from the highest weight vector,
thereby proving theorem 2. 
This vector is never a null vector, so that this statement holds
both for the Verma module and for the \irmod. Thus we have
  \be  \chil(t,\tau) = \Chil(t,\tau) 
  = \eE^{2\pi\ii tk_\Lambda} \eE^{2\pi \ii \tau \Delta_\Lambda} \, , \labl{cc1}
where $\Delta_\Lambda$ denotes the eigenvalue of $L_0$ on the highest 
weight vector.

\binternal{
One recovers the shift found in \cite{scya6}. Integrable symmetric weights
have levels $\kv$ which are multiples of $N$, $\kv=N\ell$. There is a single
symmetric weight with $\Lambdab= \ell \rhob$. It has conformal weight 
$\Delta_{\Lambda}= (N^2-1)\ell(\ell+2)/24(\ell+1)$, while the Virasoro
central charge is $c/24 = (N^2-1)\ell/24(\ell+1)$. This gives for the
modular anomaly $\Delta_{\Lambda}- c/24 = (N^2-1)\ell/24$, which is the value
of the shift given in \cite{scya6}. 
}\einternal

To show that only the highest weight vector contributes to the \omchar\ of the
Verma module, we label the positive real roots of $A_{N-1}^{(1)}$ 
in the following way. The positive roots of $A_{N-1}$ are
$\alpab ij{}:=\alphab^{(i)}+\alphab^{(i+1)}+\ldots+\alphab^{(j-1)}$ with
$1\le i<j\le N$. Then all positive real roots of $A_{N-1}^{(1)}$ are covered by
  \be \alpa ijn := \left\{ \begin{array}{lll}
  (\alpab ij{}, 0, n)  & {\rm for}& i<j\,, \\[.3em]
  (-\alpab ji{},0, n+1)& {\rm for}& i>j  \end{array} \right. \labl{alph}
with $1\le i\ne j\le N$ and $n\in\zetpluso$.

The outer automorphism acts on the positive roots as
  \be  \omT(\alpa ijn) = \alpa{i+1}{j+1}n \,; \labl{oijn}
here (as well as in some formul\ae\ below) for convenience the upper  
indices are considered as being defined only $\bmod N$.
Hence for any fixed $n$ there are exactly $N-1$ orbits of length  
$N$, which can be written as
  \be   \{ \alpa ijn \mid i-j={\rm const} \} \,, \ee
where ${\rm const}\in\{1,2,\ldots, N-1\}$.
It follows directly from the definition \erf{alph} that
  \be  \sum_{k=0}^{N-1} \alpa {i+k}{i+k+l}n = (0,0,Nn+l)   \ee
for $1\leq l\leq N-1$. Thus
the horizontal projection of the sum of the roots of each orbit vanishes,
and the grade of $\sum_{l=0}^{N-1}\omT^l(\alpa ijn)$ is $nN+j-i$.

\binternal{
The equation \erf{oijn} can be derived as follows. $\omtb$ acts on the 
horizontal projection of the simple roots as
  \be  \alphab^{(1)} \mapsto \alphab^{(2)} \mapsto \,\ldots\, \mapsto 
  \alphab^{(N-1)} \mapsto -\sum_{i=1}^{N-1} \alphab^{(i)} = - \thetab \,.  \ee
Hence for $1\leq i<j<N$ one has
  \be  \omT(\alpa ijn) \equiv \omT(\sum_{k=i}^{j-1} \alphab^{(k)}, 0, n) 
  = (\sum_{k=i+1}^{j} \alphab^{(k)}, 0, n) \equiv \alpa{i+1}{j+1}n \,, \ee
for $1\leq i<j=N$
  \be  \omT(\alpa iNn) \equiv \omT(\sum_{k=i}^{N-1} \alphab^{(k)}, 0, n) 
  = (\sum_{k=i+1}^{N-1} \alphab^{(k)} - \sum_{k=1}^{N-1}  
\alphab^{(k)}, 0, n+1)
  = (- \sum_{k=1}^{i} \alphab^{(k)}, 0, n+1) \equiv \alpa{i+1}1n  \,,\ee
for $1\leq j<i<N$
  \be  \omT(\alpa ijn) \equiv \omega(-\sum_{k=j}^{i-1}  
\alphab^{(k)}, 0, n)
  = (- \sum_{k=j+1}^{i} \alphab^{(k)}, 0, n) \equiv \alpa{i+1}{j+1}n  
 \,, \ee
and for $1\leq j<i=N$
  \be  \omT(\alpa Njn) \equiv \omT(-\sum_{k=j}^{N-1} \alphab^{(k)},  
0, n+1)
  = (-\sum_{k=j+1}^{N-1} \alphab^{(k)} + \sum_{k=1}^{N-1}  
\alphab^{(k)}, 0, n)
  = (- \sum_{k=1}^{j} \alphab^{(k)}, 0, n) \equiv \alpa1{j+1}n  \,.\ee
}\einternal

On the step operators $H^i_n$ ($n\geq 1$) associated to lightlike roots,
$\omega$ acts as $H^i_n\mapsto H^{i+1}_n$ for $1\le i\le  N-2$, while it
sends $H^{N-1}_n$ to $-\sum_{k=1}^{N-1} H^k_n$
(compare \erf{H0} and \erf{om}). Thus the linear combinations
  \be  h_n^p := \sum_{j=1}^{N-1} (\zeta^{-pj} -1) H^j_n \, , \ee
with $n\geq 1$ and $p=1,2,\ldots,N-1$, and with $\zeta=\exp(2\pi\ii/N) $
a primitive $N$th root of unity, obey
  \be \begin{array}{lll}
  \omega(h_n^p) &=& \dsum_{j=1}^{N-2} (\zeta^{-pj} -1)\, H^{j+1}_n 
  - (\zeta^{-p(N-1)} -1) \dsum_{j=1}^{N-1} H^j_n   \\ {}\\[-.7em]
  &=&\dsum_{j=1}^{N-1} (\zeta^{-p(j-1)} -1 -\zeta^p +1) H^j_n
  = \zeta^p_{} h^p_n \,,  \end{array}  \ee
i.e.\ they are eigenvectors of $\omega$ to the eigenvalue $\zeta^p$.

According to the \pbw\ theorem a basis for $\U(\g_-)$ (the subalgebra of the
universal enveloping algebra $\U(\g)$ that is generated by the step operators
corresponding to negative roots of $\g=A_{N-1}^{(1)}$) can be described 
as follows. Consider an arbitrary, but definite ordering of the generators
of $\g_-$, starting, say, with the step operators corresponding to lightlike 
roots. Then for any sequences $\vn\equiv(n(m,j))$ and 
$\vnp\equiv(n'(m,j,l))$ which take values in the non-negative integers 
and for which only finitely many elements are different from zero,
we denote by $\taujl$ the element
  \be  \taujl:= \prod_{m=1}^\infty \Llb \prod_{j=1}^{N-1}  
  (h_{-m}^j)^{n(m,j)}_{} \prod_{l=0}^{N-1} (E^{-\alpa l{l+j}m})_{}
  ^{n'(m,j,l)} \Lrb \,    \labl{st1}
of $\U(\g_-)$. Here it is to be understood that the products are ordered 
according to the chosen ordering of the basis of $\g_-$. The \pbw\ theorem 
asserts that the set
  \be   \{\, \taujl \mid \vn,\,\vnp\, \} \ee
is in fact a basis of $\U(\g_-)$.

To compute the contribution of the state $v=\taujl\cdot v_\Lambda$
to the \omchar, we consider the standard filtration of $\U(\g_-)$; 
thus we denote by $\U_p$ the subspace of $\U(\g_-)$ that is
spanned by all elements of $\U(\g_-)$ which can be written as the 
product of $p$ or less elements of $\g_-$. Now
under $\omega$, the generator $\taujl$ is mapped to
  \be   \omega(\taujl) = \prod_{m=1}^\infty \Llb \prod_{j=1}^{N-1}
  (\zeta^j h_{-m}^j)^{n(m,j)}_{} \prod_{l=0}^{N-1}  (E^{-\alpa{l+1}{l+j+1}m})
  ^{n'(m,j,l)}_{} \Lrb  \, ,  \labl{st2}
and hence $\omega$ maps the subspace $U_p$ bijectively to itself. 

Moreover, for any $p$ elements $x_i$ of $\g_-$ and any permutation $\pi$ of 
$\{1,2,...\,,p\}$ we have
  \be x_1 x_2 \ldots x_p - x_{\pi(1)} x_{\pi(2)} \ldots x_{\pi(p)}
  \in \U_{p-1} \,  \ee
(it is sufficient to check this statement only for $\pi$ a transposition, 
in which case it follows from the properties of the commutator).
Now both $\taujl$ \erf{st1} and $\omega(\taujl)$ \erf{st2} are elements 
of $\U_p$ with 
  \be p = \sum_{m,j} n(m,j) + \sum_{m,j,l} n'(m,j,l) \, , \ee
but not of $\U_{p-1}$. In computing the trace of $\tauo$ we are therefore
allowed to reorder the factors in $\omega(\taujl)$ without changing the 
value of the trace, since reordering only introduces terms in $\U_{p-1}$. 
This shows that a state $\taujl\cdot v_\Lambda$ with $\taujl$ 
of the form \erf{st1} can only contribute to the  \omchar\ if the number
$n'(m,j,l)$ is constant on any orbit, or in other words, if it does 
not depend on $l$ at all. Correspondingly, we will write $n'(m,j)$ from now on. 

To proceed, it is convenient to drop the trivial dependence of the \tcha\
of the Verma module on the central element and shift $\Chil$ by 
$\eE^{-\tau \Delta_\Lambda}$; thus we define
  \be  \Chiltau := \eE^{-\tau \Delta_\Lambda} \Chil(0,\tau) \,. \labl{chiltau}
We will show that 
  \be  \Chiltau \equiv 1 \,. \labl{=1}
To see this, first note that a vector 
$\taujl\cdot v_\Lambda$ in the Verma module $V_\Lambda$ which fulfills the 
conditions formulated in the preceding paragraphs gives a contribution 
of $\eta q^n$ to $\Chiltau$, where $q\equiv\exp(2\pi\ii\tau)$,
  \be \eta := \prod_{m,j} (\zeta^j)^{n(m,j)} \qquad\mbox{and} \qquad 
  n:= \sum_{m,j} n(m,j) \cdot m + \sum_{m,j} n'(m,j) \cdot (Nm+j)    
  \,. \labl{pmj}
The function $\Chiltau$ just keeps track of the contributions \erf{pmj} 
from all states in $V_\Lambda$.
It is convenient to combine the contributions from all powers of any fixed
generator of $\g_-$; thus any $h^j_m$ yields a contribution of a factor of
  \be 1 + \zeta^j q^m + (\zeta^j q^m)^2 + \ldots = (1-\zeta^j  q^m)^{-1} \ee
to $\Chiltau$, while any orbit characterized by $j$ and $m$  
contributes a factor of
  \be 1+ q^{Nm+j} + (q^{Nm+j})^2 +\ldots = (1-q^{Nm+j})^{-1} \,. \ee
Thus
  \be  (\Chiltau)^{-1} = \prod_{m=0}^\infty \prod_{j=1}^{N-1}
  \llb (1-\zeta^j q^m) (1-q^{Nm+j}) \lrb \,. \ee
By arranging the terms in the first product differently this can be rewritten as
  \be  (\Chiltau)^{-1} = \prod_{m=0}^\infty \left\{ \prod_{j=1}^{N-1} \Llb
  (1-q^{Nm+j}) \cdot \prod_{j'=1}^{N-1} (1-\zeta^{j'}q^{Nm+j}) \Lrb
  \cdot \prod_{j=1}^{N-1} (1-\zeta^j q^{Nm}) \right\}  \, . \labl{519}
For any fixed $m$ and $j$ the term in the square brackets evaluates to
  \be \begin{array}{ll}  \dprod_{j'=0}^{N-1} (1-\zeta^{j'}q^{Nm+j}) \!\!\!&
  = q^{N(Nm+j)} \dprod_{j'=0}^{N-1} (q^{-Nm-j} -\zeta^{j'})  \\{}\\[-.8em]&
  = q^{N(Nm+j)}(q^{-N^2m-jN} -1) = 1-q^{N(Nm+j)}  \, . \end{array} \ee
Inserting this identity into \erf{519}, we find
  \be  (\ChilTau)^{-1} = \prod_{m=0}^\infty \prod_{j=1}^{N-1}
  \llb (1-\zeta^j q^{Nm})(1-q^{N(Nm+j)} ) \lrb = (\ChilnTau)^{-1}   \,. \ee
This functional equation for $\Chil$ implies that $\ChilTau$ is constant.
Evaluating the function for $q=0$ we thus find $\ChilTau\equiv 1$,  
as was claimed in \erf{=1}.

According to the definition \erf{chiltau} of $\Chiltau$, it then follows
immediately that the \tcha\ of the Verma module is given by \erf{cc1},
and hence the proof of theorem 2 is completed.

\Sect{Modular transformations}{modtr}

One important property of the untwisted affine \lie s (and of the twisted
affine \lie s $A^{(2)}_1$ and $\tilde B^{(2)}_n$) is that 
at any fixed value \kv\ of the level, the set of \ihwm s with dominant
integral \hw s carries a unitary \rep\ of the
twofold covering \slz\ of the modular group of the torus. To be precise, this
\rep\ does not act on the characters $\chii$ as we used them in the 
previous sections, but rather on the so-called {\em modified\/}
characters $\tchi$. {}From the table \erf{LS} we read off that if $\omega$ is 
a simple current automorphism, which is the case we are considering,
then also the characters of the orbit \lie\ $\gO$ -- and hence the \omchar s 
as well --  give rise to a unitary \rep\ of $SL(2,\zett)$. 
\futnot{This fact is not as surprising as it might seem: it is known 
\Cite{fugv2} that the automorphisms induced by elements of ${\cal Z}(\g)$ 
are precisely the {\em localizable} automorphisms, and there are arguments 
\Cite{mose.} from \cft\ that such automorphisms should lead to characters 
with good modular transformation properties.\\ This is something one must 
understand better -- it gives us new insight into D-invariants, hopefully}

The modified characters are defined as
  \be \tchi_\Lambda := \eE^{-s_\Lambda \delta} \chii_\Lambda \, , \labl{tchi}
where $\delta= \sum_{i=1}^r a_i \alpha^{(i)}$ is the element of the 
weight space that was defined in \erf{delta}, and where for any 
integrable highest weight $\Lambda$ of \g\ the number $s_\Lambda$
is the so-called modular anomaly
  \be s^{}_\Lambda := \frac1{(\thetab,\thetab)} \left( 
  \frac{(\Lambdab+\rhob,\Lambdab+\rhob)}{\kv+\gv} - 
  \frac{(\rhob,\rhob)}{\gv} \right )\, . \ee
For later reference, we also remark that using the strange formula 
${(\rhob,\rhob)}/{(\thetab,\thetab)} = \gv\dim\gb /{24}$ and the eigenvalues
  \be  \bar C_2(\Lambda) = (\Lambdab+2\rhob,\Lambdab) 
  = (\Lambdab+\rhob,\Lambdab+\rhob) - (\rhob,\rhob) \ee
of the second order Casimir operator $\ctwob$ of \gb,
one can rewrite the modular anomaly as
  \be s^{}_\Lambda = \frac{ \bar C_2(\Lambda)}{(\thetab,\thetab)(\kv+\gv)} 
  - \frac{\kv\dim\gb}{24\,(\kv+\gv)} = \Delta_{\Lambdab} - \frac c{24} \,. \ee
Here in the last step we implemented our convention that $(\thetab,\thetab)=2$,
and introduced the central charge
$c:= {\kv\dim\gb}/(\kv+\gv)$ of the Virasoro algebra 
and the conformal weight $\Delta_\Lambda:=\bar C_2(\Lambda)/2(\kv+\gv)$ of
the \hw\ $\Lambda$. 

\medskip
In the present section we treat the case where also the \olie\
$\gO$ is an untwisted affine \lie; the alternative case that 
$\gO=\tilde B_n\twtw$ will be described in section \ref{s.curlb}. 
The modular anomaly of the \omchar\ of \g\ is {\em not} the one of the 
ordinary character of \g, i.e.\ $\exp(-s_\Lambda \delta)$, but 
rather the pull back of the modular anomaly of the $\gO$-character. 
Defining $\hat\delta:= \projm(\deltaO)$ and
  \be  \hat s_\Lambda := \sO_{\proj \Lambda} \equiv \sO_{\LambdaO}
  = \frac{\CO(\LambdaO)}{2(\kvO+\gvO)} - \frac{\cO}{24}  \, , \ee
we can introduce {\em modified \omchar s} by
  \be \tchil := \eE^{-\hat s_\Lambda \hat\delta} \chil  \, . \ee
They are related to the modified characters of the orbit \lie\ $\gO$
by a relation analogous to \erf{1b}:
  \be \begin{array}{ll}
  \tchil(h)\!\!\!& = \eE^{-\hat s_\Lambda \hat\delta} \chil
  = \exp\llb \sO_{\proj \Lambda} (\projm\,(\deltaO))(h)\lrb\, \chiO_{\LambdaO}
  (\Pro h) \\[.5em] 
  &= \exp\llb \sO_{\proj \Lambda}\,\deltaO(\Pro h)\lrb\, \chiO_{\LambdaO}(\Pro h) 
  = \tilde {\chiO}_{\LambdaO} (\Pro h) \, . \end{array}\ee

We will now show that the difference between the 
modular anomaly of the \omchar\ and the one of the usual character is not as
big as one might expect. In fact, they differ by a constant which only depends
on the level of the weight $\Lambda$. First, the relation $\projm(\deltaO)
=\delta$ \erf{dtrans} shows that the two modifications differ only in
the value of the modular anomaly; closer inspection shows that the 
difference $\hat s_\Lambda 
- s_\Lambda$ is only a function of the level of the weight $\Lambda$. 
Since when analysing the modular transformation behavior one has to 
restrict oneself to weights at a fixed level, this shows that the modular 
anomalies differ only a constant. This constant is precisely the `shift' 
of the conformal weights that was observed in \cite{scya6}. 

More precisely, the relation between $\hat s_\Lambda$ and  $s_\Lambda$ 
is as follows. For affine \g\ the relation \erf{avo} between the dual 
Coxeter labels of $\g$ and $\gO$ 
implies that the dual Coxeter numbers $\gvo\equiv\sumIO i\avO i$ 
of \gO\ and $\gv\equiv\sumrO i\avi$ of \g\ are related by
  \be  N\,\gvo=\gv  \,.  \labl{ngg}
Further, for any symmetric \g-weight $\lambda$ of level $\kvl$, the 
level of the weight $\lambdaO=\proj(\lambda)$ is given by
  \be  \kvol\equiv \sumiO i\avO i\lambdaO^i=\Frac1N\sumi i\avi\lambda^i
  =\Frac1N\,\kvl  \,. \labl{kvo}

To compare $\hat s_\Lambda$ and  $s_\Lambda$, we also need
a relation between the second order Casimir operators of the horizontal
projection of the weights $\Lambda$ and $\LambdaO$. 
Thus consider two symmetric weights $\lambda,\mu\in\gstar0$. The
scalar product $(\lambdab,\mub)=\sumrE{i,j} \Gb ij \lambda^i \mu^j$
of their horizontal components can be written as
  \be  \begin{array}{l}  (\lambdab,\mub)=\sumto{i,j} \lambdab^i \mub^j
  \sumno{m,n} \dfrac{N_iN_j}{N^2}\, \Gb{\omdd mi}{\omdd nj}
  + \lambda^0\, \sumto{j}\, \mub^j \,
  \sumne{m}\sumno{n} \dfrac{N_j}{N}\, \Gb{\omdd m0}{\omdd nj} \\{}\\[-2mm]
  \hsp{14} + \mu^0\, \sumto{i}\, \lambdab^i \,
  \sumno{m}\sumne{n} \dfrac{N_i}{N}\, \Gb{\omdd mi}{\omdd n0} 
  + \lambda^0\mu^0\, \sumne{m,n} \Gb{\omdd m0}{\omdd n0} \,.
  \end{array}\labl{lm1}
Further, we have
  \be  \lambda^0= \kvl-\sumre i\avi\lambda^i = \kvl-\sumto i
  N_i\,\avi\lambda^i-(N-1)\,\lambda^0 \,, \labl{l01}
which owing to the relation \erf{avo} between the dual Coxeter labels of $\g$ 
and $\gO$ can be rewritten as
  \be   \lambda^0 = \kvol-\sumto i\, \avO i\lambdao^i = \lambdao^0  
  \,.\labl{l0o}
Inserting this identity into the right hand side of \erf{lm1} and using
the formula \erf{Go2} for $\overline{\GO}$, we can express the scalar 
product \erf{lm1} entirely in terms of the horizontal subalgebra of the 
orbit \lie. We obtain (compare \cite{scya6})
  \be  (\lambdab,\mub) = N\,(\lambdaob,\muob) + \coo\,\kvol\kvom \,, \labl{lm2}
where $\kvol$ and $\kvom$  are the levels of the $\gO$-weights $\lambdao$ 
and $\muo$, respectively. 
 \futnot{The inner product on the right hand side is the
one on the weight space of the orbit algebra $\gO$.
We do not have to introduce a separate notation for this product, 
since from  the notation for the vectors from which an inner product 
is formed it is already clear which inner product is meant.}
\binternal{
More explicitly:
  \be  \begin{array}{l}  (\lambdab,\mub)= N\!\!\! \sumto{i,j}\!\! \Gobb ij\, 
  \lambdaob{}^i \muob{}^j 
  + \coo\, \llb \!\!\sumto{i,j}\! \avO i\avO j\, \lambdaob{}^i \muob{}^j
  + \lambdao{}^0\!\! \sumto{j} \avO j\,\muob{}^j 
  + \muo{}^0\!\! \sumto{i} \avO i\, \lambdaob{}^i 
  + \lambdao{}^0\muo{}^0 \lrb \\{}\\[-2mm] \hsp{10}
  = N\,(\lambdaob,\muob) + \coo\,[(\lambdaob,\ttaob)(\muob,\ttaob)
     + \lambdao{}^0\, (\muob,\ttaob) + \muo{}^0\, (\lambdaob,\ttaob) +  
  \lambdao{}^0 \muo{}^0 ] \\[2.9mm] \hsp{10}
  = N\,(\lambdaob,\muob) + \coo\,\kvol\kvom \,,  \end{array}\labl{lm2ext}
}\einternal

Then in particular, the quadratic Casimir eigenvalue of a symmetric
highest \g-weight $\Lambda$ at level $\kv$ can be written as
  \be  (\Lambdab,\Lambdab+2\rhob)=N\,(\Lambdaob,\Lambdaob+2\rhoob)
  +\coo\,\kvo(\kvo+2\gvo) \,.  \labl{cas}
(In \cite{scya6}, this formula was obtained in a different guise, which
is obtained from the present one by \erf{ngg} and the identity
  \be  \coo\,\gvo = \frac N{12}\,(d-\brev d)   \,, \labl{cgg}
where $d$ and $\brev d$ are the dimensions of the simple \lie s \gb\ 
and $\gOb$, respectively.)
 \futnot{Inspection shows that \erf{cgg} is indeed valid for all \gb.}
Dividing \erf{cas} by $2(\kv+\gv)$ and using \erf{ngg} and \erf{kvo}, we then
obtain
a simple relation between the conformal dimensions of primary fields of the
$\g$ and $\gO$ WZW \cfts\ (at levels $\kv$ and $\kvo$, respectively), namely
  \be  \cd_\Lambda = \cdo_{{\LambdaO}} + \frac1{2N^2}\,\coo\kv\,
  (1+\frac{\gv}{\kv+\gv})  \,, \labl{cd1}
or, equivalently, using \erf{cgg}
  \be  \cd_\Lambda = \cdo_{{\LambdaO}} + \frac1{24}\, \llb\, 
  \frac{\kv}{\gv}\,(d-\brev d)+c -\brev c \,\lrb   \,,\mbox{~~} \labl{cd2}
This shows that the two modular anomalies in fact only differ by a 
(level-dependent) constant:
  \be  s_\Lambda = \cd_\Lambda - \frac c{24} 
  = \cdo_{{\LambdaO}} - \frac{\brev c}{24}\, + \frac1{24}\,\frac{\kv}{\gv}
  \,(d-\brev d) = \hat s_\Lambda + \frac1{24}\,\frac{\kv}{\gv}\,(d-\brev d) 
  \,. \labl{cd3}

The analysis above reproduces in particular the results concerning
the fixed point \cfts\ that have been obtained in \cite{scya6}. Note that
in \cite{scya6} the fixed point theory has been found by looking for those
affine \lie s $\gf$ for which a relation of the form \erf{cd3} between the 
conformal dimensions of the symmetric weights of \g\ and the weights of $\gf$ 
exists. Equation \erf{cd3} shows that the orbit \lie\ $\gO$, which was defined by
a folding procedure of the Cartan matrix, fulfills precisely these requirements.
Also note that from the explicit formul\ae\ \erf{AO} and \erf{Go2} it is by
no means manifest that the symmetrized Cartan matrix of the horizontal \olie\
is the inverse of the quadratic form matrix $\overline{\GO}$ as defined in
\erf{Go2}, which coincides with the result obtained in \cite{scya6} for the 
quadratic form matrix of $\gfb$; that this is nevertheless true can thus be 
seen as a non-trivial check of the identification of $\gO$ with $\gf$.

\Sect{Twisted orbit \lie s}{curlb}

When comparing the list of orbit \lie s in \erf{LS} with the list 
of `fixed point \cfts' as presented in \cite{scya6}, for the cases involving
the simple current automorphisms of order two of $\g=C_{2n}^{(1)}$ or 
$B_{n+1}\untw$ some additional explanations are in order. In these cases the 
\olie\ is $\gO = \tilde B_n^{(2)}$, while in \cite{scya6}
the fixed point theory was conjectured to be the $C_n\untw$ WZW theory
at level $\p$ if the level of $\g$ is $\kv = 2\p+1$.
For even level the spectrum could not be matched with any known 
\cft\ apart from a few special cases.
Based on a level-rank duality of $N=2$ superconformal coset models,
an \smat\ for the spectra at even levels was conjectured in \cite{fuSc2}.

In this section we explain how these observations fit together. For odd levels
$\kv=2\p+1$ of \g, we show that the \smat\ of $\gO=\tilde B_n^{(2)}$ at level 
$\kvo=\kv$ coincides (up to sign factors which are related to certain shifts
appearing in the application to fixed point resolution) with the \smat\ of 
$\gf:=C_n\untw$ at level $\p=(\kv-1)/2$. For even levels $\kv=2\p$ we prove the 
conjecture of \cite{fuSc2} related to level-rank duality. Note that 
the level of $\tilde B_n^{(2)}$ is defined with a conventional factor of two
\cite{KAc3},
  \be \kvlo = 2 \sum_{i\in\IO} {\brev a}_i^\Vee \lambdaO^i    \,, \ee
as compared to the formula \erf{kvl} of the untwisted case; this cancels 
the factor of $1/2$ that according to \erf{kvo} is present in the relation 
between $\kvo$ and \kv.

The modular \smat\ of $\gO=\tilde B_n^{(2)}$ at level $\kvo$
is described by the \kpf\ 
  \be  S_{\LambdaO,\LambdaO'} = \ii^{|\bar\Delta_+|} \left| \frac{M^\star}
  {(\kvO +\gvO)M} \right|^{-1/2} 
  \sum_{\bar w\in \Wnull} \epsilon(\bar w) \exp\Llb -2\pi\ii\,
  \frac{(\Lambdaob+\rhoob,
  \bar w(\Lambdaobp+\rhoob))}{\kvO+\gvO} \Lrb \, . \labl{kpf}
Such a formula holds for all untwisted affine \lie s, while among the
twisted algebras it is valid only for $\tilde B_n^{(2)}$ (here and below
we employ the convention that the twisted algebra $A_1^{(2)}$ is denoted
by $\tilde B_1\twtw$ and hence is included in the $\tilde B_n^{(2)}$ series;
also recall that in \cite{KAc3} these twisted algebras are denoted by
$A_{2n}\twtw$).
The notation used in \erf{kpf} is as follows. For $\gO=\tilde B_n^{(2)}$,
$\gnull$ is the unique diagram subalgebra isomorphic to $C_n$, while for 
untwisted \aff s it is the \hsa\ generated by the zero modes. The summation is
over the Weyl group $\Wnull$ of $\gnull$, and $\lambdaob$
is the projection of the \gO-weight $\lambdaO$ to $\gnull$. 
In the prefactor of the sum, $\bar\Delta_+$ is the set of positive roots
of $\gnull$, $M$ is the translation subgroup of the Weyl group of $\gO$, and
$M^\star$ its dual lattice. We also note that the dual Coxeter number 
of $\tilde B_n^{(2)}$ is $\gvO= 2n+1$, and that the translation lattice $M$ 
of the Weyl group of $\tilde B_n^{(2)}$ is isomorphic to the 
root lattice of the simple \lie\ $B_n$. Moreover \cite[corollary 6.4.]{KAc3},
if we normalize the invariant bilinear form of \g\ such that the longest
roots have length 2, the restriction of this invariant bilinear form to
$\tilde B_n^{(2)}$ is twice the normalized form of $C_n$. 

For concreteness, from now on we consider $\gO=\tilde B_n^{(2)}$ as the 
\olie\ of $\g=B_{n+1}\untw$. The case $\g=C_{2n}\untw$ is very similar.

\subsection{Odd level}

Let us first treat the case of $B_{n+1}\untw$ at odd level $\kv=2\p+1$. 
We start by showing
that the prefactors in the \kpf\ for $\gf=C_n^{(1)}$ at level $\p$ and 
$\gO=\tilde B_n^{(2)}$ at level $2p+1$ coincide.
The powers of $\ii$ are identical because $\gfb=C_n=\gOb$, i.e.\ the 
horizontal algebras coincide. Further, for $\tilde B_n^{(2)}$, $M$ is the root 
lattice $L$ of $B_n$, while for $C_n^{(1)}$ it is the coroot lattice 
$L^\Vee$ of $C_n$; these lattices
are proportional because $B_n$ and $C_n$ are dual \lie s. To determine the 
relative normalization, we notice that the simple coroots of $C_n$ are
$\gamma^{(i)^\Vee} = 2\gamma^{(i)}$ with length squared 4 for $i=1,...\,,n-1$,
and $\gamma^{(n)^\Vee} = \gamma^{(n)}$ with length squared 2, while the simple
roots of $B_n$ are $\beta^{(i)}$ with length squared 2 for $i=1,...\,,n-1$,
and $\beta^{(n)}$ with length 1. Hence 
  \be M( \tilde B_n^{(2)}) = L(B_n) = \Frac1{\sqrt2}\, L^\Vee(C_n) 
  = \Frac1{\sqrt2}\, M(C_n) \,,\qquad
  M^\star( \tilde B_n^{(2)}) = \sqrt2\,M^\star(C_n)    \, . \ee
Finally for $C_n^{(1)}$ we have $\kvf+\gvf=\p+(n+1)=\p+n+1$, while for 
$\tilde B_n^{(2)}$, $\kvO+\gvO=(2\p+1)+(2n+1) = 2(\p+n+1)$. Taking these 
results together, we find that the prefactors coincide as claimed.

Furthermore, the terms in the exponent coincide as well. As
already seen, the denominators differ by a factor of two; this is cancelled
by a factor of two in the numerator from the different normalization of the 
invariant bilinear form. Finally, the Weyl groups of $B_n$ and $C_n$ are
isomorphic so that also the summation is the same for both cases. Thus we
conclude that the formul\ae\ for the $S$-matrices coincide. 

However, we still have to determine the precise relation between the 
weights in the two descriptions. Now clearly, the mappings of the 
symmetric integrable highest weights of \g\ to
those of $\gO$ and the mapping to weights of $\gf$ that was
considered in \cite{scya6} are different. But even the restrictions
of these maps to the isomorphic subalgebras $\gOb=C_n$ and $\gfb=C_n$ do 
not coincide; rather, the two mappings are related as follows.

A weight $\lambda=(\lambda^0,\lambda^1,\ldots,\lambda^{n+1})$ of 
$\g=B_{n+1}\untw$ is symmetric if $\lambda^0=\lambda^1$.
 \futnot{what about the grade?}
The mapping of symmetric \g-weights to weights of $\gO=\tilde B_n^{(2)}$
reads 
  \be  \lambda \mapsto \lambdaO:=(\lambda^1,\lambda^2,\ldots,\lambda^{n+1})
  \,,  \labl{lo}
or in other words, $\lambdaO^i:=\lambda^{i+1}$ for $i=\oton$. The
restriction of this mapping to the diagram subalgebra $\gOb=C_n$ of
$\tilde B_n^{(2)}$ is then given (in the conventional labelling of the
$C_n$ \dyd, i.e.\ with the $n$th node corresponding to the long simple
root) by
  \be  \lambda \mapsto \lambdaob:=(\lambda^n,\lambda^{n-1},\ldots,\lambda^1)
  \,,  \labl{lob}
i.e.\ by $\lambdaob{}^i:=\lambda^{n-i+1}$ for $i=\oneton$.
On the other hand, the mapping to weights of $\gfb=C_n$ described in
\cite{scya6} is
  \be  \lambda \mapsto \lambdafb:=(\lambda^2,\lambda^3,\ldots,\lambda^n,
  \onehalf\,(\lambda^{n+1}-1)) \,,  \labl{lfb}
i.e.\ $\lambdafb{}^i:=\lambda^{i+1}$ for $i=\onetonm$ and
$\lambdafb{}^n:=\onehalf\,(\lambda^{n+1}-1)$. Extending this map to the
affinization $\gf=C_n\untw$ of $\gfb$ at level $p$ one has
$\lambdaf{}^i=\lambdafb{}^i$, with $\lambdafb$ as defined in \erf{lfb}, for
$i=\oneton$, supplemented by
  \be  \lambdaf{}^0 = \p - \sum_{i=1}^n \lambdaf{}^i  
  = \onehalf\,((\kv-1)-2\sum_{i=2}^n \lambda^i -(\lambda^{n+1}-1) )
  = \onehalf\, (\lambda^0+\lambda^1)= \lambda^1 \, . \ee
These relations, as well as the analogous mapping from $\g=C_{2n}\untw$ to
$\tilde B_n^{(2)}$, are displayed in figure 1.
\addtocounter{figure}{1}

%figure2
\clearpage
\begin{figure}%[tbh]
\epsfysize=4cm
\begin{center} \leavevmode \epsfbox{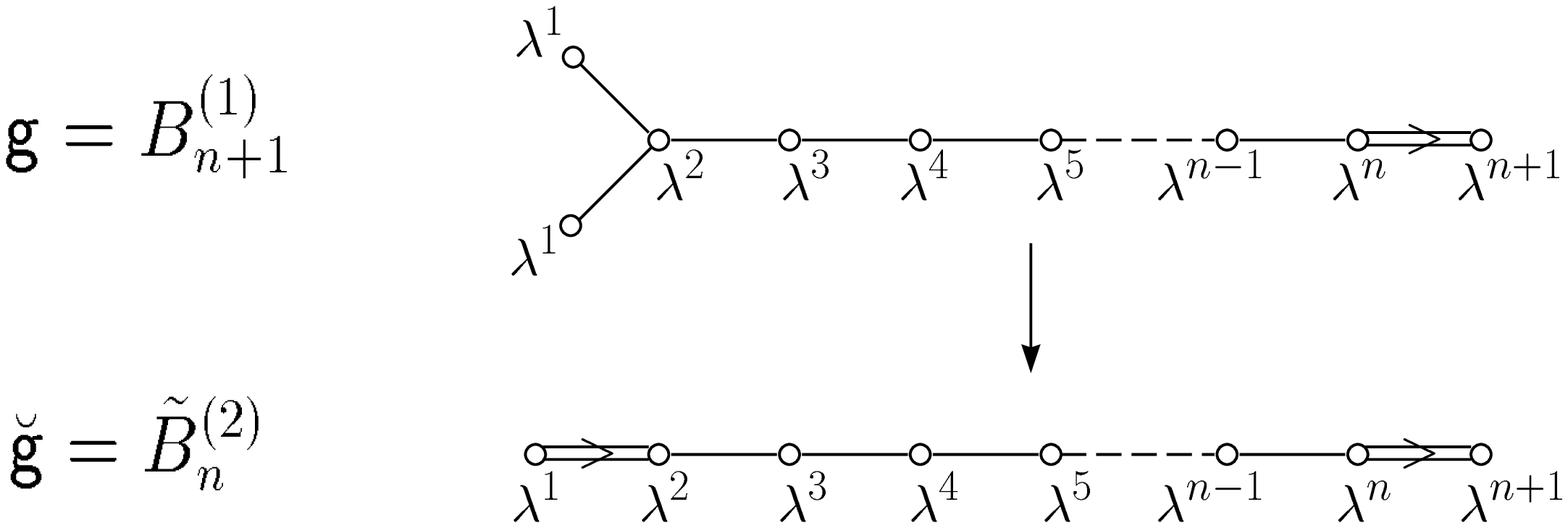} \end{center}

Figure 1a: Relation between symmetric weights (`fixed points')
of $B_{n+1}^{(1)}$ and weights of the \olie\ $\tilde B^{(2)}_n$.
\vskip 18mm

\epsfysize=3.5cm
\begin{center} \leavevmode \epsfbox{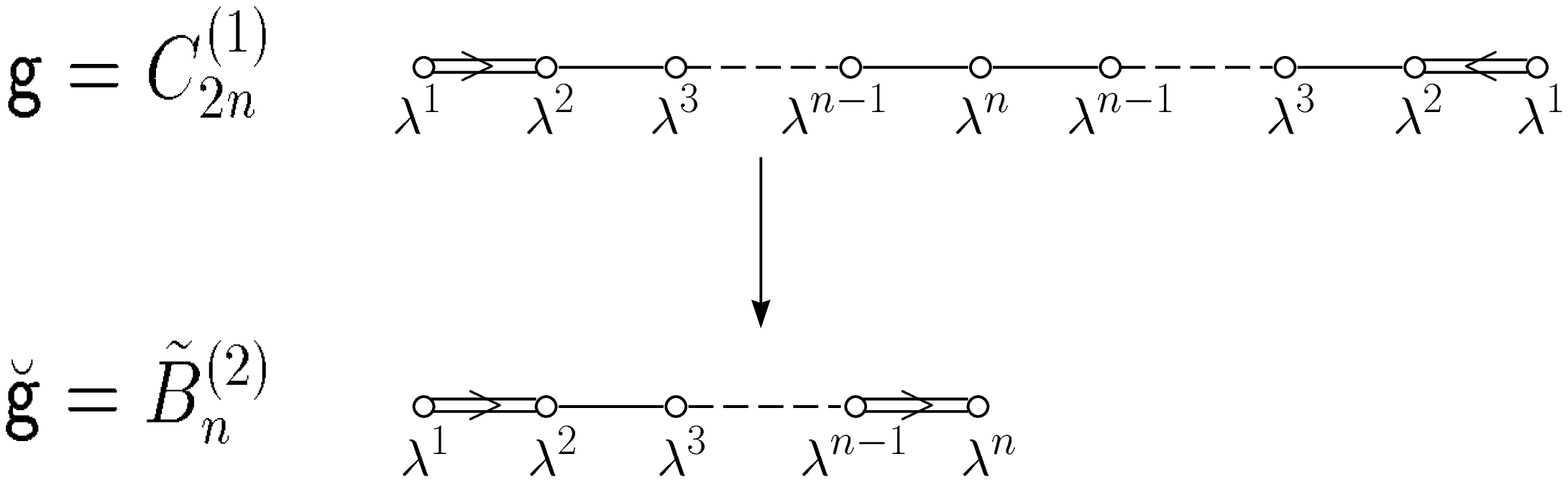} \end{center}

\vskip 3mm
Figure 1b: Relation between symmetric weights of $C_{2n}^{(1)}$ and
weights of the \olie\ $\tilde B^{(2)}_n$.
\vskip 18mm

\epsfysize=6.5cm
\begin{center} \leavevmode \epsfbox{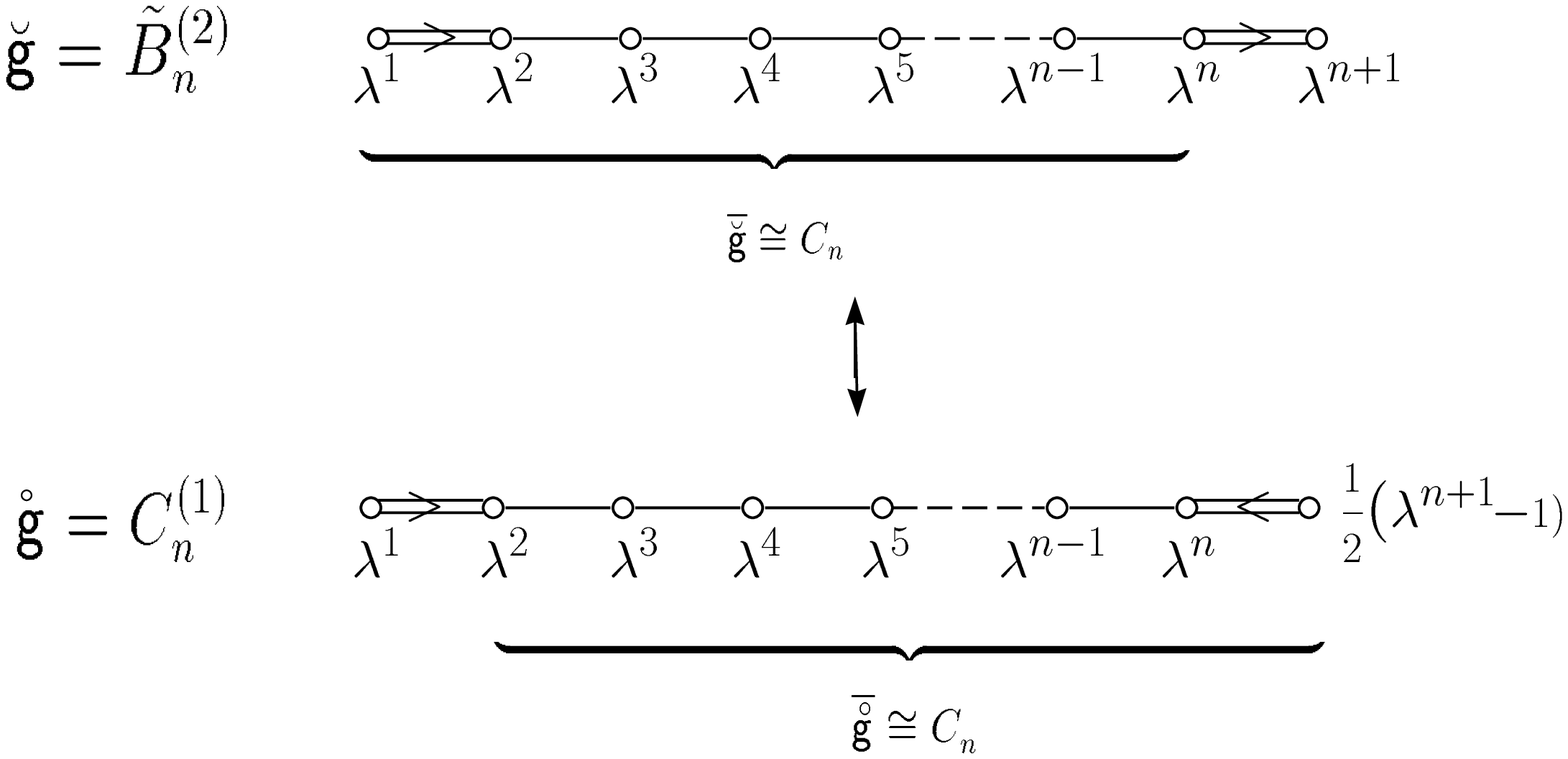} \end{center}

\vskip 3mm
Figure 1c: Map between weights of $\tilde B^{(2)}_n$ and
weights of $C_n^{(1)}$ at odd levels.
\vskip 18mm

\label{figure2} \end{figure}
\clearpage

Finally, we can also extend the map \erf{lob} to the affinization $C_n\untw$
of $\gOb$; this yields a weight $\muf$ of $C_n\untw$ with
$\muf{}^i=\lambdaob{}^i$ for $i=\oneton$ and zeroth component
  \be  \muf{}^0 = \p - \sum_{i=1}^n \muf{}^i
  = \onehalf\,((\kv-1)-2\sum_{i=1}^n \lambda^i
  = \onehalf\, (\lambda^{n+1}-1) \, . \ee
Combining these formul\ae, we learn that the $C_n\untw$-weights
$\lambdaf$ and $\muf$ are related by
  \be  \muf{}^i=\lambdaf{}^{n-i} \labl{mufl}
for $i=\oneton$. This means that $\lambdaf$ and $\muf$ are mapped to each
other by the non-trivial (simple current)
diagram automorphism of $C_n^{(1)}$.

We can now use the relation between \smat\ elements involving fields 
transforming into each other under a simple current automorphism that
was found in \cite{scya} to relate the \smat\ $\Sscya$ proposed in 
\cite{scya6} to the \smat\ $\SO$ of the orbit \lie\ $\gO$. We obtain
  \be  \SO_{\LambdaO,\LambdaOp} = (-1)^{Q(\LambdaO)+Q(\LambdaOp)
  +Q(\LambdaO_J)} \cdot \Sscya_{\!\Lambdaf,\Lambdafp}    \, , \ee
where $\LambdaO_J=\p\LambdaO_{(n)}$ is the weight of the simple current 
of $C_n^{(1)}$ and $Q(\LambdaO):= \sum_{j=1}^n j\,\LambdaO{}^j$
is the so-called monodromy charge of $\LambdaO$ \wrt the simple current,
which coincides modulo 2 with the conjugacy class of the $C_n$-weight
$\Lambdaob$. Thus the two modular $S$-matrices coincide up to sign 
factors, as claimed. These sign factors factorize into a global sign and 
signs associated to each row and column of the \smat. 

To compare this result with the description of the fixed point theory $\gf$ 
in \cite{scya6}, we note
that there the \smat\ $\Sscya$ was only defined up to a \onedim\ \rep\ of the
modular group; this allows for a global sign between $\Sscya$ and the \smat\
of $\gO$. Further, the second type of sign factors which depends on the \rep s 
can be compensated in the process of 
fixed point resolution by interchanging the role of the two
fields into which the fixed point is resolved, so that they
cannot be noticed in the fixed point resolution procedure at the level of
\rep s of the modular group either.

\subsection{Even level}

Consider now $B_{n+1}\untw$ at even level $\kv=2\p$. 
It will be convenient to describe the symmetric weights of $B_{n+1}\untw$
in terms of an orthogonal basis of the weight space of $B_{n+1}$. Thus for 
the weight $\lambda$ with Dynkin components $\lambda^i$ we introduce the numbers
  \be  l_i \equiv l_i(\lambda):= \sum_{j=i}^n \lambda^j +n+2-i 
  +\onehalf \lambda^{n+1} \ee
for $i=\onetonp$, which are the components of $\lambda+\rho$ in the 
orthogonal basis. We have $l_1>l_2>\ldots>l_{n+1} \ge1$. Furthermore, that the 
weight $\lambda$ is symmetric means that $\lambda^0=\lambda^1$, and hence
the level can be written as
  \be  2\p \equiv \lambda^0+\lambda^1+2 \sum_{j=2}^n \lambda^j + \lambda^{n+1}
   = 2 \sum_{j=1}^n \lambda^j + \lambda^{n+1} \,. \ee
This relation shows that for symmetric weights $\lambda^{n+1}$ must 
be even at even level, so that all the numbers $l_i$ are integers, and it
also implies that for a symmetric weight the number
  $   l_1 = \sum_{j=1}^n \lambda^j + \onehalf \lambda^{n+1} + n+1 =\p +n+1 $
is independent of $\lambda$. A symmetric weight can therefore be
characterized by a subset $M_{B_{n+1}}(\lambda)$
of $n$ numbers out of the set $M_{\p+n}:= \{ 1,2, \ldots,\p+n\}$. 

Let us now compute the weight with respect to the subalgebra $\gOb=C_n$
of the \olie\ $\gO$. For a $C_n$-weight $\muob=(\muo^i)$ the
components of $\muo+\rhoO$ in the orthogonal basis read
  \be m_i \equiv m_i(\muo) = \sum_{j=i}^n \muob^j + n+1 -i   \ee
for $i=\oneton$. As $\muo$ has level $\p$, we have
  $ n+\p \geq m_1 > m_2 > \ldots > m_n\ge 1$.
Thus these weights are again characterized by a subset of $n$ elements
of $M_{\p+n}$, which we denote by $M_{C_n}(\muo)$. 
The relations \erf{lo} and \erf{lob} between a symmetric $B_{n+1}\untw$-weight
$\lambda$ and the associated weights $\lambdao$ of $\tilde B_n\twtw$ and
$\lambdaob$ of $C_n$ then imply that $m_i(\lambdao)=\sum_{j=1}^{n+1-i} 
\lambda^j +n+1-i$ for $i=\oneton$, and hence
  \be  m_i(\lambdao) + l_{n-i+2}(\lambda) = \sum_{j=1}^n \lambda^j 
  + \onehalf \lambda^{n+1} + n+1 = \p+n+1  \, . \ee
This means that the sets $M_{C_n}$ and $M_{B_{n+1}}$ obtained from a weight
$\lambda$ are related by
  \be  M_{C_n}(\lambdao) = \{ \p+n+1-i \mid i\in M_{B_{n+1}}(\lambda) \} \, . 
  \labl{mcb}

With this information, we can express the \smat\ of $\gO$ as follows. In the 
prefactor ${\cal N}$ in \erf{kpf}, we now have $\kvo+\gvO= 2\p+(2n+1)= 2(\p+n+
\onehalf)$. Comparing this with the corresponding number $\p+(n+1)$ in
the prefactor for the \smat\ of $C_n^{(1)}$ at level $\p$, we see that
  \be  {\cal N}(\tilde B_n\twtw) = \left( \Frac{\p+n+1}{\p+n+1/2}\right)
  ^{\!-n/2} {\cal N}(C_n\untw)     \, .  \ee
With the known \smat\ of $C_n^{(1)}$ at level $\p$,
 \futnot{compare e.g.\ equation (5.6) of \cite{fuSc2}}
we then find
  \be  S_{\muo,\muop}(\tilde B_n\twtw) = \LLb \Frac{\p+n+1/2}{\p+n+1}\LRb 
  ^{\!n/2}\!\! \cdot S_{\muo,\muop}(C_n^{(1)}) = (-1)^{n(n-1)/2} 2^{n/2} 
  (\p+n+\onehalf)^{-n/2} \cdot\!\det_{{\scriptstyle \p\in M_{C_n}(\muo),} \atop 
  {\scriptstyle q\in M_{C_n}(\muop)}} \!\!{\cal M}_{pq}  \,, \labl{918}
where
  \be {\cal M}_{pq} := \sin \LLb  \Frac{\pi pq}{\p+n+1/2}\LRb   \labl{mmdef}
for all $p,q\in M_{\p+n}$.

This result will now be compared with the conjecture for the \smat\ 
obtained from level-rank duality. By level-rank duality, 
symmetric weights of $B_{n+1}\untw$ at level $2\p$  are mapped to a pair
of so-called `spinor non-symmetric simple current orbits' of weights of 
$D_\p\untw$ at level $2n+3$ \cite{fuSc2}; the latter are simple current orbits
which contain a $D_\p\untw$-weight $\nud$ with $\nud^\p\neq \nud^{\p-1}$. 
Again we characterize the weights $\nud$ by the components of $\nud+\rho$ 
in the orthogonal basis, i.e.\ by 
  \be  \begin{array}{l}
  n_i(\nud) := \sum_{j=i}^{\p-2} \nud^j +\onehalf\, (\nud^{\p-1} + \nud^\p) 
  +\p -i \quad {\rm for}\ i=\onetopmm\,, \\[.4em]
  n_{\p-1}(\nud) = \onehalf\,(\nud^{\p-1} + \nud^\p) +1 \,,\qquad 
  n_\p(\nud) = \onehalf\,(-\nud^{\p-1} + \nud^\p)   \, .   \end{array}  \ee
To characterize pairs of simple current orbits, we choose the unique 
representative of each pair of orbits which has $\nud^\p - \nud^{\p-1}
\in 2\zet_{\ge0}$ and $\nud^0\ge\nud^1$; for 
this representative, all $n_i$ are positive integers and $n_1>n_2 >\ldots>
n_\p\geq 0$. In fact, $n_\p\ge1$ because of $\nud^\p\neq \nud^{\p-1}$.
Moreover, as $\nud$ is spinor non-symmetric and at level $2n+3$, the integer
$n_1$ obeys
  \be n_1 = \sum_{j=1}^{\p-2} \nud^j +\onehalf\, (\nud^{\p-1} + \nud^\p) +\p -1
  = \onehalf\,(2n+3) + \onehalf\,(\nud^1-\nud^0) +\p-1 \le n+\p +\onehalf \,, \ee
implying that $n_1\le n+\p$. 
It follows that we can characterize each pair of spinor non-symmetric 
orbits by a subset $M_{D_\p}(\nud)$ of $\p$ elements of $M_{\p+n}$. 

In terms of this subset, the level-rank duality between symmetric weights
$\lambda$ of $B_{n+1}\untw$ at level $2\p$ and these orbits of $D_\p\untw$-weights
reads \cite{fuSc2}
  \be  M_{D_\p}(\nud) = \{ \p+n+1-j \mid j\not\in M_{B_{n+1}}(\lambda) \} \, . 
  \labl{mdb}
 \futnot{This is indeed a subset of $M_{\p+n}$, since $M_{B_{n+1}}(\lambda)$
always contains $l_1=\p+n+1$, so that the largest number in the complement is
$\p+n$.}
 Combining the results \erf{mcb} and \erf{mdb}, we find that the weights 
\wrt $C_n\untw$ and to $D_\p\untw$ that are associated to a symmetric weight of 
$B_{n+1}\untw$ are related by
  \be M_{D_\p}(\nud)\, \dot\cup\, M_{C_n}(\lambdao) = M_{\p+n} \,, \ee
where $\dot\cup$ denotes the disjoint union. 

In \cite{fuSc2} it was conjectured that, up to a phase, the \smat\
for the fixed point resolution for $B_{n+1}\untw$ is
  \be  \Sscya_{\!\Lambda,\Lambda'} = 2^{\p/2 -2} (\p+r-\onehalf)^{-\p/2}\cdot
  \det_{{\scriptstyle p\in M_{D_\p}(\Lambda),} \atop 
  {\scriptstyle q\in M_{D_\p}(\Lambda')}} \!\! {\cal M}_{pq} \, , \ee
with ${\cal M}_{pq}$ as defined in \erf{mmdef}. 
It is not difficult to verify that this matrix indeed coincides, up to sign
factors, with the \smat\ \erf{918} of $\gO$; to see this one has to employ 
Jacobi's theorem on determinants of submatrices of an invertible 
square matrix, together with the identities
  \be \det {\cal M} = (-1)^{(\p+n)(\p+n-1)/2} (\Frac{2\p+2n+1}4)^{(\p+n)/2}
  \,,\qquad {\cal M}^2 = \Frac{2\p+2n+1}4\, \one \,,    \ee
and the fact that ${\cal M}$ is symmetric (compare \cite{fuSc2}). 

\binternal{
Jacobi's theorem states that for any invertible square matrix $A$ and any
disjoint decompositions $I\dot\cup \bar I$ and
$J\dot\cup \bar J$ of the index set labelling the rows of the matrix, the
corresponding subdeterminants are related by
  \be  \det (A^{-1})^t_{IJ} = (-1)^{\Sigma_I+\Sigma_J} (\det A)^{-1} 
  (\det A)_{\bar I \bar J}          \, , \ee
where $\Sigma_I := \sum_{j\in I} j$.
The remaining steps are then as in \cite{fuSc2}.
}\einternal

We have thus completed the proof of the conjecture for the \smat\ that was
derived using level-rank duality. The conjecture for the \smat\ given
in \cite{fuSc2} was based on a resolution of fixed points at the level of 
\rep s of the modular group.
As we have remarked at the end of the previous subsection, any such 
conjecture is not sensitive to both a global sign of the \smat\ and to
multiplying corresponding rows and columns of the \smat\ with the same sign.
When comparing $\fS$ and $\SO$, we therefore did not pay attention to
such sign factors.

                                  \newpage

\ASect{$\Wh$ as a subgroup of $W$}{wwiso}

In subsection \ref{s.ww} we have already seen that $\mO_{ij}$ is a
divisor of $\hat m_{ij}$. In this appendix we show that $\hat m_{ij}$ is a
divisor of $\mO_{ij}$, or in other words, that
  \be       (\wh_i\wh_j)^{\mO_{ij}} =\eins \,, \ee
in the cases where $\mO_{ij}\in\{2,3,4,6\}$, i.e.\ when $\AOc ij\AOc ji
\in\{0,1,2,3\}$. Together, it then follows that $\hat m_{ij}=\mO_{ij}$
also in the cases; this completes the proof that the Weyl group $\WO$ 
of $\gO$ is isomorphic to the subgroup $\Wh$ of $W$.

\begin{figure}[tbh]
\epsfysize=10cm \begin{center} \leavevmode
\epsfbox{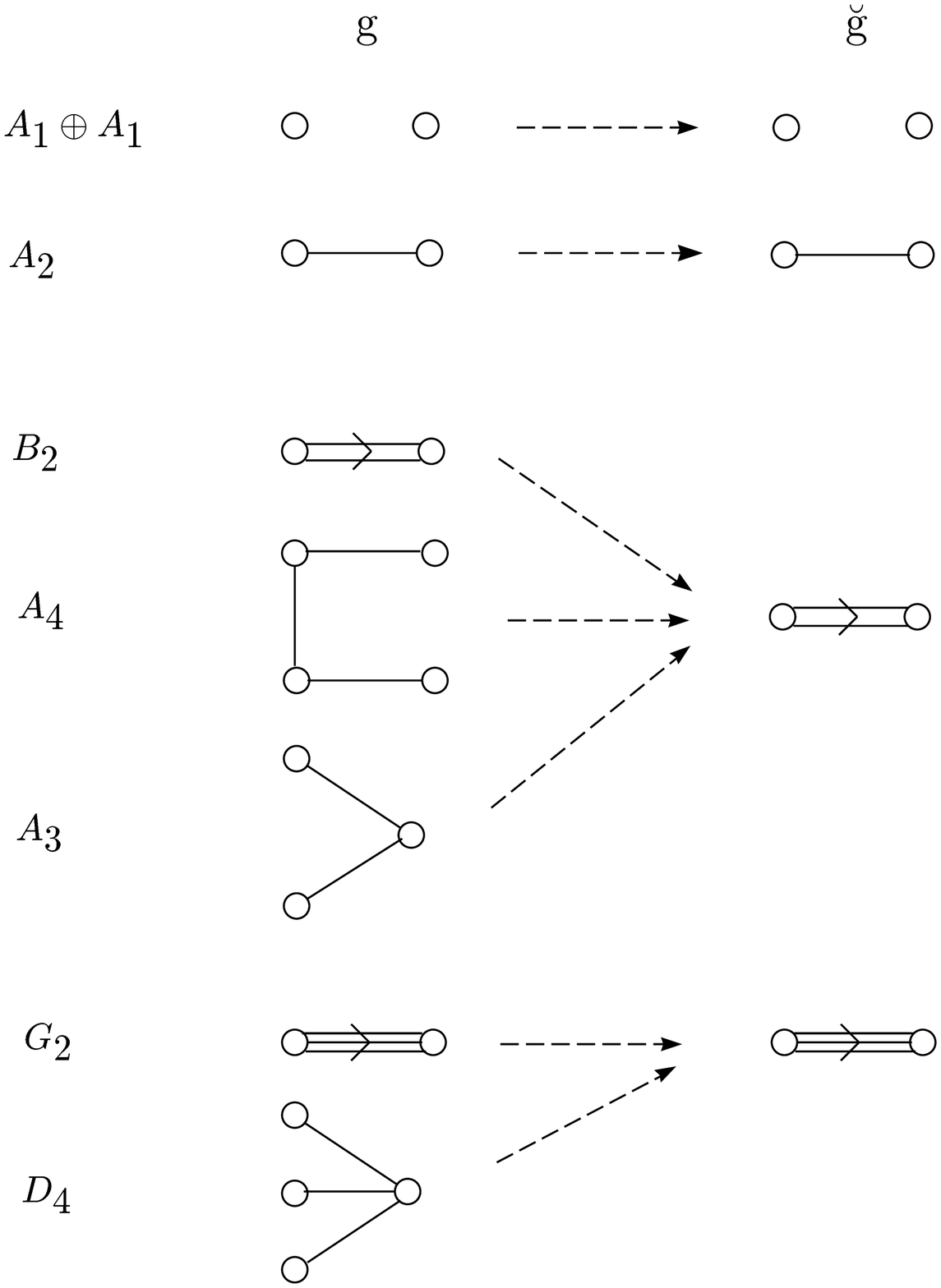}
\caption{The foldings of \dyd s with $\AOc ij\AOc ji\le3$.}
\label{figure1} \end{center} \end{figure}

Recall that the generators $\wh_i$ are defined by 
\erf{whi} and \erf{whi2} for $s_i=1$ and $s_i=2$, respectively. We will 
deal with the various cases separately; the corresponding restriction of 
the \dyd\ of $\g$ to the orbits of $i$ and $j$ is depicted in figure
\ref{figure1}.

Consider first the case $\mO_{ij}=2$. Then we have
  \be 0 = \AOc ij\AOc ji= s_i s_j\,\Frac{N_iN_j}{N^2}
  \sum_{l,l'=0}^{N-1} \Ac{\omD^l i}j \Ac{\omD^{l'}j}i \,. \labl{a1}
As $i\ne\omdd lj$ for all $l$,
all terms in the sum are non-negative; this implies that 
$\Ac{\omD^l i}j \Ac{\omD^{l'}j}i=0$ for all $l,l'$. Assume now that there is a 
value of $l$ such that $\Ac{\omD^l i}j$ is different from zero (i.e.\
negative). Then the fact that $\omD$ is an automorphism of $A$ implies that also
$\Ac i{\omD^{-l}j}<0$; since $A$ is a Cartan matrix, it follows that also
$\Ac{\omD^{-l} j}i<0$. This in turn implies that the term
$\Ac{\omD^l i}j \Ac{\omD^{-l}j}i$ gives a positive contribution to
$\AOc ij\AOc ji$, which is in contradiction with \erf{a1}.
Thus we learn that $\Ac{\omD^l i}j$ has to vanish for all $l$. Using again
the fact that $\omD$ is an automorphism of the Cartan matrix, we then find that
also $\Ac{\omD^l i}{\omD^{l'} j}$ vanishes for all values of $l$ and $l'$, which
implies that $w_{\omD^l i} w_{\omD^{l'} j}=  w_{\omD^{l'} j} w_{\omD^l i}$.
This relation implies that also $\whi \hat w_j = \hat w_j\whi$,
which shows that $\mh_{ij}=2=\mO_{ij}$ in this case.
(We also see that for $N_i=N_j$ the restriction of the \dyd\ of $A$ to 
the orbits of $i$ and $j$ consists out of $N_i$ copies of the situation 
$A_1\oplus A_1$ of figure \ref{figure1}, and analogously for $N_i\ne N_j$.)

In the remaining cases $\AOc ij\AOc ji\in\{1,2,3\}$, we need
either $\AOc ij=-1$ or $\AOc ji=-1$. Since the labels $i$ and $j$ appear
symmetrically in the definition of $\mO_{ij}$, 
we can assume without loss of generality that $\AOc ij =-1$. The relation 
$\AOc ij = s_i \sum_{l=0}^{N_i-1}\Ac{\omD^l i}j$ for $\AO$ then
implies that $s_i = 1$. Moreover, for any representative $\omdd mj$
of the orbit of $j$, the Cartan matrix element $\Ac{\omD^li}{\omdd mj}$ 
is different 
from zero only for a single value of $l$, for which it is equal to $-1$. 

Let us first deal with the case $s_j =2$. Then the product
  \be \AOc ij\AOc ji = 2 \sum_{l,l'}\Ac{\omD^l i}j \Ac{\omD^{l'} j}i \labl{a2}
is even, so that the only case we have to analyze is when it is equal to two,
and hence $\mO_{ij}=4$. As in subsection \ref{s.ww} we can assume that $N=2$;
then $s_j=2$ implies that $N_j=2$. Assume now first that $N_i=1$, i.e.\
$\omD i = i$; this implies that $\AOc ji= 2(\Ac ji +\Ac{\omD j}i)=  2(\Ac ji 
+ \Ac j{\omD i}) = 4\Ac ji$. Thus if $\AOc ji$ is non-zero, it is in fact
$\le-4$, which implies that $\AOc ij\AOc ji\geq 4$. This does not belong to the
cases we are investigating here, and hence we can assume that $N_i=2$. 

Now $s_j=2$ means that $\Ac j{\omD j} = \Ac{\omD j}j= -1$, while $s_i=1$ 
tells us that $\Ac i{\omD i} =\Ac{\omD i}i= 0$. Further, because of
$\AOc ij=\Ac ij+\Ac\omdi j = -1$, we can assume without loss 
of generality that $\Ac{\omD i}j=0$ and $\Ac ij =-1$.
The automorphism property of $\omD$ then implies that 
$\Ac i{\omD j} =0$ and $\Ac{\omD i}{\omD j}=-1$. As $A$ is a Cartan 
matrix, we then also have $\Ac j{\omD i}=0 =\Ac{\omD j}i$. 
To determine the matrix elements $\Ac ji=\Ac{\omD j}{\omD i}$ (which because
of $\Ac ij\ne0$ are non-zero), we observe that
$\AOc ji= 2(\Ac ji+\Ac{\omD j}i) = 2\Ac ji$ must be $\ge-3$ in order to 
yield $\AOc ij\AOc ji\leq 3$; thus $\Ac ji=\Ac{\omD j}{\omD i} =-1$, and we
are in situation $A_4$ of figure \ref{figure1}.

%We will label the two nodes as follows: $1:=i$, $2:=j$, $3:=\omD j$ and 
%$4:=\omD i$ so that the following relations hold:
%\be w_i w_{i+1} w_i = w_{i+1} w_i w_{i+1} \quad\mbox{and}\quad
%    w_i w_j = w_j w_i \quad\mbox{else} \, . \ee
%We then have to show that 
%\be \wh_a := w_1 w_4 = w_4 w_1     \quad\mbox{and}\quad 
%\wh_b := w_2 w_3 w_2 = w_3 w_2 w_3   \labl{a4}
%fulfill the relation
%\be (\wh_a \wh_b)^2 = (\wh_b \wh_a)^2 \, . \ee
%This follows in fact by applying the relations \erf{a4} repeatedly:
%\be \begin{array}{lll}
%(\wh_a \wh_b)^2 & = & \wf 14232 \wf 14232 \\
%&=&  \wf 41323 \, \wf 14232 
% =   \wf 43121 \, \wf 34232 \\
%&=&  \wf 43212 \, \wf 34323 
% =   \wf 43212 \, \wf 43423 \\
%&=&  \wf 43421 \, \wf 23423
% =   \wf 34321 \, \wf 23243 \\
%&=&  \wf 34321 \, \wf 32343
% =   \wf 34321 \, \wf 32434 \\
%&=&  \wf 34323 \, \wf 12434
% =   \wf 34232 \, \wf 12434 \\
%&=&  \wf 32432 \, \wf 12434
% =   \wf 32431 \, \wf 21434 \\
%&=&  \wf 32143 \, \wf 24314
% =   \wf 32143 \, \wf 42314 \\
%&=&  \wf 32134 \, \wf 32314
% =   \wf 32314 \, \wf 32314 \\
%&=&  (\wh_b \wh_a)^2 \, . 
%\end{array}  \ee

Having found these Cartan matrix elements, we know that
  \be  w_i w_j w_i = w_j w_i w_j \,, \qquad
   w_\omdi w_\omdj w_\omdi = w_\omdj w_\omdi w_\omdj \,, \qquad
   w_j w_\omdj w_j = w_\omdj w_j w_\omdj  \ee
and
  \be  w_i w_\omdj = w_\omdj w_i \,, \qquad w_j w_\omdi = w_\omdi w_j 
   \,, \qquad w_i w_\omdi = w_\omdi w_i \,. \ee
Applying these relations repeatedly, we obtain
  \be \begin{array}{ll} 
    (\wf i\omdi j\omdj j)^2 \!\!
  &= \wf \omdi i\omdj j\omdj  \, \wf i\omdi j\omdj j 
   = \wf \omdi \omdj iji      \, \wf \omdj \omdi j\omdj j      \\[.3em]
  &= \wf \omdi \omdj jij      \, \wf \omdj \omdi \omdj j\omdj  
   = \wf \omdi \omdj jij      \, \wf \omdi \omdj \omdi j\omdj  \\[.3em]
  &= \wf \omdi \omdj \omdi ji \, \wf j\omdj \omdi j\omdj  
   = \wf \omdj \omdi \omdj ji \, \wf j\omdj j\omdi \omdj       \\[.3em]
  &= \wf \omdj \omdi \omdj ji \, \wf \omdj j\omdj \omdi \omdj  
   = \wf \omdj\omdi\omdj j\omdj\,\wf ij\omdi \omdj \omdi       \\[.3em]
  &= \wf \omdj \omdi j\omdj j \, \wf ij\omdi \omdj \omdi  
   = \wf \omdj j\omdi \omdj i \, \wf ji\omdi \omdj \omdi       \\[.3em]
  &= \wf \omdj ji\omdi \omdj  \, \wf j\omdi \omdj i\omdi  
   = \wf \omdj ji\omdi \omdj  \, \wf \omdi j\omdj i\omdi       \\[.3em]
  &= \wf \omdj ji\omdj \omdi  \, \wf \omdj j\omdj i\omdi  
   = \wf \omdj j\omdj i\omdi  \, \wf \omdj j\omdj i\omdi       \\[.3em]
  &= (\wf j\omdj ji\omdi )^2  \,.
  \end{array}  \ee
Thus the generators $\wh_i:=w_iw_\omdi$ and $\wh_j:=w_jw_\omdj w_j$ of $\Wh$ satisfy
$(\wh_i\wh_j)^2=(\wh_j\wh_i)^2$, or what is the same,
  \be  (\wh_i\wh_j)^4=\id \,, \ee
which is the relation we need, since $\AOc ij\AOc ji= 2$.

Let us now turn to the case $s_j=1$. As the number $t$ of those values 
of $l$ for which $\Ac{\omD^l j}{\omdd mi}$ is non-zero is the same for any 
representative $\omdd mi$ of the orbit of $i$, we then find that
  \be \AOc ji = \sum_{l=0}^{N-1}\Ac{\omD^l j}i \leq-t  \, , \ee
which shows that $t$ can only have the values $1,\;2$ or $3$. Note that we
still have $\AOc ij=-1$, so that $N_i= N_j /t$ for each of these values of $t$. 

We can now classify the possible situations through the restriction of the
\dyd\ of \g\ to the orbits of $i$ and $j$. For $t=1$, we have
$N_i=N_j$, and the restriction of the \dyd\ to the two
relevant orbits consists of $N_i$ disconnected copies of the \dyd\ of
either $A_2$, $B_2\equiv C_2$, or $G_2$, according to whether 
$\AOc ij\AOc ji$ is $1,\;2$ or $3$. These algebras have also been used 
to denote the corresponding folding in figure \ref{figure1}. For $t=2$, 
there is only one possibility which satisfies all required constraints, 
namely that one has $N_i$ disconnected 
copies of the \dyd\ of $A_3$, such that the middle node lies on the orbit of 
$i$ while the two extremal nodes lie on the orbit of $j$.
Finally, for $t=3$ there are $N_i$ disconnected copies of the \dyd\ of 
$D_4$, with the middle node lying on the orbit of $i$ and the three extremal 
nodes on the orbit of $j$; this corresponds to the last case in figure
\ref{figure1}.

We will deal with the different values of $t$ consecutively. All cases 
with $t=1$ can be treated simultaneously. In these cases the orbits of 
$i$ and $j$ have the same length. We can therefore label the simple 
reflections in $W$ associated to the elements of these orbits
as follows. We define $r_l:= w_{\omD^l i}$ for $l=1,...\,, N_i$, and then set
$r'_l := w_{\omD^{l'}j}$, with $l'$ chosen such that $r'_l$ commutes with
all $r_l$ for $l\neq l'$. With this notation we have
  \be  r_l r'_m = r'_m r_l \quad\mbox{for}\ l\neq m\, , \qquad
  (r_l r'_l)^{\mO_{ij}} = \eins \, . \ee
Moreover, it follows from $s_i=s_j=1$ that for all $l,m$ the reflections
of the same `type' commute, $r_lr_m=r_mr_l$ and $r'_lr'_m=r'_mr'_l$.
Using these relations, we find that
  \be (\wh_i \wh_j)^{\mO_{ij}} 
  = (\Prod{l=0}{N_i-1} r_l \cdot \Prod{m=0}{N_j-1} r'_m)^{\mO_{ij}}
  = \prod_{l=0}^{N_j-1}(r_l r'_l)^{\mO_{ij}} = \eins  \ee
as required.

Next consider the case $t=2$. Then we have $N_i$ commuting
copies of $A_3$. By similar arguments as in the $t=1$ case, we
can restrict our attention to just one of these copies, 
% In this case we choose a
% numbering of the three simple roots under consideration such that 
% \be w_1 w_2 = w_2 w_1, \quad
%     w_0 w_i w_0 = w_i w_0 w_i \quad\mbox{for} \quad i=1,2  \, .\ee
% We have then $\wh_0:=w_0$ and $\wh_1:= w_1 w_2$, and want to show that
% $(\wh_0\wh_1)^2 = (\wh_1\wh_0)^2$. This is indeed the case:
% \be \begin{array}{lll}
% (\wh_0\wh_1)^2 &=& \wfs 012012 
%  =  \wfs 012021 
%  =  \wfs 010201 \\
% &=& \wfs 101201
%  =  \wfs 102101
%  =  \wfs 102010 \\
% &=& \wfs 120210
%  =  (\wh_1\wh_0)^2 \,.\end{array}  \ee
and we can assume that the labelling is such that
the relevant representatives of the orbits of $i$ and $j$ are $i$ and 
$j$ themselves together with $\omdj$. Then we have the relations
  \be  w_j w_\omdj = w_\omdj w_j \,,\qquad
  w_i w_j w_i = w_j w_i w_j \,, \qquad
  w_i w_\omdj w_i = w_\omdj w_i w_\omdj \,, \ee  
which imply in particular
  \be \begin{array}{ll}
     (w_i w_j w_\omdj)^2 \!\!
  &= \wfs ij\omdj i\omdj j = \wfs iji\omdj ij = \wfs jij\omdj ij \\[.4em]
  &= \wfs ji\omdj jij      = \wfs ji\omdj iji = \wfs j\omdj i\omdj ji
   = (w_j w_\omdj w_i)^2 \,. \end{array}  \ee
Thus $\wh_i:=w_i$ and $\wh_j:=w_jw_\omdj$ satisfy
$(\wh_i\wh_j)^2=(\wh_j\wh_i)^2$, i.e.\ $(\wh_i\wh_j)^4=\id$ as required
by $\AOc ij\AOc ji = t =2$.

Finally, for $t=3$ the calculation is similar, though somewhat lengthier. 
There are $N_i$ commuting copies of $D_4$, and we
can restrict ourselves to one of these copies, with the
labels of its nodes being $i$ for the middle node and $j$, $\omdj$ and
$\omdd2j$ for the others. The associated simple reflections of $W$ satisfy
$w_i w_{\omdd mj} w_i = w_{\omdd mj} w_iw_{\omdd mj}$ for $m=0,1,2$, and
$w_{\omdd lj}w_{\omdd mj}=w_{\omdd mj}w_{\omdd lj}$ for $l,m=0,1,2$;
repeated use of these relations yields
 % \be \begin{array}{ll}
 %    (w_0 w_1 w_2 w_3)^3 \!\!
 % &= \wfs 012030 \, \wfs 120213  \\
 % &= \wfs 012030 \, \wfs 102013  \\
 % &= \wfs 012031 \, \wfs 012013  \\
 % &= \wfs 021013 \, \wfs 012013  \\%%
 % &= \wfs 020103 \, \wfs 012013  \\
 % &= \wfs 202103 \, \wfs 012013  \\%%
 % &= \wfs 202103 \, \wfs 020103  \\
 % &= \wfs 201202 \, \wfs 302103  \\
 % &= \wfs 201020 \, \wfs 302103  \\
 % &= \wfs 210120 \, \wfs 302103  \\%%
 % &= \wfs 210123 \, \wfs 032103  \\
 % &= \wfs 120123 \, \wfs 021030  \\%%
 % &= \wfs 120130 \, \wfs 201030  \\
 % &= \wfs 120130 \, \wfs 210130  \\%%
 % &= \wfs 120310 \, \wfs 120130  \\
 % &= \wfs 120301 \, \wfs 020130  \\
 %  = (w_1 w_2 w_3 w_0)^3 \,. \end{array}  \ee
  \be \begin{array}{l}
     (w_i w_j w_\omdj w_\omddj)^3 
   = \wfs ij\omdj i\omddj i     \, \wfs j\omdj i\omdj j\omddj   \\[.2em] \hsp{9}
   = \wfs ij\omdj i\omddj i     \, \wfs ji\omdj ij\omddj       %\\[.2em] \hsp{9}
   = \wfs ij\omdj i\omddj j     \, \wfs ij\omdj ij\omddj        \\[.2em] \hsp{9}
   = \wfs i\omdj iji\omddj      \, \wfs ij\omdj ij\omddj       %\\[.2em] \hsp{9}
   = \wfs \omdj i\omdj ji\omddj \, \wfs i\omdj iji\omddj        \\[.2em] \hsp{9}
   = \wfs \omdj ij\omdj i\omdj  \, \wfs \omddj i\omdj ji\omddj %\\[.2em] \hsp{9}
   = \wfs \omdj iji\omdj i      \, \wfs \omddj i\omdj ji\omddj  \\[.2em] \hsp{9}
   = \wfs \omdj jij\omdj \omddj \, \wfs i\omddj \omdj ji\omddj %\\[.2em] \hsp{9}
   = \wfs j\omdj ij\omddj i     \, \wfs \omdj iji\omddj i       \\[.2em] \hsp{9}
   = \wfs j\omdj i\omddj ji     \, \wfs j\omdj ij\omddj i      %\\[.2em] \hsp{9}
   = \wfs j\omdj i\omddj ij     \, \wfs i\omdj ij\omddj i       \\[.2em] \hsp{9}
   = (w_j w_\omdj w_\omddj w_i)^3 \,. \end{array}  \ee
Thus $\wh_i:=w_i$ and $\wh_j:=w_jw_\omdj w_\omddj$ satisfy
$(\wh_i\wh_j)^3=(\wh_j\wh_i)^3$, i.e.\ $(\wh_i\wh_j)^6=\id$, which is
again the required Coxeter relation for $\AOc ij\AOc ji = t =3$.

\newpage %\vskip4em

  \newcommand{\wb}{\,\linebreak[0]} \def\wB {$\,$\wb}
  \newcommand{\Bi}[1]    {\bibitem{#1}}
  \newcommand{\Erra}[3]  {\,[{\em ibid.}\ {#1} ({#2}) {#3}, {\em Erratum}]}
  \newcommand{\BOOK}[4]  {{\em #1\/} ({#2}, {#3} {#4})}
  \newcommand{\vypf}[5]  {\ {\sl #5}, {#1} [FS{#2}] ({#3}) {#4}}
  \renewcommand{\J}[5]     {\ {\sl #5}, {#1} {#2} ({#3}) {#4} }
\def\jf            {J.\ Fuchs}
 \def\anop  {Ann.\wb Phys.}
 \def\foph  {Fortschr.\wb Phys.}
 \def\hepa  {Helv.\wb Phys.\wB Acta}
 \def\ijmp  {Int.\wb J.\wb Mod.\wb Phys.\ A}
 \def\jopa  {J.\wb Phys.\ A}
 \def\npbF  {Nucl.\wb Phys.\ B\vypf}
 \def\npbp  {Nucl.\wb Phys.\ B (Proc.\wb Suppl.)}
 \def\nuci  {Nuovo\wB Cim.}
 \def\nupb  {Nucl.\wb Phys.\ B}
 \def\phlb  {Phys.\wb Lett.\ B}
 \def\comp  {Com\-mun.\wb Math.\wb Phys.}

 \def\A       {Algebra}
 \def\alg     {algebra}
 \def\Be     {{Berlin}}
 \def\BIR    {{Birk\-h\"au\-ser}}
 \def\Ca     {{Cambridge}}
 \def\CUP    {{Cambridge University Press}}
 \def\furu    {fusion rule}
 \def\GB     {{Gordon and Breach}}
 \newcommand{\inBO}[7]  {in:\ {\em #1}, {#2}\ ({#3}, {#4} {#5}),  p.\ {#6}}
 \def\Infdim  {Infinite-dimensional}
 \def\NY     {{New York}}
 \def\Q       {Quantum\ }
 \def\qg      {quantum group}
 \def\Rep     {Representation}
 \def\SV     {{Sprin\-ger Verlag}}
 \def\syms    {sym\-me\-tries}
 \def\wzw     {WZW\ }

 \version\versionno 
\messs\messs\message{ While this paper is rather long, there are still some} 
\message{ technical details which are not presented, but may be}
\message{ useful to the reader. If you wish to include these details,}
\message{ put the parameter `extension' to 1; this is done by}
\message{ un-commenting the line 73 of the LaTeX source file.}\messs\messs
\end{document}